\documentclass[a4paper,twocolumn,11pt,unpublished]{quantumarticle}
\pdfoutput=1
\usepackage[utf8]{inputenc}
\usepackage[english]{babel}
\usepackage[table]{xcolor}
\usepackage[T1]{fontenc}
\usepackage{amsmath}
\usepackage{array}
\usepackage{subfigure}
\usepackage{amssymb}
\usepackage{tikz}
\usepackage{diagbox} 
\usepackage{lipsum}
\usepackage{graphicx}
\usepackage{epsfig}
\usepackage{epstopdf}
\usepackage{braket}
\usepackage{comment}
\usepackage{hyperref}
\usepackage{amsthm}
\usepackage{enumitem}
\usepackage{dsfont}
\usepackage{bbold}
\usepackage{bm}
\usepackage{eurosym}
\usepackage{diagbox}
\usepackage[numbers,sort&compress]{natbib}
\usepackage[ruled,vlined]{algorithm2e}

\newcommand{\be}{\begin{equation}}
\newcommand{\ee}{\end{equation}}
\newcommand{\ba}{\begin{aligned}}
\newcommand{\ea}{\end{aligned}}

\newcommand{\R}{\mathbb{R}}
\newcommand{\bc}{\begin{center}}
\newcommand{\ec}{\end{center}}
\newcommand{\beq}{\begin{equation}}
\newcommand{\eeq}{\end{equation}}
\newcommand{\beqq}{\begin{equation*}}
\newcommand{\eeqq}{\end{equation*}}
\newcommand{\beqa}{\begin{align}}
\newcommand{\eeqa}{\end{align}}
\newcommand{\barr}{\begin{array}}
\newcommand{\earr}{\end{array}}
\newcommand{\bi}{\begin{itemize}}
\newcommand{\ei}{\end{itemize}}

\newcommand{\E}{\mathbb{E}}
\newcommand{\C}{\mathbb{C}}

\newtheorem{lem}{Lemma}
\newtheorem{theo}{Theorem}

\newtheorem{pb}{Problem}

\newtheorem{protocol}{Protocol}
\DeclareMathOperator{\poly}{poly\,}

\DeclareMathOperator{\Tr}{Tr}

\DeclareMathOperator{\N}{\mathbb{N}}
\DeclareMathOperator{\I}{\mathbb{I}}

\definecolor{quantum}{HTML}{53257F}
\hypersetup{
    colorlinks = true,
    linkcolor = quantum,
    citecolor = quantum
    }

\begin{document}

\title{An efficient quantum state verification framework and its application to bosonic systems}

\author{Varun Upreti}
\affiliation{DIENS, \'Ecole Normale Sup\'erieure, PSL University, CNRS, INRIA, 45 rue d’Ulm, Paris, 75005, France}
\author{Ulysse Chabaud}
\email{ulysse.chabaud@inria.fr}
\orcid{0000-0003-0135-9819}
\affiliation{DIENS, \'Ecole Normale Sup\'erieure, PSL University, CNRS, INRIA, 45 rue d’Ulm, Paris, 75005, France}

\maketitle

\onecolumn

\begin{abstract}
Modern quantum devices are highly susceptible to errors, making the verification of their correct operation a critical problem. Usual tomographic methods rapidly become intractable as these devices are scaled up. In this paper, we introduce a general framework for the efficient verification of large quantum systems. Our framework combines robust fidelity witnesses with efficient classical post-processing to implement measurement back-propagation. We demonstrate its usefulness by focusing on the verification of bosonic quantum systems, and developing efficient verification protocols for large classes of target states using the two most common types of Gaussian measurements: homodyne and heterodyne detection. Our protocols are semi-device independent, designed to function with minimal assumptions about the quantum device being tested, and offer practical improvements over previous existing approaches. Overall, our work introduces efficient methods for verifying the correct preparation of complex quantum states, and may be used for calibrating large quantum devices, witnessing quantum properties, supporting demonstrations of quantum computational speedups and enhancing trust in quantum computations.
\end{abstract}


\tableofcontents
\twocolumn
\section{Introduction}\label{sec:introduction}
The promise of quantum computers to outperform their classical counterparts for certain computational problems, known as ``quantum computational speedup'' or ``quantum supremacy'' \cite{preskill2012quantumcomputingentanglementfrontier,Ronnow2014,Harrow2017}, has given rise to a very active field of research: several sub-universal quantum models \cite{Aaronson2013,Terhal2004,Shepherd2009,Bremner2011,Boixo2018,Mezher2019} have been introduced to demonstrate quantum speedup in the near term, together with experimental efforts to build the corresponding quantum devices \cite{Harrow2017,Giordani2018,Arute2019,Wang2019,Zhong2020,madsen2022quantum}.

A demonstration of quantum computational advantage consists of three parts: a computational problem provably hard for a classical computer to solve, a quantum device that efficiently solves the problem and a verification that the quantum device is working correctly. Once such a quantum device is built, verifying its correct functioning becomes imperative: quantum devices are subject to inevitable environmental interactions that introduce errors \cite{Preskill1998}. Although fault-tolerant quantum computing offers a potential remedy, it demands a substantial resource overhead \cite{knill2005}. Moreover, the possibility of the quantum device behaving adversarially cannot be overlooked, e.g., in the context of delegated quantum computing \cite{Leichtle2024}. Consequently, it is crucial to devise methods that rigorously ensure that quantum devices function as intended. This task is usually referred to as \textit{quantum verification} or \textit{quantum certification} \cite{eisert2020quantum}.

The task of verification can be abstractly modeled by a \textit{prover} and a \textit{verifier}. The prover holds a powerful device, a quantum device here, which is untrusted, whereas the verifier holds a trusted device with limited computational power. The verifier assigns a specific computational task to the prover and evaluates the prover's trustworthiness, potentially through several rounds of interaction. It is important to note that although this verification model originates from a cryptographic interaction between two parties, it can similarly represent the calibration or certification of a physical experiment, where the experiment acts as the prover and the experimentalist serves as the verifier. Picking a noise model for the experiment then amounts to making assumptions on the prover.

What defines a good quantum verification protocol depends of course on the exact setting, but it is desirable to minimize the complexity of the verifier, the complexity of the prover, the assumptions made on the prover's behavior, and the number of samples required \cite{Eisert2020}. In addition, for quantum speedup experiments, a meaningful figure of merit for verification is based on the total variation distance between a target probability distribution and the true distribution describing the functioning of the device, as it bears operational significance. Assumptions can be made on the prover, such as restricting its computational capabilities or assuming it follows a specific behavior. A typical assumption is that the different copies of the quantum state produced by the experiment are i.i.d.\ (independent and identically distributed). This i.i.d.\ assumption can be removed, by paying the cost of an associated overhead in the number of samples required for verification \cite{Renner2007,Gheorghiu2019,Zhu2019PRL,Zhu2019PRA}. 

In quantum physics experiments, verification is done via classical post-processing of measurement outcomes, drawn from an output probability distribution describing the experiment. General-purpose techniques for verification such as quantum tomography rapidly become impractical for large quantum systems because of their resource requirements \cite{d2003quantum,artiles2005invitation}. Moreover, other verification protocols in which the verifier can only perform classical computations either rely on strong assumptions \cite{Shepherd2009,KahanamokuMeyer2023forgingquantumdata,Yung2020} or introduce complex protocols that prevent their practical use in current and near-term experiments \cite{Mahadev2018,Brakerski2018,Brakerski2020}. 

For approaches that require fewer resources, one may opt for partial verification techniques \cite{Boixo2018,Arute2019,Spagnolo2014,Drummond2021,Aaronson2013_2}. In these methods, instead of evaluating the total variation distance, specific characteristics of the experimental probability distribution are examined. These techniques are contingent upon certain assumptions regarding the operational mechanisms of the quantum device \cite{Ferracin2019}.  Alternatively, one may require the verifier to possess only basic quantum computational abilities, such as measuring single quantum systems. In this scenario, the verifier's objective is to assess how closely the state generated by the prover aligns with a predetermined target state in terms of trace distance or (in)fidelity, which in turn provides a bound on the similarity between the actual and target probability distributions in total variation distance. Fidelity estimation can be done in several ways: for graph states, the fidelity can be estimated both in the i.i.d.\ and non-i.i.d.\ case by measuring the expectation values of the stabilizers of the target graph state over the obtained experimental state \cite{Hayashi2015stabilizertesting}. Alternatively, the verification problem can be framed as a hypothesis testing problem, where we want to test whether the obtained experimental state is the desired target state or if it is at least $\epsilon$-away from it in (in)fidelity \cite{Hayashi2009,Pallister2018,Zhu2019PRL,Zhu2019PRA,Liu2021}. Alternatively, the fidelity between the target state and experimental state can be estimated through the application of a fidelity witness, which serves as a lower bound on the fidelity between the two states \cite{chabaud2020building,Chabaud2021efficient,sater2025efficient}.
 
While verification of discrete-variable (DV) quantum systems such as qubit-based systems is extensively studied, the verification of bosonic or continuous-variable (CV) quantum systems is relatively less established. These systems have emerged as promising candidates for quantum computing \cite{Gottesman2001,Knill2001,menicucci2006universal,madsen2022quantum}, driven by experimental breakthroughs such as the generation of large-scale entangled states \cite{yokoyama2013ultra} and remarkable error correction capabilities \cite{sivak2023}. In the context of bosonic quantum systems, Gaussian state preparation and measurements are basic operations: they have been the subject of considerable theoretical investigation \cite{Adesso2014} and lead to computations that can be simulated classically \cite{Bartlett2002}; moreover, they can be implemented experimentally on a large scale \cite{Yokoyama2013,Yoshiakawa2016,Ferraro2005}. Nevertheless, there remains a significant gap concerning efficient verification protocols based on such Gaussian measurements. Existing protocols \cite{Aolita2015,Chabaud2021efficient,Liu2021,wu2021efficient} suffer from experimental impracticality either due to exponential scaling for non-Gaussian target states \cite{Aolita2015} or limited robustness \cite{Chabaud2021efficient,wu2021efficient}, i.e., the verification method used require a high value of the true fidelity between the target and experimental state to guarantee any meaningful information.

In this work, we introduce a general framework for the efficient verification of large quantum systems based on robust fidelity witnesses and efficient classical post-processing. We apply this framework to bosonic quantum systems and derive semi-device independent verification protocols based on Gaussian homodyne or heterodyne measurements, which are both efficient and robust, addressing current limitations of existing protocols. These protocols generalise and provide a robust alternative to the methods in \cite{Chabaud2021efficient} for verifying the output states of quantum speedup experiments such as Boson sampling, and can be used for efficient verification of a large class of bosonic states not known to be efficiently verifiable before (see Table \ref{tab:verifiable_classes}).

\begin{table*}[ht]
\centering
\renewcommand{\arraystretch}{1.5} 

\begin{tabular}{
|m{4.4cm}|
>{\raggedright\arraybackslash}m{2.3cm}|
>{\raggedright\arraybackslash}m{2.8cm}|
>{\raggedright\arraybackslash}m{2.5cm}|
>{\raggedright\arraybackslash}m{2.8cm}|
}
\hline
\diagbox[width=4.8cm,height=2.8cm]
{\textbf{Measurement}}
{\textbf{Class of states}} & \textbf{Boson sampling states} & \textbf{Orthogonal Boson sampling states} & \textbf{$\log$-doped passive states} & \textbf{If yes, does it extend to non-iid states?} \\ \hline
Balanced Heterodyne & $\checkmark$ \cite{Chabaud2021efficient} & $\checkmark$ \cite{Chabaud2021efficient}  & $\checkmark$ [This work] & $\checkmark$ \cite{Chabaud2021efficient} \\ \hline
Synchronous Homodyne & $\times$  & $\checkmark$ [This work] & $\times$  & $\times$ \\ \hline
Noisy Balanced Heterodyne & $\checkmark$ [This work] & $\checkmark$ [This work] & $\checkmark$ [This work] & $\checkmark$ [This work] \\ \hline
Noisy Synchronous Homodyne 
& $\times$ & $\checkmark$ for $\eta > 1/2$ [This work]  & $\times$ & $\times$ \\ \hline
\end{tabular}

\caption{Classes of efficiently verifiable quantum states via fidelity witnessing. 
\textit{Boson sampling states} refer to the class of states produced by applying 
a passive Gaussian unitary to tensor product of finite superpositions of Fock states, 
\textit{Orthogonal Boson sampling states} is the subset of Boson sampling states 
where the applied passive Gaussian unitary is also orthogonal 
(see section \ref{sec:prelims}), and $\log$\textit{-doped passive states} refers 
to a subset of $t$-doped states defined in section 
\ref{sec_verification_doped}. The efficient verification of Boson sampling states 
(and consequently Orthogonal Boson sampling states) was introduced in 
\cite{Chabaud2021efficient}. In this work, we provide efficient verification 
protocols for orthogonal Boson sampling states using synchronous homodyne 
(and noisy synchronous homodyne) measurements (note that the estimator function 
we use to derive these protocols were derived in \cite{MauroDAriano2003}), 
as well as provide efficient verification protocols for $\log$ depth 
$t$-doped passive states using the multi-mode fidelity witness and heterdyne 
measurements. Finally, we also extend all the heterodyne measurement protocols 
to the case of noisy heterodyne measurements.}

\label{tab:verifiable_classes}
\end{table*}
 
The rest of the paper is organized as follows:
we provide preliminary material in section \ref{sec:prelims}, and detail our framework for efficient verification in section \ref{sec:framework}. This framework is based on robust fidelity witnesses, and we introduce a general family of such witnesses in section \ref{sec:robust_fidelity_witness}, of which we study the robustness from both analytical and numerical points of view.
Then, we apply our framework to derive new efficient verification protocols for bosonic systems: in section \ref{sec:homodyne_sampling}, we first introduce efficient verification protocols based on homodyne detection and existing fidelity witnesses; then, in section \ref{sec:heterodyne_sampling}, we describe verification protocols with heterodyne sampling utilizing our new fidelity witnesses, and show in particular that they allow for verifying a large class of non-Gaussian states known as doped Gaussian states \cite{AnnaMele2024}. As a byproduct, in section \ref{sec:homvshet} we give a detailed comparison of heterodyne and homodyne detection in the context of continuous-variable quantum verification protocols. Finally, we conclude in section \ref{sec:conclusion} with open directions.


\section{Preliminaries}
\label{sec:prelims}

We refer the reader to \cite{NielsenChuang} for background on quantum information theory tools used in this work, and to \cite{Braunstein2005,Ferraro2005,Weedbrook2012} for continuous-variable quantum information material.

Hereafter, the sets $\N, \R$ and $\C$ are the set of natural, real, and complex numbers respectively, with a * exponent when $0$ is removed from the set, $m \in \N^*$ denotes the number of modes and $\{\ket{n} \}$ is the single-mode Fock basis. Single-mode displacement operators and single-mode squeezing operators are defined as $\hat{D}(\alpha) = e^{\alpha \hat{a}^\dagger - \alpha^* \hat{a}}$ and $\hat{S} (\xi) = e^{\frac12 (\xi \hat{a}^2 - \xi^* \hat{a}^{\dagger 2})}$ respectively, where $\alpha, \xi \in \C$ and $\hat{a},\hat{a}^\dagger$ denote the annihilation and creations operator, respectively \cite{Weedbrook2012}.  Finally, we use bold math notations for vectors, multimode states, and multi-index notations. For all $\boldsymbol{\beta,\xi} \in \C^m$, we write $\hat{D}(\boldsymbol{\beta}) = \otimes_{i=1}^m \hat{D}(\beta_i)$ and $\hat{S}(\boldsymbol{\xi}) = \otimes_{i=1}^m \hat{S}(\xi_i)$.
Passive linear quantum operations play an important role in the paper. A passive linear quantum operation is a Gaussian transformation which preserves the total energy or photon number of a system. Common examples include beam splitters and phase shifters.

The position and momentum operators are referred to as the quadrature operators, and are given by
\begin{eqnarray}
    \hat{q} &=& \frac{1}{\sqrt{2}} (\hat{a} + \hat{a}^\dagger), \\
    \hat{p} &=& \frac{1}{i\sqrt{2}} (\hat{a} - \hat{a}^\dagger),
\end{eqnarray}
where $\hat{q},\hat{p}$ refer to the position and momentum quadrature operator respectively. They follow the canonical commutation relation $[\hat{q},\hat{p}] = i\I$ (with the convention $\hbar = 1$). Finally, a rotated quadrature is given by
\begin{equation}
    \hat X_{\theta} = \frac{1}{\sqrt{2}}\left(\hat{a}e^{-i\theta} + \hat{a}^\dagger e^{i\theta}\right).
\end{equation}
Operations that transform a Gaussian state into a Gaussian state are called Gaussian operations. The most common Gaussian measurements are heterodyne detection, which projects the state onto a coherent state and correspond to sampling from the Husimi $Q$ function; and homodyne detection, which correspond to the measurement of a rotated quadrature operator.

The fidelity gives a measure of the closeness of two quantum states $\rho$ and $\sigma$ and is defined as
\begin{equation}
    F(\rho,\sigma) = \mathrm{Tr} \left[\sqrt{\sqrt{\sigma}\rho \sqrt{\sigma}}\right]^2.
\end{equation}
Despite it not being apparent, $F(\rho,\sigma)$ is symmetric in $\rho$ and $\sigma$. When one of the states is pure, the expression reduces to
$F(\ket{\psi},\rho) = \mathrm{Tr}[\ket{\psi}\!\bra{\psi}\rho] = \bra{\psi}\rho\ket{\psi}$
and in particular when both states are pure, $F(\ket{\psi},\ket{\phi}) = |\braket{\phi|\psi}|^2$.

The total variation distance between two probability distributions $P$ and $Q$ over a continuous sample space $\Omega$ is defined as
\begin{equation}
    d_{TV}(P,Q) = \frac12 \underset{x \in \Omega}{\int} |P(x) - Q(x)| dx.
\end{equation}

The trace distance between two states $\rho,\sigma$ is given by 
\begin{equation}
    D(\rho,\sigma) = \frac12 \mathrm{Tr}(|\rho - \sigma|),
\end{equation}
and satisfies $D(\rho,\sigma)\!=\! \underset{\hat{O}}{\mathrm{max}}\,||P_{\rho}^{\hat{O}} - P_{\sigma}^{\hat{O}}||_\mathrm{tvd}$. Here, the probability distribution $\smash{P_{\rho}^{\hat{O}}}$ (and similarly $\smash{P_{\sigma}^{\hat{O}}}$) arises from measuring the observable $\hat{O}$ on the states $\rho$ (reps.\ $\sigma$), with the maximum total variation distance taken over all observables. Two states being close in trace distance thus indicates that quantum computations using either state as input will yield nearly identical results from a statistical standpoint. Finally, the fidelity and the trace distance are related by the Fuchs-van de Graaf inequalities \cite{Fuchs1999}
\begin{equation}
    1 - \sqrt{F(\rho,\sigma)} \leq D(\rho,\sigma) \leq \sqrt{1 - F(\rho,\sigma)}.
\end{equation}
Therefore, a lower bound on the fidelity between two states gives an upper bound on the trace distance between them.

The efficiency of fidelity estimation protocols between a measured state and a target state depend on the number of measurements performed. The fidelity can be written as a classical expectation value of estimator functions over these measurement outcomes, and with a finite number of samples $N$, we use Hoeffding's inequality \cite{Hoeffding1963} to estimate the closeness of the average estimator values calculated from these measurement outcomes to the expectation value of the estimators.
\begin{lem}[Hoeffding's inequality]\label{lem_hoeffding}
     Let \(\epsilon > 0\), let \(n \geq 1\), let \(z_1, \ldots, z_n\) be i.i.d.\ complex random variables from a probability density \(D\) over \(\mathbb{R}\), and let \(g : \mathbb{C} \rightarrow \mathbb{R}\) be a bounded function and let \(G \geq \max g(z) - \min g(z)\). Then
\begin{eqnarray}
&&\mathrm{Pr}\left(\left|\frac{1}{N} \sum_{i=1}^N g(z_i) - \mathbb{E}_{z \leftarrow D} [g(z)] \right| \geq \epsilon\right)  \nonumber \\ && \hspace{30mm} \leq 2 \exp\left(-\frac{2N\epsilon^2}{G^2}\right).
\end{eqnarray}
\end{lem}

\section{Verification framework}
\label{sec:framework}

In this section, we introduce a general framework for efficient verification of quantum states, which we base our protocols on. 

Let us first define the verification problem that we deal with in this paper:
\begin{pb}\label{pb:verification_problem}
    Given the description of a target state $\ket{\psi}$ and access to copies of an unknown state $\rho$, \textsf{Verify}($\ket{\psi},c,s,\delta$) is defined as the problem of outputting \textbf{Accept} with probability at least $1-\delta$ if
    \begin{equation}
        F(\rho,\ket{\psi}) > 1 - c,
    \end{equation}
    and outputting \textbf{Reject} with probability at least $1-\delta$ if
    \begin{equation}
        F(\rho,\ket{\psi}) < 1 - s,
    \end{equation}
    promised that one of the two holds, where $F$ is the fidelity and where $c$ and $s$ are the completeness and soundness parameters, respectively, with $0\le c\le s\le 1$.
\end{pb}

 Note that for $c=s$, \textsf{Verify}($\ket{\psi},c,s,\delta$)  corresponds to the problem of fidelity estimation, $c < s$ corresponds to fidelity witnessing and finally $c>0$ implies that the certification is robust.

Throughout the main text, we assume that the different copies are i.i.d.\ (independent and identically distributed), and note that this assumption can be removed at the cost of a sample overhead. We give a detailed discussion of this point while describing the verification protocols (see also Appendix~\ref{proof_no_iid}).

\begin{figure*}[t]
    \centering\includegraphics[width=\linewidth]{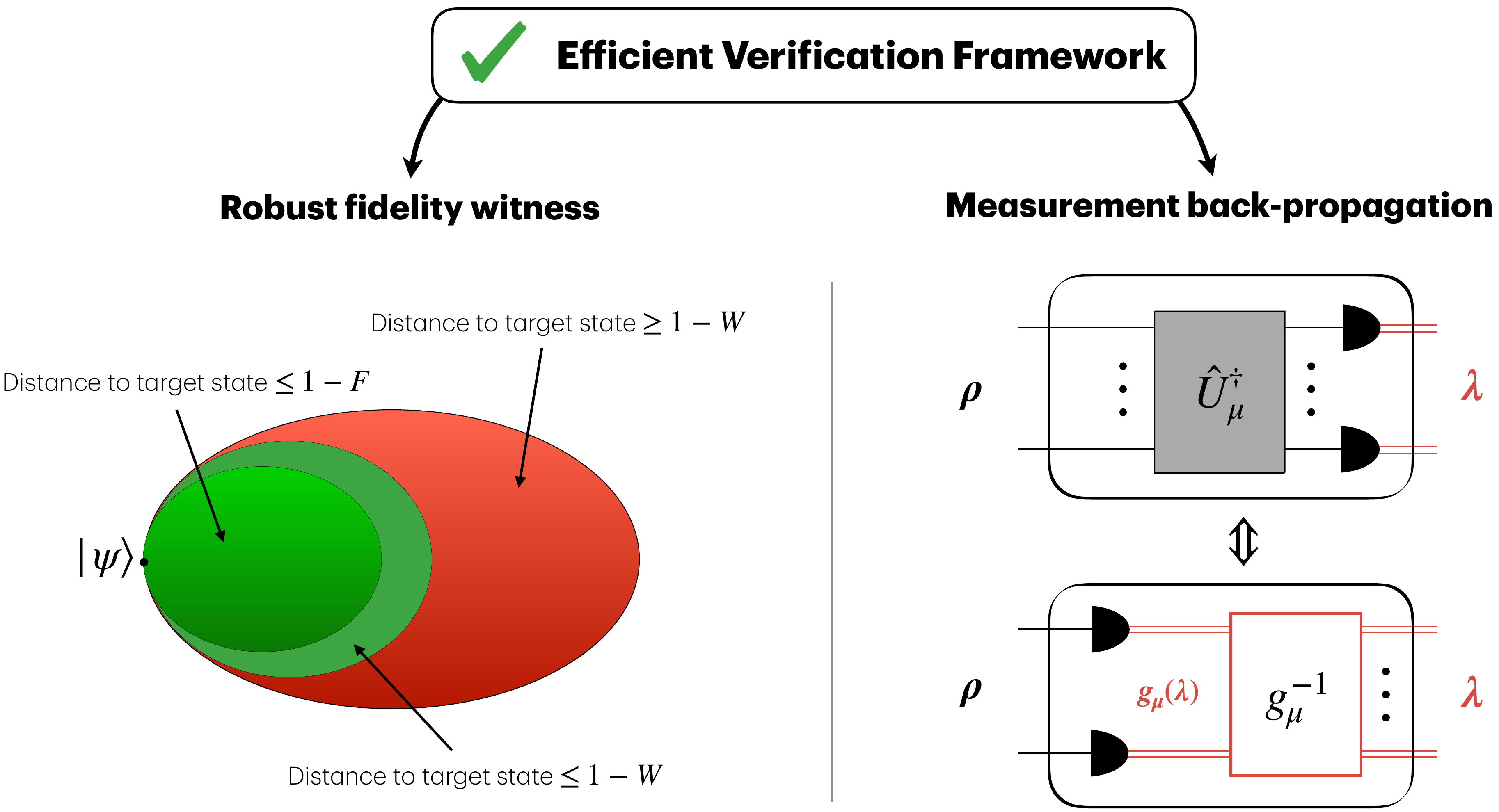}
    \caption{Concepts behind our efficient verification framework. Any verification protocol combining efficient and robust fidelity witnesses for a class of target states with measurement back-propagation for emulating quantum operations before the measurements can be used to efficiently verify a larger class of target states. The right hand side depicts the information retrieved by estimating a fidelity witness. The left hand side represents measurement back-propagation: in black are quantum operations and measurements and in red classical measurement outcomes and operations.}  \label{image_verfication_schematics}
\end{figure*}

In general, verifying the correct preparation of a state $\ket{\psi}$ through tomography requires exponentially many measurements in the number of subsystems of the state \cite{d2003quantum}. However, quantum verification methods focusing on fidelity witnessing and specific classes of states may achieve exponentially better sample complexity, as we demonstrate hereafter. Here, we want to verify the fidelity of an unknown state $\rho$ with a target state $\ket{\psi}$, by making measurements whose positive operator-valued measure (POVM) elements are denoted $\{\Pi_\lambda\}_\lambda$. Our efficient verification framework is designed for these fidelity-witness-based verification problems and combines two ingredients (see Figure \ref{image_verfication_schematics}):

\begin{enumerate}
    \item\label{enum:witness}\textbf{Efficient and robust fidelity witness:} For an unknown state $\rho$ and a class of target states $\{\ket{\psi}\}$, we require a fidelity witness $W(\rho,\psi)$ that is efficiently computable using samples from the measurement described by $\{\Pi_\lambda\}_\lambda$ and that is robust, i.e., that satisfies an equation of the form:
    \begin{equation}\label{eqn_generalized_fidelity_witness}
        f(F) \leq W(\rho,\psi) \leq F(\rho,\psi),
    \end{equation}
    where $f$ is a continuous function such that $f(1) = 1$. The inequality on the right-hand side indicates that the fidelity witness is indeed a lower bound on the fidelity $F(\rho,\psi)$, whereas the inequality on the left-hand side gives the range of fidelity where the fidelity witness provides meaningful information, i.e., where $W(\rho,\psi) \geq 0$.
    \item\label{enum:postproc}\textbf{Measurement back-propagation:} To extend the class of states which can be verified efficiently, we use efficient classical post-processing of measurement outcomes to simulate undoing the effects of a unitary operation before the measurement. Denoting such a set of unitaries as $\{\hat{U}_{\mu}\}_\mu$, this is possible when the POVM elements $\Pi_\lambda$ of the measurement device satisfy the following condition:
    \begin{equation}
        \hat{U}_{\mu}\Pi_\lambda \hat{U}_{\mu}^\dag=\Pi_{g_\mu(\lambda)}, 
    \end{equation}
    where $g_\mu(\lambda)$ is an invertible function of $\lambda$ $\forall \mu$, which can be computed efficiently.
\end{enumerate} 
The first point allows us to verify efficiently the correct preparation of a target state $\ket{\psi}$, by estimating the value of the fidelity witness using measurement outcomes from the POVM $\{\Pi_\lambda\}_\lambda$ and consequently, finding the lower bound on the fidelity between the target and experimental state. The second point allows us to extend the efficient verification to all the states of the form $\hat{U}_\mu\ket{\psi}$ with classical post-processing, since
\begin{equation}
    \mathrm{Tr} [\rho \Pi_{g_\mu(\lambda)}]=\mathrm{Tr}[\hat{U}_{\mu}^\dag\rho \hat{U}_{\mu}\Pi_\lambda].
\end{equation} 
This means that measuring $\rho$ with the POVM $\{\Pi_\lambda\}_\lambda$ and post-processing the measurement outcomes as $\lambda\mapsto g_\mu(\lambda)$ effectively emulates the application of the quantum operation $\hat U_\mu^\dag$ on $\rho$ before the measurement, and reduces the verification of $\hat{U}_\mu\ket{\psi}$ to the verification of $\ket{\psi}$.

This framework is applicable to both discrete-variable and continuous-variable quantum systems and relies on the availability of robust fidelity witnesses. In the next section, we introduce a family of such fidelity witnesses, for arbitrary \textit{separable} target states. Then, in the rest of the paper, we focus on continuous-variable verification protocols combining these new fidelity witnesses with efficient classical post-processing to efficiently verify large classes of continuous-variable quantum states, and overcome limitations of existing Gaussian verification protocols.

\section{Robust fidelity witnesses}
\label{sec:robust_fidelity_witness}

In this section, we introduce a family of fidelity witnesses for arbitrary separable target states and provide analytical and numerical analyses of their robustness.

\subsection{Fidelity witnesses based on \texorpdfstring{$k$}{k}-mode fidelities}

We consider a $m$-mode quantum systems, and we assume that $m$ is divisible by $k$. For two $m$-mode quantum states $\rho$ and $\ket{\psi}$, let us define the following multimode fidelity witness for an $m$-mode system:
\begin{equation}
  W^{(k)} (\rho,\psi):= 1 -  \underset{i = 1}{\overset{m/k}{\sum}}[1 - F (\rho_{i;k},\ket{\psi_{i;k}} )],
    \label{eq3.1}
\end{equation}
where $\boldsymbol{\sigma}_{i;k}:=\mathrm{Tr}_{\{1,\dots,m\} \backslash \{ki - (k-1),\dots,ki\}} [\boldsymbol{\sigma}]$ denotes the reduced state over the $i^{th}$ subset of $k$ modes $\{ki - (k-1),\dots,ki\}$.
Note that the choice of modes is for brevity, and one can similarly define witnesses based on any other partition of $m$ modes into subsets of $k$ modes, or even based on a partition into subsets of modes of different sizes. Setting $k=1$ retrieves a witness based on single-mode fidelities previously introduced in \cite{Chabaud2021efficient}, while setting $k=m$ retrieves the $m$-mode fidelity. 

Intuitively, one may think that the witness in Eq.~(\ref{eq3.1}) provides a better bound on the fidelity as $k$ increases, since we are losing less information by tracing out some of the modes. This intuition is formalized by the following result:


\begin{lem}[Robustness of fidelity witnesses]\label{lemma_mm_fid_witness_2}
Let $\rho$ and $\ket{\psi}$ be states over $m$ subsystems. Then for all $k\in\{1,\dots,m\}$,
\begin{equation}
        W^{(1)} (\rho,\psi) \leq W^{(k)} (\rho,\psi),
        \label{eqn_better_k_mode}
    \end{equation}
\noindent and
\begin{equation}
    1 - \frac{m}{k} (1 - F(\rho,\ket{\psi})) \leq W^{(k)} (\rho,\psi) \leq F(\rho,\ket{\psi}).
    \label{eqn_inequality_k-mode}
\end{equation}
\end{lem}
\noindent We direct the reader to section \ref{proof_lemma_2} of the appendix for the proof. 

Eq.~(\ref{eqn_inequality_k-mode}) shows that $W^{(k)} (\rho,\psi)$ gives us a lower bound of the fidelity between $\rho$ and $\ket{\psi}$, whereas Eq.~(\ref{eqn_better_k_mode}) proves $W^{(k)} (\rho,\psi)$ is a better lower bound of the fidelity than $W^{(1)} (\rho,\psi)$. In addition $W^{(k)}(\rho,\psi)$ requires fidelity $F(\rho,\ket{\psi}) > 1 - k/m$ to guarantee meaningful information, i.e., $W^{(k)}(\rho,\psi) > 0$. Therefore, $W^{(k)} (\rho,\psi)$ is more robust than $W^{(1)} (\rho,\psi)$ for $k>1$. 
On the other hand, $W^{(k)} (\rho,\psi)$ requires the estimation of $k$-mode fidelities, which has sample complexity exponential in $k$ in general.

Given this trade-off between robustness and sample complexity, it is beneficial in practical applications to identify scenarios where employing a multimode fidelity witness offers a clear advantage. We address this problem through numerical examples in the next section.

\subsection{Numerical analysis}
\label{sec:num_analysis}

In this section, we illustrate the robustness of the fidelity witnesses defined in Eq.~(\ref{eq3.1}) through numerical analysis of three examples. In the first example, the deviation of the tested state from the target state is essentially random and aimed at demonstrating the utility of the different fidelity witnesses in the presence of the most common source of noise in bosonic architectures---photon loss---and under incorrect classical post-processing. In contrast, the remaining two examples explore structured deviations to highlight the witness robustness. These structured cases illustrate why grouping the correct modes is beneficial when picking a $k$-mode fidelity witness, when prior knowledge of the state's imperfections is available (for example, when there are systematic errors on specific subsystems).

\medskip

\textbf{Example 1:} We start by the following example: the target state $\ket{\psi}=\hat{{U}}\ket1^{\otimes6}$ is chosen as the output state of a $6$-mode Boson sampling experiment, i.e., a Fock state passing through a lossless passive linear interferometer described by the unitary $\hat{{U}}$. The tested state is a lossy and noisy version of the target state: the prepared tensor product of Fock states $\ket 1^{\otimes 6}$ undergoes single-mode losses, represented as a six-mode channel $\mathcal L_{\bm\eta}$, where $\eta$ is a loss parameter (the lossless case corresponding to $\eta=1$). In addition, during classical post-processing, we apply a slightly perturbed interferometer $\hat{{{U}}}'^\dagger$ instead of the ideal $\hat{{U}}$, such that a small random passive linear unitary imperfection $\hat{{V}} :=\hat{{U}}^\dagger \hat{{{U}}}'\approx \hat\I$ still acts on the state. Consequently, after the classical post-processing step, instead of achieving the target state $\ket{\psi}$, we obtain the noisy state $\hat{{V}}\mathcal L_{\bm\eta}(\ket{0}\bra{0}^{\otimes 6}) L_{\bm\eta}^\dag\hat{{V}}^\dag$. 

\begin{figure}
    \centering
    \includegraphics[width=\linewidth]{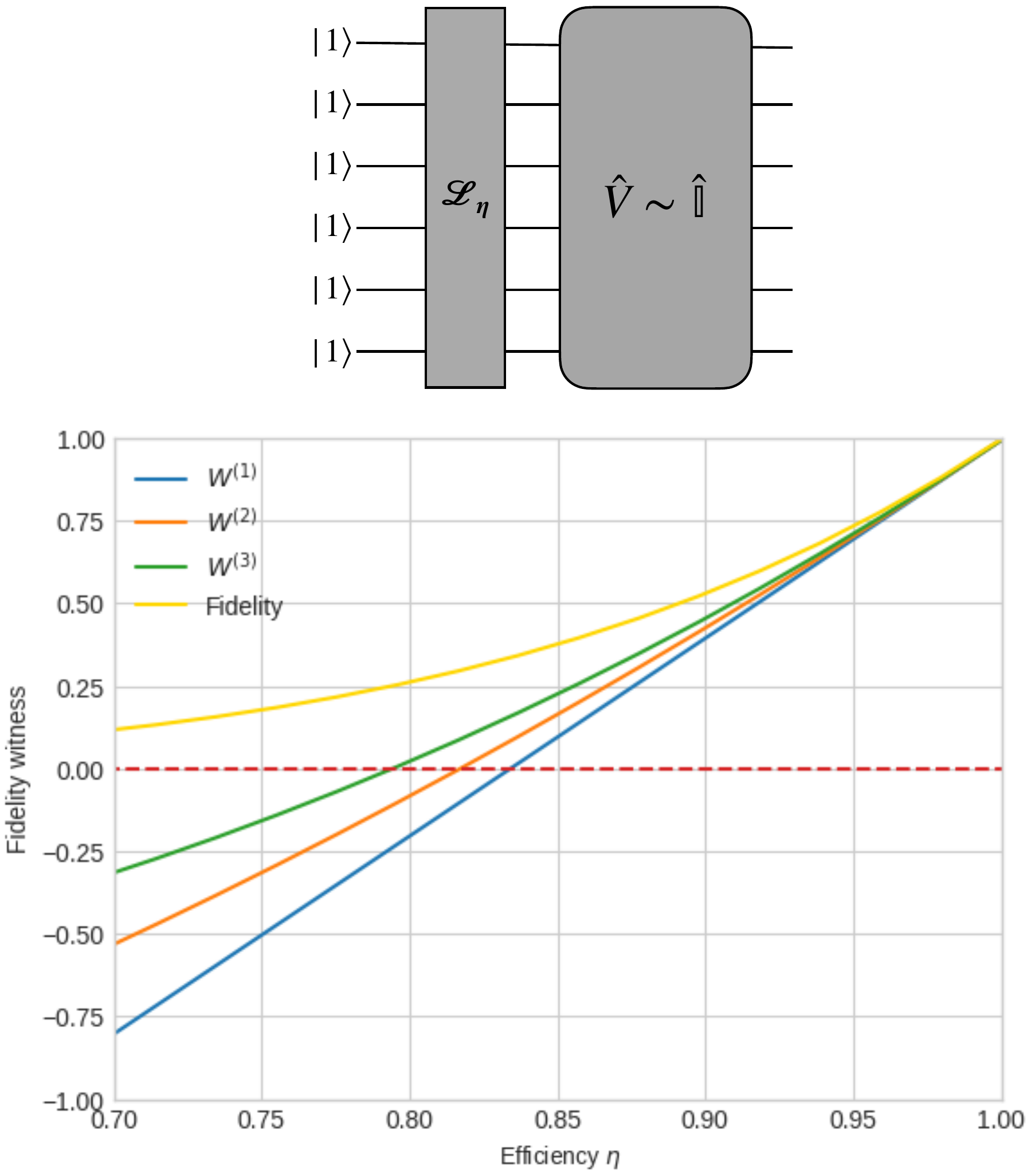}
    \caption{(Top) Quantum circuit considered in the analysis given in Example 1 of section \ref{sec:num_analysis}. 
    (Bottom) Fidelity witnesses based on single-mode, two-mode, and three-mode fidelities as a function of the loss parameter $\eta$. For each value of $\eta$, we average over different choices of unitary interferometers deviations $\hat{V}$ that are close to identity. For a fidelity witness to be useful, it needs to take positive values. The fidelity witness based on  two-mode fidelities is useful for the largest range of single-mode loss parameter values, followed by the witness based on two-mode fidelities, and finally by the one based on single-mode fidelities.}
    \label{image_num_fidelity_1}
\end{figure}

Varying the loss parameter $\eta$ from $0$ to $1$ and for each value of $\eta$, averaging over a family of unitary interferometer imperfections $\hat{V}$ that are close to identity, we analyze the performance of single-mode, two-mode, and three-mode fidelity witnesses as a function of $\eta$.
The results are presented in Figure \ref{image_num_fidelity_1}. As expected, the fidelity witness based on three-mode fidelities remains positive over a larger range of $\eta$, meaning it remains effective over a wider span of losses compared to the witnesses based on two-mode and single-mode fidelities, respectively.

This example illustrates that, in cases with single-mode losses followed by random close-to-identity interferometers, the utility of the three-mode or two-mode fidelity witnesses marginally surpasses that of the single-mode fidelity witness.
Because of the random deviation, the pairing of the modes does not affect the quality of the witnesses. However, in scenarios where systematic errors are present, different pairings of modes can yield different information (and thus distinct fidelity witness values), due to the structure of the deviation of the tested state from the target state. This feature can be useful to characterize the noise structure within a device and exploit it for efficient verification. To illustrate this point, we consider two additional examples.

\begin{figure}[t]
        \centering
        \includegraphics[width=\linewidth]{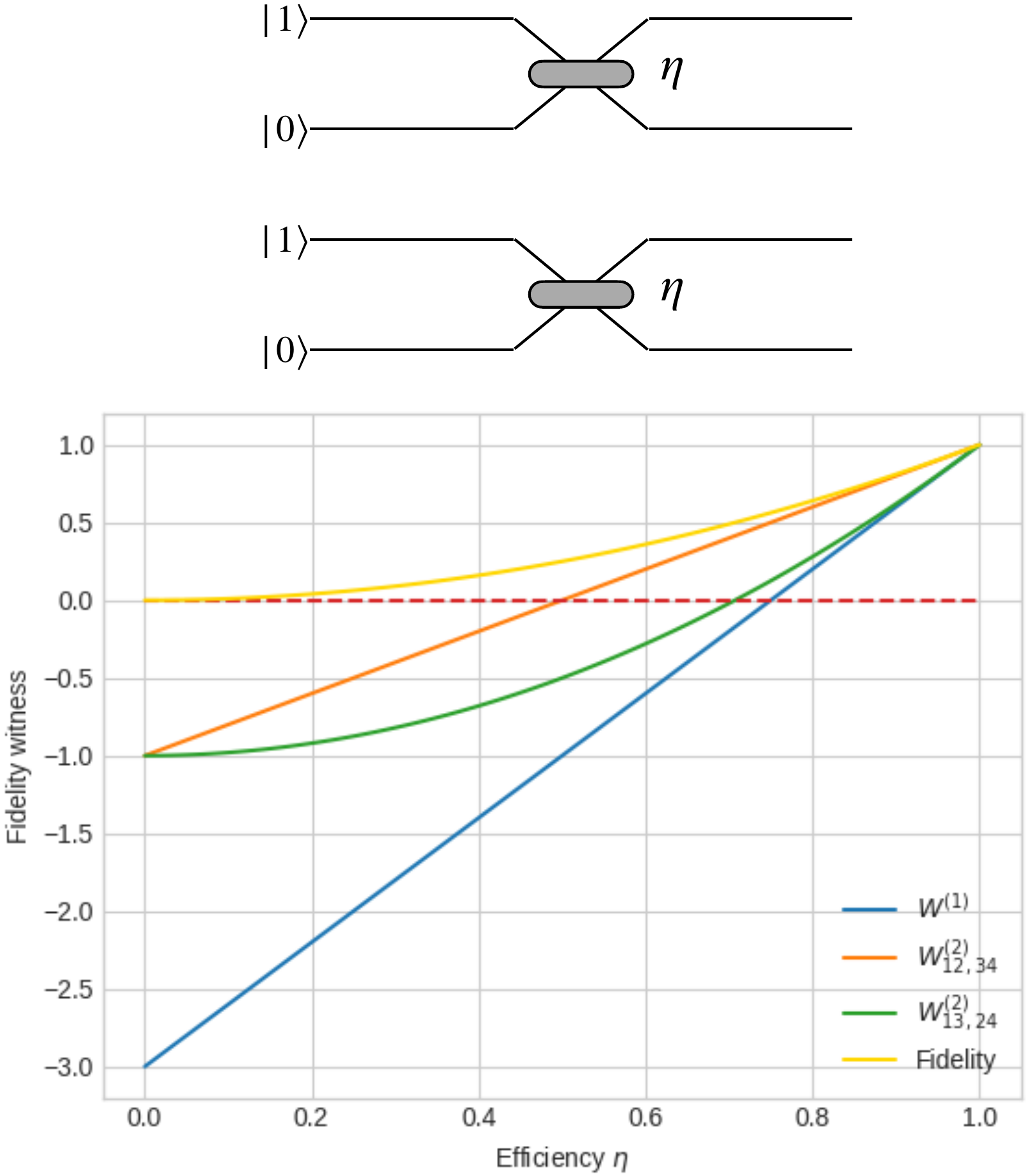}
        \caption{(Top) Quantum circuit considered in the analysis given in Example 2 of section \ref{sec:num_analysis}. (Bottom) Comparison of the two-mode fidelity witnesses for the circuit given. The two-mode fidelity witness where the modes $1,2$ and $3,4$ are paired up is the closest to the actual value of fidelity, mimicking the noise structure.}
        \label{scheme_case_1}
\end{figure}

\medskip

\textbf{Example 2:} We consider a two-mode state, $\ket{1010}$, passing through two beamsplitters with efficiency $\eta$, applied between modes 1 and 2, and modes 3 and 4. With the target state being $\ket{1010}$ ($\eta = 1$), we compare the actual value of fidelity between the input state and the output state of this circuit with fidelity witnesses based on two-mode fidelities for pairs (1,2) and (3,4) ($W^{(2)}_{12,34}$), and for pairs (1,3) and (2,4) ($W^{(2)}_{13,24}$), and on single-mode fidelities (Figure \ref{scheme_case_1}). When modes affected by correlated noise are grouped as pairs, the two-mode fidelity witnesses align more closely with the actual fidelity. This suggests that selecting pairs which accurately capture the correlation structure of the noise yields a better fidelity lower bound. In scenarios where pairing all correlated modes is not feasible, choosing fidelity witnesses that capture as much entanglement as possible remains the optimal approach.

\begin{figure}[t]
        \centering
        \includegraphics[width=\linewidth]{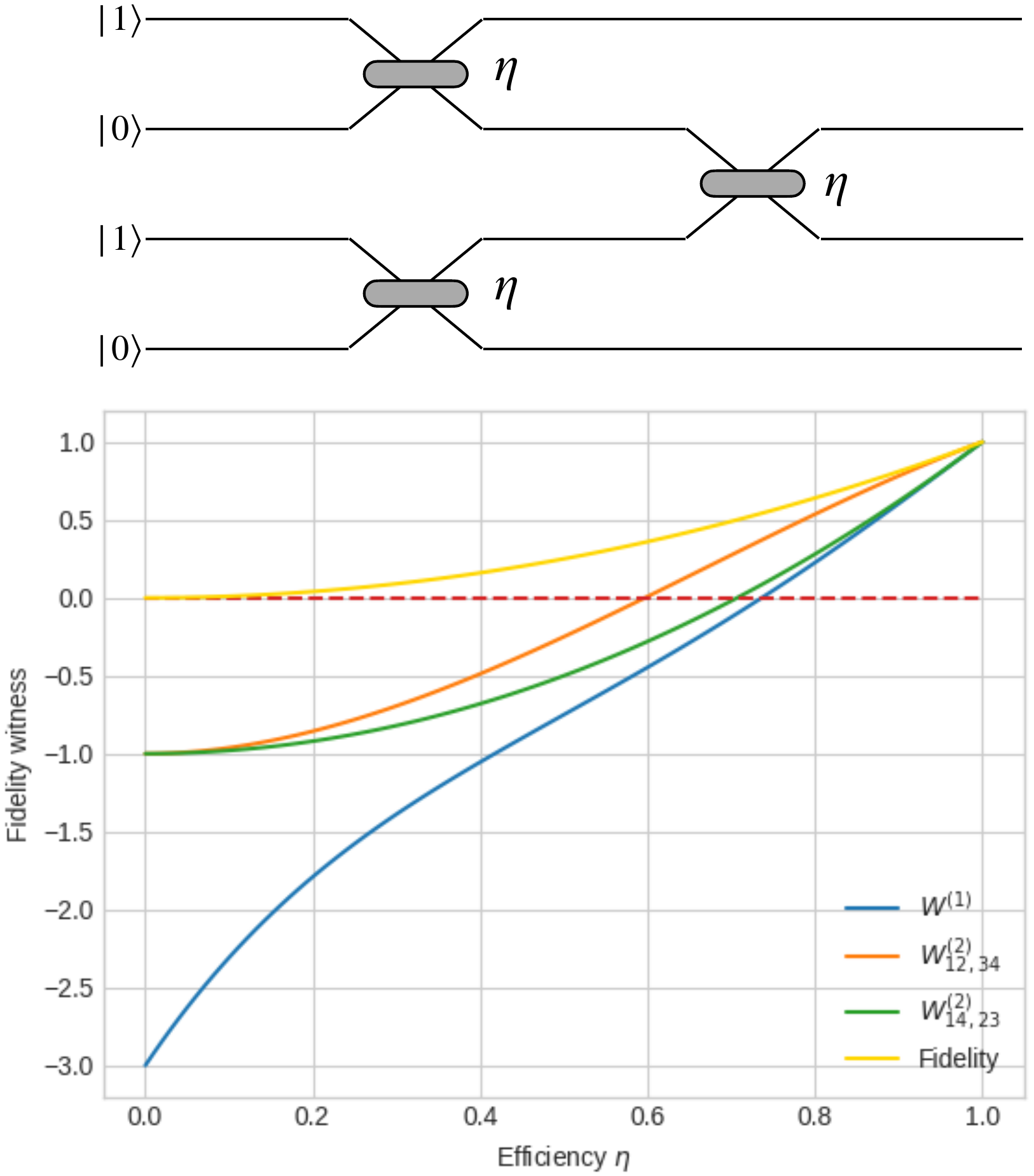}
        \caption{(Top) Quantum circuit considered in the analysis given in Example 3 of section \ref{sec:num_analysis}. (Bottom) Comparison of the two-mode fidelity witnesses and single-mode fidelity witness for the circuit given. Two-mode fidelity witness which pairs more entangled modes gives a better estimate of the fidelity.}
        \label{scheme_case_2}
\end{figure}

\medskip

\textbf{Example 3:} The last example we consider closely mirrors the previous one, with the input state changed from $\ket{1010}$ to $\ket{1001}$, and an additional beamsplitter entangling modes 2 and 3. With the target state being $\ket{1001}$ ($\eta = 1$), we compare the actual value of fidelity between the target state and output state of this circuit with the fidelity witnesses based on two-mode for pairs (1,2) and (3,4) ($W^{(2)}_{12,34}$), and for pairs (1,4) and (2,3) ($W^{(2)}_{14,23}$), and on single-mode fidelities (Figure \ref{scheme_case_2}). In this case, there is correlated noise between modes 1 and 2, modes 3 and 4, as well as between modes 2 and 3. The two-mode fidelity witness that pairs the more strongly correlated modes, namely $W^{(2)}_{12,34}$, provides a better estimate of the fidelity.


In what follows, we illustrate the applicability of our efficient verification framework introduced in section \ref{sec:framework}, by combining the witnesses introduced in this section with efficient classical post-processing for measurement back-propagation. We consider the verification of bosonic systems based on continuous-variable quantum measurements, namely homodyne detection (section \ref{sec:homodyne_sampling})-- where we extend the verification protocols of \cite{Chabaud2021efficient} to include homodyne detection and understand the limitations of using heterodyne measurement protocols for verification--and heterodyne detection (section \ref{sec:heterodyne_sampling}), where we derive verification protocols using $k$-mode fidelity witness and demonstrate how these fidelity witness allow for efficient verification of a new class of non-Gaussian states-doped Gaussian states.


\section{Verification protocols with homodyne detection}
\label{sec:homodyne_sampling}

\subsection{Homodyne detection}
Homodyne detection is a single-mode Gaussian measurement that mathematically corresponds to projection on the rotated quadrature operator  $\hat{X}_{\theta} = \frac{1}{\sqrt{2}}\left(\hat{a}e^{-i\theta} + \hat{a}^\dagger e^{i\theta}\right)$, and homodyne measurement on a state $\rho$ corresponds to sampling from the probability distribution
\begin{equation} \label{equation_P_rho}
    P_\rho (x,\theta) = \bra{x}_\theta \rho \ket{x}_\theta,
\end{equation}
where $\ket{x}_\theta$ is an eigenstate of the quadrature operator $\hat{X}_{\theta}$. Experimentally, homodyne detection is implemented by mixing an incoming signal beam with a strong local oscillator beam at the same frequency, typically using a beamsplitter. The mixed beams then interfere, and the resulting signal is measured by photodetectors. The phase and amplitude of the signal beam are inferred by adjusting the phase of the local oscillator and analyzing the difference in photodetector outputs. The positive operator valued-measure (POVM) elements of a tensor product of single-mode homodyne detection over $m$ modes of quadrature operator at positions $\boldsymbol{x} = \{x_1,\ldots,x_m \} \in \R^m$ rotated at the same angle $\theta$ are given by
\begin{equation}
    \Pi_{\boldsymbol{x}} = \ket{\boldsymbol{x}}_\theta\!\bra{\boldsymbol{x}}_\theta,
\end{equation}
where $\ket{\boldsymbol{x}}_\theta$ is the eigenstate of the quadrature operator $\hat{\boldsymbol{X}}_{\theta} = \frac{1}{\sqrt{2}}\left(\hat{\boldsymbol{a}}e^{-i\theta} + \hat{\boldsymbol{a}}^\dagger e^{i\theta}\right)$. Since all angles are equal, we call this multimode measurement \textit{synchronous homodyne detection}.

\subsection{Single-mode homodyne fidelity estimation}
We now describe a protocol to estimate the fidelity between an experimental state and a single-mode target state with a finite support over the Fock basis, also known as core states \cite{chabaud2020stellar}.
We introduce homodyne estimator functions as
\begin{equation}
\label{eqn_homodyne_estimatorfkl}
    f_{lk} (x) e^{i(l-k)\theta},
\end{equation}
where 
\begin{eqnarray} \label{equation_f_lk}
    \nonumber f_{lk} (x) &:=& 4x u_k(x) v_l (x) - 2\sqrt{k+1} u_{k+1} (x) v_l (x) \\& & - 2 \sqrt{l+1} u_k(x) v_{l+1} (x),
\end{eqnarray}
with $u_j(x)$ and $v_j (x)$ the normalizable and unnormalizable eigenfunctions of the harmonic oscillator with eigenvalue $j$, respectively, whose expressions are given in section \ref{normal_unnormal_HO} of the appendix. The notable property of these estimator functions is \cite{MauroDAriano2003}:
\begin{equation}\label{equation_perfect_homodyne_estimation}
    \rho_{kl} = \underset{ P_{\rho}(x,\theta)}{\E}[f_{lk}(x) e^{i(l-k) \theta} ],
\end{equation}
where the angle $\theta$ is drawn uniformly randomly. 
In particular, since homodyne detection corresponds to sampling from the probability distribution function $P_\rho (x,\theta)$ given by Eq.~(\ref{equation_P_rho}), Eq.~(\ref{equation_perfect_homodyne_estimation}) allows the estimation of any element $k,l$ of the density matrix by computing the mean of the function $(x,\theta) \mapsto f_{lk}(x) e^{i(l-k) \theta} $ over these homodyne samples, for random choices of homodyne angles $\theta$. 

For any single-mode core state $\ket{C} = \sum_{n = 0}^{c-1} c_n \ket{n}$, where $c \in \N^*$ and $x,\theta \in \R$, we define the function

\begin{equation} \label{eqn_homodyne_estimator}
    g^{(\mathrm{hom})}_{\ket C}(x,\theta) = \underset{0 \leq k,l \leq c - 1}{\sum}c_k^* c_l f_{lk}(x) e^{i(l-k) \theta}.
\end{equation}
Computing the mean of $g^{(\mathrm{hom})}_{\ket C}$ over homodyne samples of some unknown state $\rho$ gives us an estimate of the fidelity with the state $\ket C$. 
This leads to the following single-mode fidelity estimation protocol using homodyne detection, assuming that the different copies of $\rho$ are $i.i.d.$, i.e., independent and identically distributed.
\begin{protocol}[Single-mode fidelity estimation with homodyne detection]\label{protocol_homodyne_single_fidelity}
Let $\ket{C} = \sum_{n = 0}^{c-1} c_n \ket{n}$ be a core state, where $c \in \N^*$. Let also $N \in \N^*$. Let $\rho^{\otimes(N+1)}$ be N+1 copies of an unknown single-mode (mixed) quantum state.
\begin{enumerate}
    \item Measure $N$ copies of $\rho$ with homodyne detection at a phase $\theta$ (sampled by measuring the phase of the local oscillator), obtaining the measurement outcomes $x^{(1)},\dots,x^{(N)}; \theta^{(1)}, \dots, \theta^{(N)} \in \R$.
    \item Compute the mean $F_C$ of the function $(x,\theta) \mapsto g^{(\mathrm{hom})}_{\ket C}(x,\theta)$ (given by Eq.~(\ref{eqn_homodyne_estimator})) over the measurement outcomes $(x^{(1)},\theta^{(1)}),\dots,(x^{(N)},\theta^{(N)}) \in \R^2$.
    \item $F_C$ gives the estimate of the fidelity between the one remaining copy of $\rho$ and $\ket{C}$.
\end{enumerate}
\end{protocol}
\noindent The following theorem summarizes the efficiency of this protocol:
\begin{theo}[Efficiency of Protocol~\ref{protocol_homodyne_single_fidelity}] \label{thrm_single_mode_homodyne}
    Let $\epsilon,\delta > 0$. With the notations of Protocol \ref{protocol_homodyne_single_fidelity}, the estimate of $F_C(\rho)$ is $\epsilon$-close to $F (\rho,\ket{C})$ with probability 1 - $\delta$ whenever $N \geq N_1$ with \\
\begin{equation}
    N_1 = \mathcal{O}\left( \left(\frac{C^\frac{10}{3}}{\epsilon}\right)^{2} \log \left(\frac{1}{\delta}\right) \right),
\end{equation}
where $C$ is the support size of the target core state.
\end{theo}
We direct the reader to section \ref{proof_thm_3.1} of the appendix for the proof, which combines Eq.~(\ref{equation_perfect_homodyne_estimation}) with Hoeffding's inequality \cite{Hoeffding1963}. 

Since $N_1$ scales inverse polynomially in $\epsilon$ and logarithmically in $1/\delta$, Protocol \ref{protocol_homodyne_single_fidelity} is efficient for fidelity estimation between an unknown single-mode quantum state and a core state. Since core states form a dense subset (for the trace norm) of the set of normalized single-mode pure quantum states, given any target normalized single-mode pure state $\ket{\psi}$, we can use our fidelity estimation protocol by targeting instead a truncation of the state $\ket{\psi}$ in the Fock basis in order to estimate the fidelity of any single-mode continuous-variable quantum state with the state $\ket{\psi}$ using homodyne detection. While of similar nature, our protocol differs from the one recently studied in \cite{Gandhari2023}, in that we perform direct fidelity estimation while their bounds apply to reconstructing a full density matrix up to some cutoff in Fock basis.

In a cryptographic scenario, the i.i.d.\ assumption cannot be assumed, that is, the state $\rho^{N+1}$ sent by the quantum device may not be of the form $\rho^{\otimes(N+1)}$. Protocol \ref{protocol_homodyne_single_fidelity} can be adapted to such a scenario using standard de Finetti reductions and random permutations of physical systems \cite{Renner2007,Gheorghiu2019}. From a similar method used in \cite{Chabaud2021efficient} and in section \ref{proof_no_iid} of the appendix for heterodyne detection, one may expect that this would result in a polynomial overhead in the number of samples required for a given precision \cite{Renner2009}, likely preventing the protocol from being experimentally realizable in the near term. Therefore we do not delve into it hereafter, focusing instead on practical verification methods, under the i.i.d.\ assumption.

\subsection{Multimode homodyne verification}

In this section, we show that Protocol \ref{protocol_homodyne_single_fidelity} can be extended to estimate a multimode fidelity witness, i.e., estimate a tight lower bound on the fidelity as explained in Eq.~(\ref{eqn_generalized_fidelity_witness}), for multimode target states of the form $\hat{D}(\boldsymbol{\beta})\hat{O} \underset{i=1}{\overset{m}{\bigotimes}} \ket{C_i}$, where $\ket{C_i}$ is a core state in the $i^{th}$ mode, and where $\hat{O}$ is an $m$-mode passive linear transformation with an associated $2m \times 2m$ block diagonal \textit{orthogonal} symplectic matrix $S_O$, which describes its action on the quadrature operators. We also detail later why specifically this class of state is verifiable by homodyne detection using the current protocols.

Given an $m$-mode output state $\rho$ and a target state $\ket{\psi} = \ket{\psi_1}\otimes \ldots \otimes \ket{\psi_m}$, we use the witness in Eq.\ (\ref{eq3.1}) for $k=1$ (the reason for not using a multi-mode fidelity witness will be clear after Lemma \ref{lemma_prop_heterodyne}). This witness, originally introduced in \cite{Chabaud2021efficient}, is based on single-mode fidelities and given by 
\begin{equation} \label{eqn_1_mode_fw}
    W (\rho,\psi) := 1 - \underset{i=1}{\overset{m}{\sum}} (1 - F(\rho_i,\psi_i)),
\end{equation}
where $F(\rho_i,\psi_i)$ is the fidelity between $\ket{\psi_i}$ and the reduced state $\rho_i = \mathrm{Tr}_{\{ 1,\ldots, m \} \backslash \{ i\}} [\rho]$.

By Lemma \ref{lemma_mm_fid_witness_2} with $k=1$ (see also \cite[Lemma 2]{Chabaud2021efficient}), 
\begin{equation}
            1 - m(1 - F(\rho,\psi)) \leq W (\rho,\psi) \leq F(\rho,\psi).
    \label{eq1.1}
    \end{equation}
\noindent Therefore, $W$ is a tight lower bound on the fidelity. Furthermore, $W$ is a function of only single-mode fidelities. Being able to estimate single-mode fidelities with single-mode pure states using homodyne detection thus allows us to witness the multimode fidelity of an $m$-mode output experimental state $\rho$ with any target pure product state using homodyne detection, by using Protocol \ref{protocol_homodyne_single_fidelity} for each subsystem in parallel. 

We make use of the properties of homodyne detection to extend the class of target states for which tight fidelity witnesses can be efficiently obtained:
an orthogonal passive linear transformation followed by a displacement operator before homodyne detection is equivalent to performing the homodyne detection directly, then classically post-processing the samples obtained. Formally,
\begin{lem}[Back-propagation of synchronous homodyne measurement] \label{lemma_prop_homodyne}
    Let $\boldsymbol{\beta} \in \C^m$ and let $\hat{V} = \hat{D}(\boldsymbol{\beta})\hat{O}$, where $\hat{O}$ is an $m$-mode passive linear transformation with the associated $m \times m$ block diagonal orthogonal unitary matrix $S_O$. Let $\boldsymbol{x'} = (\mathrm{S}_O)_1 \boldsymbol{x} + \boldsymbol{\beta}_\theta$, where $\boldsymbol{\beta}_\theta$ is the component of $\beta$ along the direction defined by $\theta$ in the phase space ($ \boldsymbol{\beta}_\theta  = \mathrm{Re}(\boldsymbol{\beta}) \cos(\theta) + \mathrm{Img}(\boldsymbol{\beta}) \sin(\theta) $) and $(\mathrm{S}_O)_1$ is expressed by 
\begin{equation}\label{eqn_S_O_main}
\mathrm{S}_O = \left[ {\begin{array}{cc}
   (\mathrm{S}_O)_1 & \boldsymbol{0} \\
   \boldsymbol{0} & (\mathrm{S}_O)_1 \\
  \end{array} } \right].
\end{equation}
Then,
\begin{equation}
    \Pi_{\boldsymbol{x'}_\theta} = \hat{V} \Pi_{\boldsymbol{x}_\theta} \hat{V}^\dagger,
\end{equation}
where
\begin{equation}
    \Pi_{\boldsymbol{x}_\theta} = \ket{\boldsymbol{x}}_\theta\bra{\boldsymbol{x}}_\theta
\end{equation}
is the POVM  for synchronous homodyne detection at angle $\theta$. $\ket{\boldsymbol{x}}_\theta$ is the eigenstate of the quadrature operator $\boldsymbol{X}_{\theta} = \frac{1}{\sqrt{2}}\left(\boldsymbol{a}e^{-i\theta} + \boldsymbol{a}^\dagger e^{i\theta}\right)$.
\end{lem}
\noindent We direct the reader to section \ref{proof_lemma_3.2} of the appendix for the proof. An important implication is that a certain class of quantum operations (orthogonal passive linear operations) before a synchronous homodyne detection can be inverted by doing classical post-processing on the samples of the synchronous homodyne detection. In particular, for such a transformation $\hat{V}$, if a multimode pure product state $\otimes_{i=1}^m \ket{\psi_i}$ can be efficiently verified using synchronous homodyne detection, then the state $\hat{V} \otimes_{i=1}^m \ket{\psi_i}$ can be efficiently verified using synchronous homodyne detection and efficient classical post-processing. We note that the class of operations that can be inverted using homodyne detection is narrower as compared to the invertible class of operations with heterodyne detection \cite{Chabaud2021efficient}. Furthermore, since we are required to measure all the modes at the same angle for inversion of operation to work, multi-mode fidelity witnesses are tomographically incomplete, and we are forced to restrict to single-mode fidelity witnesses.

Lemma \ref{lemma_prop_heterodyne} leads us to the following multimode homodyne verification protocol:
\begin{protocol}\label{prot_multimode_fidelity_estimation_homodyne}
(Multimode verification with synchronous homodyne detection). Let $c_1,\dots, c_m \in \N^*$. Let $\ket{C_i} = \sum_{n = 0}^{c_i - 1} c_{i,n} \ket{n}$ be a core state, for all $i \in \{1,\dots,m\}$. Let $\hat{O}$ be an $m$-mode orthogonal passive linear transformation with associated $m \times m$ block diagonal orthogonal matrix $\mathrm{S}_O$, given by Eq.~(\ref{eqn_S_O_main}), and $\boldsymbol{\beta} \in \C^m$. We write $\ket{\psi} = \hat{D}(\boldsymbol{\beta}) \hat{O} \bigotimes_{i=1}^m \ket{C_i}$ the $m$-mode target pure state. Let $N \in \N^*$ and let $\rho^{\otimes (N+1)}$ be $N+1$ copies of an unknown $m$-mode (mixed) quantum state $\rho$.
\begin{enumerate}
    \item Measure all $m$ subsystems of $N$ copies of $\rho$ with synchronous homodyne detection at angle $\theta_k$, chosen randomly, with $k \in \{1,\dots,N\}$, obtaining the vectors of measurement outcomes $\boldsymbol{x'}_1,\dots,\boldsymbol{x'}_N \in \R^m$.
    \item For all $k \in \{1,\dots,N\}$, compute the vectors $\boldsymbol{x}_k = (\mathrm{S}_O)_1^{-1} (\boldsymbol{x'}_k - \boldsymbol{\beta}_{\theta_k})$, where $\boldsymbol{\beta}_\theta  = \mathrm{Re}(\boldsymbol{\beta}) \cos(\theta) + \mathrm{Im}(\boldsymbol{\beta}) \sin(\theta)$
    is the component of $\boldsymbol{\beta}$ along the $\theta$ direction. We write $\boldsymbol{x}_k = (x_{1k},\dots,x_{mk}) \in \R^m$.
    \item For all $i \in \{1,\dots, m \}$, compute the mean $F_{C_i}$ of the function $(x,\theta) \mapsto g^{(\mathrm{hom})}_{\ket{C_i}}(x,\theta)$ (Eq. (\ref{eqn_homodyne_estimator})) over the measurement outcomes $x_{i1},\dots,x_{iN} \in \R$.
    \item Compute the fidelity witness estimate $W_{\psi} = 1 - \sum_{i=1}^m (1 - F_{C_i})$.
    \item \textbf{Accept} the remaining state $\rho$ if $W_{\psi} > 1 - 3\epsilon/4$ and $\textbf{Reject}$ if $W_{\psi} < 1 - \epsilon(m^2 + 1/4)$.
\end{enumerate}
\end{protocol} 
\noindent$W_{\psi}$ obtained using Protocol \ref{prot_multimode_fidelity_estimation_homodyne} provides a lower bound on the fidelity between any $m$-mode (mixed) quantum state $\rho$ and $\ket{\psi}$. The efficiency of the protocol is summarized as follows:
\begin{theo}[Efficiency of Protocol \ref{prot_multimode_fidelity_estimation_homodyne}]\label{thm_multimode_homodyne}
Protocol \ref{prot_multimode_fidelity_estimation_homodyne} solves \textsf{Verify}($\ket{\psi},c,s,\delta$) (Problem \ref{pb:verification_problem}) for the class of target states
\begin{equation}
    \ket{\psi} = \hat{D}(\boldsymbol{\beta}) \hat{O} \underset{i=1}{\overset{m}{\bigotimes}} \ket{C_i},
\end{equation}
with
\begin{eqnarray}\label{eq1}
    N &=& \mathcal{O}\left( \frac{m^2}{\varepsilon^2} C^{\frac{20}{3}} \mathrm{log} \left(\frac{m}{\delta} \right) \right), \nonumber \\
    c &=& \varepsilon, \nonumber \\
    s &=& m\varepsilon,
\end{eqnarray}
where $C = \underset{i \in \{1,\dots,m \}}{\mathrm{max}} c_i$, where $c_i$ is the support size for the core state in the mode i.
\end{theo}
\noindent We direct the reader to section \ref{proof_thm_3.2} in the appendix for the proof. 

Protocol \ref{prot_multimode_fidelity_estimation_homodyne} provides an efficient method to witness the fidelity between an unknown $m$-mode (mixed) quantum state $\rho$ and $\ket{\psi}$, where $\ket{\psi} = \hat{D}(\boldsymbol{\beta}) \hat{O} \underset{i=1}{\overset{m}{\bigotimes}} \ket{C_i}$ using homodyne sampling. Note that the restriction on the class of verifiable state stems from the fact that only a certain class of unitary gates are invertible via homodyne measurements and classical post-processing (Lemma \ref{lemma_prop_homodyne}). In the next section, we detail the modification to the verification protocol in case the homodyne detections are imperfect.

\subsection{Imperfect homodyne detection}

Protocols \ref{protocol_homodyne_single_fidelity} and \ref{prot_multimode_fidelity_estimation_homodyne} are described under the assumption of unit efficiency of the homodyne detection. For an arbitrary quantum efficiency $\eta$, $f_{lk} (x)$ and $P_\rho (x,\theta)$ should be modified as \cite{MauroDAriano2003}:
\begin{eqnarray}\label{eqn_noisy_estimator}
    f^\eta_{kl}(x)&=& e^{i\frac{\pi}{2}(l-k)} \sqrt{\frac{k!}{l!}} \nonumber \\ && \hspace{-30mm} \times \int_{-\infty}^{\infty} dt |t| e^{\frac{1-2\eta}{2\eta} t^2 - i2tx} t^{(l-k)} L_k^{(l-k)} (t^2),
\end{eqnarray}
and 
\begin{equation}
    \begin{aligned}
        P_\rho^\eta (x,\theta) &= \sqrt{\frac{2\eta}{\pi (1 - \eta)}} \\
        &\times\int_{-\infty}^{\infty} dx' \, e^{-\frac{2\eta}{1-\eta}(x-x')^2} P_\rho(x', \theta),
    \end{aligned}
\end{equation}
where $L^a_n$ are the generalized Laguerre polynomials given by 
\begin{equation}
    L^a_n (x) = \frac{e^x x^{-a}}{n!} \frac{d^n}{dx^n} (e^{-x}x^{n+a}).
\end{equation}
Then, both protocols still work in the case of homodyne detection with non-unit quantum efficiency, by using the noisy estimators in Eq.~(\ref{eqn_noisy_estimator}). Note that Eq.~($\ref{eqn_noisy_estimator}$) imposes an efficiency $\eta>\frac12$ for the estimator to be well-defined.


\section{Verification protocols with heterodyne detection}
\label{sec:heterodyne_sampling}

\subsection{Heterodyne detection}

Heterodyne detection, also known as double homodyne, dual-homodyne, or eight-port homodyne detection \cite{Ferraro2005}, constitutes a single-mode Gaussian measurement that projects onto (unnormalized) coherent states. In mathematical terms, performing single-mode heterodyne detection on a state $\rho$ involves sampling from the Husimi $Q$ phase-space quasiprobability distribution \cite{Yuen1980}:
\begin{equation}
    Q_\rho(\alpha) = \frac{1}{\pi} \bra{\alpha}\rho\ket{\alpha},
\end{equation}
where $\ket{\alpha} = e^{-\frac12 |\alpha|^2} \sum_{n \geq 0} \frac{\alpha^n}{\sqrt{n!}} \ket{n}$ is the coherent state of amplitude $\alpha \in \C$. Experimentally, heterodyne measurement is implemented by mixing the state to be measured with vacuum through a beam splitter and doing a double homodyne detection, where the state to be measured is split into two using a beam splitter and two homodyne detections are implemented to measure perpendicular quadratures, with outcomes $x$ and $p$, and get the complex number $\alpha = x + i p$ as the output.

On the other hand, an \textit{unbalanced heterodyne measurement} is implemented by unbalancing the beam splitter splitting the state into two and changing the phase of the local oscillator, and allows the measurement of squeezed coherent states in phase space. Mathematically, the POVMs of unbalanced heterodyne detection over $m$ modes with unbalancing parameters $\boldsymbol{\xi} = \{\xi_1,\ldots, \xi_m \} \in \C^m$ are given by
\begin{equation}\label{eqn_unbalanced_heterodyne_projector}
    \Pi_{\boldsymbol{\alpha}}^{\boldsymbol{\xi}} = \frac{1}{\pi^m} \ket{\boldsymbol{\alpha,\xi}} \!\bra{\boldsymbol{\alpha,\xi}},  
\end{equation}
for all $\boldsymbol{\alpha} = \{\alpha_1,\ldots,\alpha_m \} \in \C^m$. Here $\ket{\boldsymbol{\alpha,\xi}} = \bigotimes_{i = 1}^m \ket{\alpha_i,\xi_i}$ is a tensor product of squeezed coherent states $\ket{\alpha_i,\xi_i}$. Writing $\xi = r e^{i\theta}$, the unbalancing parameter is related to the optical setup by $r = \log\left(\frac{T}{R}\right)$, where $T$ and $R$ are the reflectance and transmittance of the unbalanced beam splitter, respectively, with the phase of the local oscillator being $-\frac{\theta}{2}$ \cite{Chabaud2021efficient}.

\subsection{k-mode heterodyne fidelity estimation}
\label{sec:khetfidest}

We now elaborate on how we can estimate the fidelity of a $k$-mode system using heterodyne detection, by generalizing heterodyne estimator functions from \cite{chabaud2020building,Chabaud2021efficient}.

Given an output experimental state $\rho$, the individual density matrix elements are given as
\begin{equation}
    \rho_{mn} = \underset{\alpha \in \C}{\int} {Q}_{\rho} (\alpha) P_{mn} (\alpha) d\alpha,
\end{equation}
where ${P_{mn}}$ is the Glauber--Sudarshan $P$ function of the operator $\ket{n}\bra{m}$ and ${Q}_{\rho}$ is the Husimi $Q$ function of $\rho$. Since the Glauber--Sudarshan $P$ function is often singular, we use a regularised version of the $P$ function as our heterodyne estimator function. It is defined as
\begin{equation}
    g_{\boldsymbol{mn}}^{\boldsymbol{p} (\mathrm{het})} (\boldsymbol 
  {\alpha},\tau) = g_{m_1 n_1}^{p_1 (\mathrm{het})} (\alpha_1,\tau) \cdots g_{m_k n_k}^{p_k (\mathrm{het})} (\alpha_k,\tau),
\end{equation}
where $g_{m n}^{p(\mathrm{het})} (\alpha, \tau)$ is given by  \begin{eqnarray}\label{eqn_heterodyne_estimator}
      \nonumber g_{m,n}^{p (\mathrm{het})} (z,\tau) &=& \underset{j=0}{\overset{p - 1}{\sum}} (-1)^j \tau^j f_{m+j,n+j} (z,\tau) \\
      && \times \sqrt{\binom{m+j}{m} \binom{n+j}{n}},
\end{eqnarray}
where for all $k,l$,
\begin{equation}
    f_{k,l} (z,\tau) := \frac{1}{\tau^{1+\frac{k+l}{2}}} e^{\left(1 - \frac{1}{\tau} \right)zz^*} \mathcal{L}_{l,k} \left( \frac{z}{\sqrt{\tau}} \right),
\end{equation}
with
\begin{eqnarray}
      \mathcal{L}_{l,k} (z) &=& e^{zz^*} \frac{(-1)^{k+l}}{\sqrt{k! l!}} \frac{\partial^{k+l}}{\partial z^k \partial z^{*l}} e^{- zz^*} \\
    &&\hspace{-10mm} = \sum_{p=0}^{\mathrm{min}(k,l)}\frac{\sqrt{k!l!}(-1)^p}{p!(k-p)!(l-p)!} z^{l-p} z^{*k-p}.
\end{eqnarray}
Here $p_i$ and $\tau$ are free parameters, for $i \in \{1,\ldots,k\}$. 

Unlike the homodyne estimator functions from section \ref{sec:homodyne_sampling}, these regularised heterodyne estimators are no longer unbiased, however their bias can be controlled analytically:
\begin{lem}[Biased heterodyne estimators] \label{lemma_density_estimator}
    Let $k\in\mathbb N^*$, let $p_1,\ldots,p_k \in \N^*$, let $m_1,n_1,\ldots,m_k,n_k \in \N$, $\boldsymbol{m,n} = \{m_1,n_1,\ldots,m_k,n_k \}$ and let $0 < \tau \leq 1$. Let $\rho = \sum_{\boldsymbol{m,n} = \boldsymbol{0}}^{\boldsymbol{\infty}  }\rho_{\boldsymbol{m,n}} \ket{\boldsymbol{m}}\!\bra{\boldsymbol{n}} $. Then,

    \begin{equation}
        \begin{aligned}
    &\left| \rho_{\boldsymbol{m,n}} - \underset{\boldsymbol{\alpha} \leftarrow Q_{\rho}}{\E}\left[g_{m_1 n_1}^{p_1 (\mathrm{het})} (\alpha_1,\tau) \ldots g_{m_k n_k}^{p_k (\mathrm{het})} (\alpha_k,\tau) \right]  \right|  \\&\quad\quad\quad\quad\quad\quad\quad\quad\leq \varepsilon_{\mathrm{bias}}(\bm{p},k) .
        \end{aligned}
    \end{equation}
\noindent Here $\varepsilon_{\mathrm{bias}}(\bm{p},k)$ is a number independent of $\rho$ which can be brought arbitrarily close to zero by tuning the free parameters, whose expressions is given in section \ref{proof_lemma_2.3} of the appendix.
\end{lem}

\noindent We direct the reader to section \ref{proof_lemma_2.3} of the appendix for the proof. 

Since sampling from the Husimi $Q$ function corresponds to heterodyne detection, Lemma \ref{lemma_density_estimator} allows us to estimate the individual elements of a $k$-mode density matrix using heterodyne detection by computing the mean of the function $\boldsymbol{z} \mapsto g_{\boldsymbol{mn}}^{\boldsymbol{p} (\mathrm{het})} (\boldsymbol{z},\tau)$ over these samples, while controlling analytically the approximation error. The result generalises the bound obtained for the heterodyne estimator in \cite{Chabaud2021efficient} in the case $k=1$. We can control the approximation error by controlling the value of the free parameters $p_1,\dots,p_k,\tau$.

Therefore, given an experimental state $\rho$ and target core state $\ket{C} = \underset{\boldsymbol{0} \leq \boldsymbol{n} \leq \boldsymbol{c -1}}{\sum} c_{\boldsymbol{n}} \ket{\boldsymbol{n}}$ where $\boldsymbol{c} = \{c_1,\ldots,c_k\} \in (\N^*)^k$, $\boldsymbol{n} = \{n_1, \ldots, n_k \}$ and for all $\boldsymbol{p} = \{p_1,\dots,p_k\} \in (\N^*)^k$, and $0 < \tau \leq 1$, we define the function

\begin{equation}
    g^{(\mathrm{het})}_{\mathrm{k-est}}(\boldsymbol{\alpha},\tau, \boldsymbol{p}) = \underset{\boldsymbol{0} \leq \boldsymbol{m,n} \leq \boldsymbol{c - 1}}{\sum}  c_{\boldsymbol{m}}^* c_{\boldsymbol{n}} g_{\boldsymbol{m n}}^{\boldsymbol{p}} (\boldsymbol{\alpha},\tau),
     \label{mm_fidelity_estimate_eqn}
\end{equation}
for all $\boldsymbol{\alpha} = \{\alpha_1,\ldots,\alpha_k\} \in \C^k$. Computing the mean of $g^{(\mathrm{het})}_{\mathrm{k-est}}(\boldsymbol{\alpha},\tau, \boldsymbol{p})$ over $N$ heterodyne detections gives us an estimate of the fidelity between the experimental state and the target core state. Based on this, we describe a protocol for $k$-mode fidelity estimation using heterodyne sampling, with the assumption that the different copies sent by the quantum device are i.i.d.:

\begin{protocol}\label{protocol_k_mode_fidelity}
(k-mode fidelity estimation using heterodyne detection). Let $c_1,\dots,c_k \in \N^*$ and let $\ket{C} = \underset{\boldsymbol{0} \leq \boldsymbol{n} \leq \boldsymbol{c -1}}{\sum} c_{\boldsymbol{n}} \ket{\boldsymbol{n}}$ be a core state over $k$ modes. Let also $N \in \N^*$ and $\boldsymbol{p} = \{p_1,\dots,p_k\} \in (\N^*)^k$ and let $0 < \tau < 1$ be free parameters. Let $\rho^{\otimes (N+1)}$ be N+1 copies of an unknown k-mode (mixed) quantum state $\rho$.
\begin{enumerate}
    \item Measure $N$ copies of $\rho$ with homodyne detection, obtaining the measurement outcomes $\boldsymbol{\alpha}^{(1)},\dots,\boldsymbol{\alpha}^{(N)} \in \C^k$.
    \item For $ i \in \{1,\dots,k\}$ Compute the mean of the function $\boldsymbol{\alpha} \mapsto g^{(\mathrm{het})}_{\mathrm{k-est}}(\boldsymbol{\alpha},\tau, \boldsymbol{p})$ over the measurement outcomes $\boldsymbol{\alpha}^{(1)},\ldots,\boldsymbol{\alpha}^{(N)} \in \R^k$
    \item The mean gives an estimate of the fidelity $F_C^{(k-est)} (\tau)$. 
\end{enumerate}
\end{protocol}
\noindent $F_C^{(k-est)} (\tau)$ gives an estimate of the fidelity between the remaining copy of $\rho$ and the target state. The efficiency of the protocol is summarized by the following result:
\begin{theo}[Efficiency of Protocol \ref{protocol_k_mode_fidelity}] \label{theorem_k_mode_fidelity_est}
Let $\lambda > 0$. With the notations of Protocol \ref{protocol_k_mode_fidelity}, the statistical average of $g^{(\mathrm{het})}_{\mathrm{k-est}}(\boldsymbol{\alpha},\tau, \boldsymbol{p})$ over $N$ samples, $F_C^{\mathrm{k-est}}(\tau)$, is $\epsilon$-close to the real fidelity $F(\rho,\ket{\boldsymbol{C}})$ with the probability $1- \delta$, with $\epsilon$ and $\delta$ given by

\begin{equation}\label{optimization_equation_k-mode_fidelity_1}
     \epsilon = \lambda + \epsilon_{(\mathrm{bias})}(\boldsymbol{C},\boldsymbol{p},\tau),
\end{equation}
\begin{equation}\label{optimization_equation_k-mode_fidelity_2}
    \delta = 2\exp\left ( - \frac{N\lambda^2}{2(R(\boldsymbol{C},\boldsymbol{p},\tau)^2})\right),
\end{equation}
where $\lambda$ represents the statistical error, $\epsilon_{(\mathrm{bias})}(\boldsymbol{C},\boldsymbol{p},\tau)$ represents the biased error of the estimator independent of $\rho$ which can be made arbitrarily close to zero by tuning the free parameters $\boldsymbol{p}$ and $\tau$, and $R(\boldsymbol{C},\boldsymbol{p},\tau)$ represents the range of the estimator $g^{(\mathrm{het})}_{\mathrm{k-est}}(\boldsymbol{\alpha},\tau, \boldsymbol{p})$. The expressions of $\epsilon_{(\mathrm{bias})}(\boldsymbol{C},\boldsymbol{p},\tau)$ and $R(\boldsymbol{C},\boldsymbol{p},\tau)$ are given in section \ref{proof_theorem_2.1} of the appendix.

\end{theo}
We direct the reader to section \ref{proof_theorem_2.1} of the appendix for the proof, which combines Lemma \ref{lemma_density_estimator}. Therefore, after implementing protocol \ref{protocol_k_mode_fidelity}, we can verify if the remaining copy of the state $\rho$ is close to the target state (in trace distance) with high probability. The value of the free parameters $\boldsymbol{p}$ and $\tau$ can be optimized according to the experiment, and obeying the constraint imposed by Eqs.~(\ref{optimization_equation_k-mode_fidelity_1}) and (\ref{optimization_equation_k-mode_fidelity_2}) to minimize the number of measurements needed to be done to obtain the required precision. Note that the term $R(\boldsymbol{C},\boldsymbol{p},\tau)$ scales roughly as $R^k$, where $R$ is the range of the single-mode estimator $g_{mn}$, therefore there is an exponential growth in the number of samples required to achieve a desired level of precision in terms of $k$ (we prove this heuristically in section \ref{proof_theorem_2.1} of the appendix). This is expected as tomography generally scales exponentially in the number of modes.

We detail how to remove the i.i.d.\ assumption in section \ref{proof_no_iid} of the appendix. The underlying idea is that with a limited number of samples, if we choose to measure certain samples while disregarding others, we can treat the overall state as permutation invariant. Note that the permutation of physical systems applies to the multiple rounds of the verification protocol, i.e., involving multiple \(m\)-mode states. The essential takeaway from this proof is that removing the i.i.d.\ assumption produces an overhead in terms of the number of samples we need to take to certify the state (to see it from Theorem \ref{thm_noiid_multi}, consider the extra error term, and the higher chance of the error not being bounded). While the protocol is efficient from a theoretical standpoint, combining this with the exponential overhead in the number of samples that comes from using a multimode fidelity estimator, the cost of removing the i.i.d assumption becomes unrealistic in any near-term experimental scenario.
\subsection{Robust multimode heterodyne verification}
The $k$-mode fidelity estimation protocol described in the previous section allows us to verify a large class of $m$-mode states, by combining it with the $k$-mode fidelity witness introduced in Eq.~(\ref{eq3.1}), together with the property of heterodyne measurements described by the following Lemma, first proved in \cite{Chabaud2021efficient}:
\begin{lem}[Back-propagation of heterodyne measurement \cite{Chabaud2021efficient}] \label{lemma_prop_heterodyne}
    Let $\boldsymbol{\beta,\xi} \in \C^m$ and let $\hat{V} = \hat{S}(\boldsymbol{\xi}) \hat{D} (\boldsymbol{\beta}) \hat{U}$, where $\hat{U}$ is an $m$-mode passive linear transformation with $m \times m$ unitary matrix $U$. For all $\boldsymbol{\gamma} \in \C^m$, let $\boldsymbol{\alpha} = U^\dagger (\boldsymbol{\gamma} - \boldsymbol{\beta})$. Then,

    \begin{equation}
        \Pi^{\boldsymbol{\xi}}_{\boldsymbol{\gamma}} = \hat{V} \Pi^{\boldsymbol{0}}_{\boldsymbol{\alpha}} \hat{V}^\dagger
    \end{equation}
where the expression for $\Pi^{\boldsymbol{\xi}}_{\boldsymbol{\gamma}}$ is given by Eq.~(\ref{eqn_unbalanced_heterodyne_projector}). 
\end{lem}
An important consequence of Lemma \ref{lemma_prop_heterodyne} is that the specific type of quantum operations $\hat V$ described in the Lemma performed prior to a balanced heterodyne measurement can be effectively reversed by adjusting the heterodyne measurement to be appropriately unbalanced and then applying classical operations to the resulting samples and hence, combined with Protocol \ref{protocol_k_mode_fidelity}, this allows us to verify the class of states described by
\begin{equation}\label{eqn_certifiable_states}
    \left( \underset{i=1}{\overset{m}{\bigotimes}} \hat{G}_i \right)\! \hat{U} \!\left(\underset{i=1}{\overset{m/k}{\bigotimes}}  \ket{\boldsymbol{C_i}} \right)\! = \hat{S} (\boldsymbol{\xi}) \hat{D} (\boldsymbol{\beta}) \hat{U} \!\left(\underset{i=1}{\overset{m/k}{\bigotimes}}  \ket{\boldsymbol{C_i}} \right)
\end{equation}
constitute the class of efficiently verifiable states, where $\ket{\boldsymbol{C_i}}$ is a $k$-mode core state and $i \in \{1,\ldots,m/k \}$ are blocks of $k$-modes. We now describe the protocol used to verify the class of states described by Eq.~(\ref{eqn_certifiable_states}):

\begin{protocol}\label{protocol_multimode_k_fidelity witness}
    (Multimode state verification with heterodyne detection using k-mode fidelity witness). Let \(\bm c_1, \ldots, \bm c_{m/k} \in (\mathbb{N}^{*})^k\). Let \(\ket{\bm C_i} = \sum_{\bm n=0}^{\bm c_i-1} c_{i,\bm n} |n\rangle\) be a $k$-mode core state, for all \(i \in \{1, \ldots, m/k\}\). Let $\hat{U}$ be an \(m\)-mode passive linear transformation with the associated \(m \times m\) unitary matrix \(U\), and let $\boldsymbol{\beta,\xi} \in \C^m$. We write \(|\psi\rangle = S(\boldsymbol{\xi})D(\boldsymbol{\beta}) \hat{U} \otimes_{i=1}^m \ket{\bm C_i}\) as the \(m\)-mode target pure state. Let \(N \in \mathbb{N}^*\) and let \(p_1, \ldots, p_m \in \mathbb{N}^*\) and \(0 < \tau_1, \ldots, \tau_{m/k} < 1\) be free parameters. Let \(\rho^{\otimes N+1}\) be \(N + 1\) copies of an unknown \(m\)-mode (mixed) quantum state \(\rho\).

\begin{enumerate}
    \item Measure all \(m\) subsystems of \(N\) copies of \(\rho\) with unbalanced heterodyne detection with unbalancing parameters $\boldsymbol{\xi} =$ \( \{\xi_1, \ldots, \xi_m\}\), obtaining the vector of measurement outcomes \( \boldsymbol{ \gamma^{(1)}, \ldots, \gamma^{(N)}}\in \mathbb{C}^m\).
    \item For all \(k \in \{1, \ldots, N\}\), compute the vectors $\boldsymbol{\alpha}^{(k)} = U^\dagger (\boldsymbol{\gamma}^{(k)} - \boldsymbol{\beta})$. We write $\boldsymbol{\alpha}^{(k)} = (\alpha_1^{(k)},\ldots,\alpha_m^{(k)} ) \in \C^m$
    \item Partition the $m$ modes into $m/k$ blocks of k-mode states. For each of the blocks $i \in \{1,\ldots,m/k \}$, calculate the mean $F_{C_i}^{(k-est)}(\tau_i)$ of the function $\boldsymbol{z} \mapsto g_{\mathrm{het}_i}^{k-est} (\boldsymbol{z},\tau_i, \boldsymbol{p}_i)$ (defined in Eq.~(\ref{mm_fidelity_estimate_eqn}), where $\boldsymbol{p}_i$ is the vector containing the free parameters $p$ for the $k$ modes in the block), over the samples $\boldsymbol{\alpha_i^{(1)},\ldots,\alpha_i^{(N)}} \in \C^k$, using Protocol \ref{protocol_k_mode_fidelity}.
    \item Compute the fidelity witness estimate \(W_\psi^{(k)} = 1 - \sum_{i=1}^{m/k} (1 - F_{C_i}^{(k-est)})\).
    \item \textbf{Accept} the remaining state $\rho$ if $W_{\psi} > 1 - 3\epsilon/4$ and $\textbf{Reject}$ if $W_{\psi} < 1 - \epsilon(m^2/k^2 + 1/4)$.
\end{enumerate}
\end{protocol}

$W_{\psi}^{(k)}$ gives an estimate of the lower bound on the fidelity between unknown quantum state $\rho$ and the target state $S(\boldsymbol{\xi})D(\boldsymbol{\beta}) \hat{U} \left(\underset{i=1}{\overset{m/k}{\bigotimes}}  \ket{\boldsymbol{C_i}}\right)$. The efficiency of the Protocol is summarized by the following Theorem:

\begin{theo}[Efficiency of Protocol \ref{protocol_multimode_k_fidelity witness}]\label{theorem_mm_witness_inequality}
    Protocol \ref{protocol_multimode_k_fidelity witness} solves \textsf{Verify}($\ket{\psi},c,s,\delta$) (Problem \ref{pb:verification_problem}) for the class of target states
\begin{equation}
    \ket{\psi} = S(\boldsymbol{\xi})D(\boldsymbol{\beta}) \hat{U} \left(\underset{i=1}{\overset{m/k}{\bigotimes}}  \ket{\boldsymbol{C_i}}\right),
\end{equation} 
with
\begin{eqnarray}
    N &=& {\mathcal{O}} \left(\frac{(K_1 m^{K_2})^{k}}{\varepsilon^{K_2 k}} \log \left(\frac{m}{k\delta} \right)\right), \nonumber \\
    c &=& \varepsilon, \nonumber \\
    s &=& \frac{m}{k}\varepsilon,
\end{eqnarray}
where $K_1,K_2 > 0$ are constants depending on the support size of the core target state and given in section \ref{proof_theorem_2.2} of the Appendix.
\end{theo}
\noindent We direct the reader to section \ref{proof_theorem_2.2} of the appendix for the proof. Note that the restriction on the class of verifiable state stems from the fact that only a certain class of unitary gates are invertible via heterodyne measurements and classical post-processing (Lemma \ref{lemma_prop_heterodyne}). Therefore, this protocol gives us the method of certifying the quantum state produced by an experiment/quantum device, by grouping $k$ modes together. Note that for $k$ = 1, the certification can be done in polynomial time  \cite{Chabaud2021efficient}. Whereas for any other $k$, the number of samples required to get a desired precision scale exponentially in $k$, but we get a tighter lower bound on the fidelity. Therefore, it becomes useful to recognize the cases where single-mode fidelity witnesses are close enough to the actual fidelity, and where it is useful to take $k > 1$, as discussed in section \ref{sec:num_analysis}. In addition, the $k$-mode fidelity witness allows us to efficiently verify a class of non-Gaussian states not efficiently verifiable by single-mode fidelity witness, as described in the section below.
\subsection{Verification of doped Gaussian states}\label{sec_verification_doped}

Building on $k$-mode fidelity witness, this section details the protocol to verify a class of \textit{doped Gaussian states}, which are non-Gaussian states prepared by applying Gaussian unitaries and at most $t$ non-Gaussian $\kappa$-local unitaries on the vacuum state \cite{AnnaMele2024}. However, since the classical post-processing only applies to a specific class of Gaussian unitaries (Lemma \ref{lemma_prop_heterodyne}), we restrict the Gaussian unitary gates being applied to the input vacuum state (when preparing the doped Gaussian state) to be displacement operators and passive linear unitary gates (we call them \textit{displaced passive linear unitary gates}). In particular, this protocol builds on the modified version of Theorem 6 of \cite{AnnaMele2024}, which we detail below, with a visualization for the Theorem given in Figure \ref{t-doped-gaussian_image_theorem}:

\begin{figure*}[ht]
    \centering
    \includegraphics[width =\linewidth]{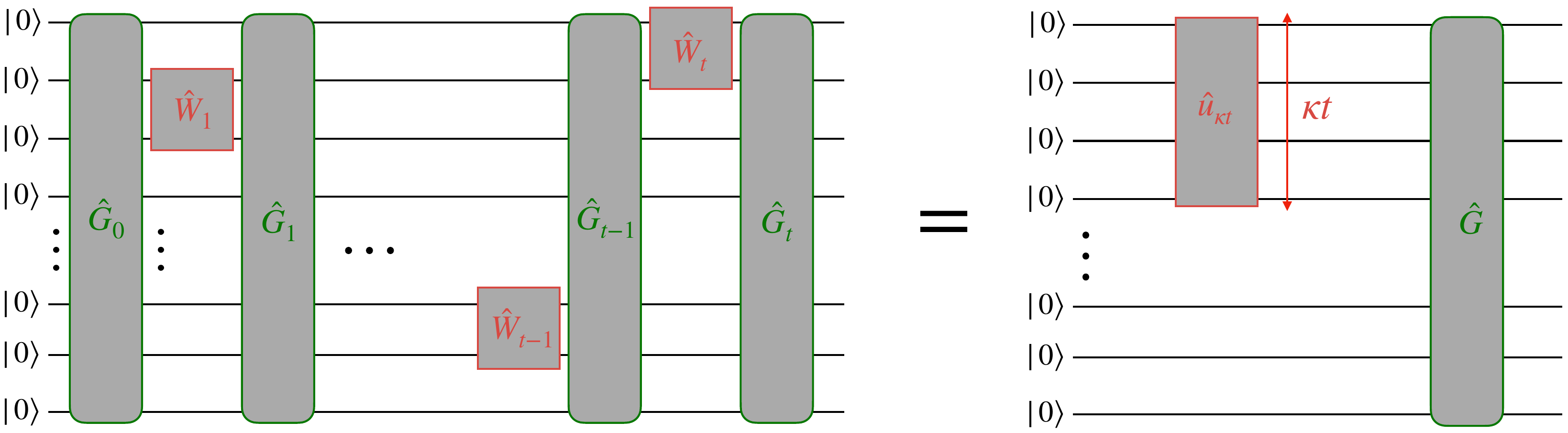}
    \caption{ Pictorial representation of a \(t\)-doped Gaussian state from \cite{AnnaMele2024}. By definition, a \(t\)-doped Gaussian state vector \(\ket{\psi} \) is prepared by applying Gaussian unitaries \(G_0, \ldots, G_t\) (green boxes) and at most \(t\) non-Gaussian \(\kappa\)-local unitaries \(W_1, \ldots, W_t\) (red boxes) to the \(n\)-mode vacuum. $G_0,\ldots,G_t$ are restricted to passive linear transformations and displacement operators for the classical post-processing to work. The figure also shows the decomposition proved in Theorem \ref{t-doped_Gaussian_thm}, which establishes that all the non-Gaussianity in \(\ket{\psi}\) can be compressed in a localized region consisting of \(\kappa t\) modes via a Gaussian unitary \(G\). 
}
    \label{t-doped-gaussian_image_theorem}
\end{figure*}

\begin{theo}\label{t-doped_Gaussian_thm}
(Non-Gaussianity compression in t-doped Gaussian unitaries and states \cite{AnnaMele2024}). If \( n \geq \kappa t \), any n-mode t-doped Gaussian unitary \( U_{\kappa,t} \) with the Gaussian unitaries restricted to passive linear transformations and displacement operators can be decomposed as
\begin{equation}
    \hat{U}_{\kappa,t} = \hat{G}(\hat{u}_{\kappa t} \otimes \mathbb{I}_{n-\kappa t}) \hat{G}_{\text{passive}},
\end{equation}
for some suitable displaced passive linear unitary \( \hat{G} \), passive linear unitary \( \hat{G}_{\text{passive}} \), and \(\kappa t \)-mode (non-Gaussian) unitary \( \hat{u}_{\kappa t} \). In particular, any n-mode t-doped Gaussian state can be decomposed as
\begin{equation}\label{eq:doped_gaussian_states}
    \ket{\psi} = \hat{G} \left( \ket{\phi_{\kappa t}} \otimes \ket{0}^{(n - \kappa t)} \right)
\end{equation}
for some displaced passive linear unitary gate \( \hat{G} \) and \( \kappa t \)-mode (non-Gaussian) state \( \ket{\phi_{\kappa t}} \).

\end{theo}
\noindent A visual representation of this Theorem is given in Figure \ref{t-doped-gaussian_image_theorem}.

For the circuits where $\ket\phi$ has a finite  support size in the Fock space (note that if $\ket{\phi}$ has bounded energy, we can always approximate it with a state of finite support size), efficient verification of such doped Gaussian states is a direct application of Protocol $\ref{protocol_multimode_k_fidelity witness}$ together with Theorem $\ref{t-doped_Gaussian_thm}$. The details of the protocol are given in section \ref{sec:verify_t-doped} of the appendix, where it is shown that verification is efficient for $\kappa t=O(\log m)$.  Note that the efficient verification of such states is not possible with single-mode fidelity witness, since the input state to the displaced passive linear unitary gate is not a tensor product of single-mode states. Furthermore, we cannot verify the correct preparation of the state by simply learning the output experimental state, since the learning is only efficient for $\kappa t = \mathcal O(1)$ \cite{AnnaMele2024}, in contrast to our verification protocol that is efficient up to $\kappa t = \mathcal O(\log m)$. Finally, we note that if there are single-mode squeezing operators at the end of displaced passive-linear unitary gates in Eq. (\ref{eq:doped_gaussian_states}), we can still classically post-process the measurement outcomes and efficient verification is still possible for $\kappa t = \mathcal O(\log m)$.

\subsection{Imperfect heterodyne detection}

Our heterodyne verification protocols have been described under the assumption of unit efficiency of the detector. For an arbitrary quantum efficiency $\eta$, we show in section \ref{noisy_hetero} of the appendix that the single-mode heterodyne estimator $g_{m,n}^{p (\mathrm{het})}$ and the Husimi $Q$ function should be modified as:
\begin{eqnarray}
g_{k,l}^{p(het)}(z,\eta,\tau)&:=&\sum_{j=0}^{p-1}(-1)^jf_{k+j,l+j}(z)t^j \nonumber \\ && \hspace{2mm}\times\sqrt{\binom{j+k}k\binom{j+l}l},
\label{eqn_noisy_heterodyne_estimator}
\end{eqnarray}
where $t=1-\eta+\tau \eta^2$, and $Q^\eta(\alpha):=\frac1{\eta}W(\frac\alpha{\sqrt{\eta}},1-2\eta^{-1})$ where for all $s<0$,
\begin{equation}
W(\alpha,s)=-\frac{2}{1+s}\int_{\beta\in\mathbb C}Q(\beta)e^{\frac2{1+s}|\alpha-\beta|^2}\frac{d^2\beta}\pi.\label{eqn_smoothened_Q}
\end{equation}

Then, heterodyne verification protocols still work in the case of heterodyne detection with non-unit quantum efficiency, by using the noisy estimator in Eq.~(\ref{eqn_noisy_heterodyne_estimator}) and smoothened Husimi $Q$ function given by Eq.~(\ref{eqn_smoothened_Q}). Unlike noisy homodyne verification, the estimator functions are well-defined for all possible values of $\eta>0$.


\section{Comparing homodyne and heterodyne verification}\label{sec:homvshet}

After describing protocols for state verification using homodyne and heterodyne detection, we are in a position to compare these two Gaussian measurements in terms of their utility for state verification. For multimode state verification using witnesses based on single-mode fidelities, the protocol using homodyne sampling is more efficient for its verifiable states (unbiased estimators and quadratic polynomial dependence on $\epsilon$ for homodyne sampling protocol compared to biased estimators and a higher polynomial dependence for heterodyne sampling). However, undoing quantum operations in classical post-processing is only possible for orthogonal passive linear unitaries in homodyne sampling, as compared to the possibility of classical post-processing all passive linear unitaries with heterodyne sampling. Therefore, heterodyne sampling is applicable to a wider range of efficient verification scenarios as compared to homodyne sampling.

Furthermore, since classical post-processing with homodyne sampling requires all the modes to be measured at the same angle (which we refered to as \textit{synchronous homodyne detection}), the sampling over multiple modes is tomographically incomplete. Therefore, estimation of witnesses based on $k$-mode fidelities is only possible with synchronous homodyne detection when $k=1$, whereas, as illustrated in section \ref{sec:robust_fidelity_witness}, we can come up with protocols utilizing fidelity witness with heterodyne sampling for state verification based on $k$-mode fidelities with $k>1$. The use of such fidelity witnesses allows for the efficient verification of certain doped Gaussian states \cite{AnnaMele2024} using heterodyne sampling, while as far as we know it is not possible with single-mode fidelity witness and consequently, homodyne sampling.

Finally, for noisy measurement devices with quantum efficiency $\eta$, noisy unbiased homodyne estimators are only well-defined for $\eta > 1/2$, whereas noisy biased heterodyne estimators are defined for all $\eta>0$.

\section{Conclusion}
\label{sec:conclusion}

In this paper, we have introduced a general verification framework, which combines efficient and robust fidelity witnesses for tensor product target states, together with classical post-processing implementing measurement back-propagation.
We have also introduced new families of such fidelity witnesses which outperform existing ones in terms of robustness, and identified both analytically and numerically a trade-off between the sample complexity and the robustness of those witnesses.

To illustrate the applicability of our framework, we have focused on continuous-variable verification of multimode bosonic quantum states, using synchronous homodyne detection as well as heterodyne detection. Our protocols are based on (ideal or lossy) homodyne and heterodyne estimators, and allow us to efficiently verify new classes of highly entangled and non-Gaussian continuous-variable quantum states such as doped Gaussian states, with rigorous confidence intervals on the fidelity between the target and experimental state (see Table \ref{tab:verifiable_classes}).

Our protocols have been described under the i.i.d.\ assumption. This mimics an experimental situation, where different copies of the quantum state are reasonably independent of each other. Since we are mainly concerned with experimental situations, our i.i.d.\ assumption is justified. However, to make our protocols useful for quantum cryptography scenarios in which the quantum state to be verified is being prepared by a potential adversarial prover, one needs to derive a protocol without the i.i.d.\ assumption. We have sketched how to achieve this by modifying our Protocol \ref{protocol_k_mode_fidelity} in section \ref{proof_no_iid} of the appendix. 

The certificates obtained through our verification protocols are robust bounds on the fidelity. As such, they can be used for calibrating large quantum devices at the state preparation stage, but also for rigorously witnessing quantum properties of experimental states: by targeting a pure state with a desired quantum property, such as entanglement or non-Gaussianity, a sufficiently large fidelity provides a witness of that property \cite{guhne2009entanglement,chabaud2020certification}.
Furthermore, as fidelity provides a bound on total variation distance, these certificates can be used to support demonstrations of quantum computational speedups in an hardware-agnostic way, and serve as the basis for rigorous, hardware-agnostic quantum benchmarks of quantum computers.

A natural follow-up direction is to identify other measurements beyond heterodyne and synchronous homodyne for which our verification framework is applicable, i.e.\ which are compatible with measurement backpropagation. In section \ref{sec:characteristic_function} of the appendix, we show how our verification framework can be extended to verify bosonic quantum states by directly measuring their characteristic function. 
We anticipate this verification protocol to be especially useful in superconducting platforms, where characteristic function measurements are readily available \cite{eickbusch2022fast}.

As we have uncovered various applications of our verification framework in the continuous-variable setting, it would be interesting to investigate its applications to the efficient verification of discrete-variable quantum states\footnote{Since the first version of this work, our verification framework has led to efficient protocols for the verification of a large class of intractable discrete-variable quantum states \cite{sater2025efficient}.}. In this setting, other families of robust fidelity witnesses are readily available \cite{hangleiter2017direct,huang2024certifying} and could be combined with classical post-processing implementing measurement back-propagation to obtain efficient verification protocols of large classes of discrete-variable quantum states. We leave this prospect for future work.



\section*{Acknowledgements}

U.C. acknowledges inspiring discussions with A.\ Eriksson, S.\ Gasparinetti, G.\ Ferrini, A.\ Ferraro, G.\ Roeland, M.\ Walschaers, D.\ Hangleiter, H.\ Ollivier, F.\ A.\ Mele, and S.\ F.\ E.\ Oliviero. We acknowledge funding from the European Union’s Horizon Europe Framework Programme (EIC Pathfinder Challenge project Veriqub) under Grant Agreement No. 101114899.


\bibliographystyle{linksen}
\bibliography{biblio}

@misc{Preskill1998,
  author = {Preskill, John},
  title = {Lecture notes for a course on quantum computation},
  year = {1998},
  note = {Unpublished lecture notes.},
  url = {http://www.theory.caltech.edu/people/preskill/ph229}
}

@article{preskill2012quantumcomputingentanglementfrontier,
      title={Quantum computing and the entanglement frontier}, 
      author={John Preskill},
      year={2012},
      eprint={1203.5813},
      archivePrefix={arXiv},
      primaryClass={quant-ph},
      doi={https://doi.org/10.48550/arXiv.1203.5813}, 
}

@article{eickbusch2022fast,
  title={Fast universal control of an oscillator with weak dispersive coupling to a qubit},
  author={Eickbusch, Alec and Sivak, Volodymyr and Ding, Andy Z and Elder, Salvatore S and Jha, Shantanu R and Venkatraman, Jayameenakshi and Royer, Baptiste and Girvin, Steven M and Schoelkopf, Robert J and Devoret, Michel H},
  journal={Nature Physics},
  volume={18},
  number={12},
  pages={1464--1469},
  year={2022},
  publisher={Nature Publishing Group UK London},
  doi={10.1038/s41567-022-01776-9}
}

@article{Zhu2019PRL,
  title = {Efficient Verification of Pure Quantum States in the Adversarial Scenario},
  author = {Zhu, Huangjun and Hayashi, Masahito},
  journal = {Phys. Rev. Lett.},
  volume = {123},
  issue = {26},
  pages = {260504},
  numpages = {6},
  year = {2019},
  month = {Dec},
  publisher = {American Physical Society},
  doi = {10.1103/PhysRevLett.123.260504}
}

@article{Hayashi2009,
doi = {10.1088/1367-2630/11/4/043028},
year = {2009},
month = {apr},
publisher = {},
volume = {11},
number = {4},
pages = {043028},
author = {Hayashi, Masahito},
title = {Group theoretical study of LOCC-detection of maximally entangled states using hypothesis testing},
journal = {New Journal of Physics}
}

@article{Liu2021,
  title = {Efficient verification of entangled continuous-variable quantum states with local measurements},
  author = {Liu, Ye-Chao and Shang, Jiangwei and Zhang, Xiangdong},
  journal = {Phys. Rev. Res.},
  volume = {3},
  issue = {4},
  pages = {L042004},
  numpages = {7},
  year = {2021},
  month = {Oct},
  publisher = {American Physical Society},
  doi = {10.1103/PhysRevResearch.3.L042004},
  url = {https://link.aps.org/doi/10.1103/PhysRevResearch.3.L042004}
}

@article{Pallister2018,
  title = {Optimal Verification of Entangled States with Local Measurements},
  author = {Pallister, Sam and Linden, Noah and Montanaro, Ashley},
  journal = {Phys. Rev. Lett.},
  volume = {120},
  issue = {17},
  pages = {170502},
  numpages = {5},
  year = {2018},
  month = {Apr},
  publisher = {American Physical Society},
  doi = {10.1103/PhysRevLett.120.170502},
  url = {https://link.aps.org/doi/10.1103/PhysRevLett.120.170502}
}

@article{Hayashi2015stabilizertesting,
  title = {Verifiable Measurement-Only Blind Quantum Computing with Stabilizer Testing},
  author = {Hayashi, Masahito and Morimae, Tomoyuki},
  journal = {Phys. Rev. Lett.},
  volume = {115},
  issue = {22},
  pages = {220502},
  numpages = {5},
  year = {2015},
  month = {Nov},
  publisher = {American Physical Society},
  doi = {10.1103/PhysRevLett.115.220502},
  url = {https://link.aps.org/doi/10.1103/PhysRevLett.115.220502}
}

@article{Zhu2019PRA,
  title = {General framework for verifying pure quantum states in the adversarial scenario},
  author = {Zhu, Huangjun and Hayashi, Masahito},
  journal = {Phys. Rev. A},
  volume = {100},
  issue = {6},
  pages = {062335},
  numpages = {38},
  year = {2019},
  month = {Dec},
  publisher = {American Physical Society},
  doi = {10.1103/PhysRevA.100.062335},
  url = {https://link.aps.org/doi/10.1103/PhysRevA.100.062335}
}

@article{hangleiter2017direct,
  title={Direct certification of a class of quantum simulations},
  author={Hangleiter, Dominik and Kliesch, Martin and Schwarz, Matthias and Eisert, Jens},
  journal={Quantum Science and Technology},
  volume={2},
  number={1},
  pages={015004},
  year={2017},
  publisher={IOP Publishing},
  doi={10.1088/2058-9565/2/1/015004}
}

@article{Harrow2017,
  author = {Harrow, Aram and Montanaro, Ashley},
  title = {Quantum computational supremacy},
  journal = {Nature},
  volume = {549},
  pages = {203--209},
  year = {2017},
  doi = {10.1038/nature23458}
}

@article{Ronnow2014,
  author    = {T. F. R{\o}nnow and Z. Wang and J. Job and S. Boixo and S. V. Isakov and D. Wecker and J. M. Martinis and D. A. Lidar and M. Troyer},
  title     = {Defining and detecting quantum speedup},
  journal   = {Science},
  volume    = {345},
  pages     = {420--424},
  year      = {2014},
  doi       = {10.1126/science.1252319}
}

@article{Terhal2004,
  author    = {B. M. Terhal and D. P. DiVincenzo},
  title     = {Adaptive quantum computation, constant depth quantum circuits and Arthur-Merlin games},
  journal   = {Quantum Information \& Computation},
  volume    = {4},
  pages     = {134--145},
  year      = {2004},
  eprint    = {arXiv:0205133},
doi = {https://doi.org/10.48550/arXiv.quant-ph/0205133}
}

@article{guhne2009entanglement,
  title={Entanglement detection},
  author={G{\"u}hne, Otfried and T{\'o}th, G{\'e}za},
  journal={Physics Reports},
  volume={474},
  number={1-6},
  pages={1--75},
  year={2009},
  publisher={Elsevier},
  doi={10.1016/j.physrep.2009.02.004}
}

@article{Shepherd2009,
  author    = {D. Shepherd and M. J. Bremner},
  title     = {Temporally unstructured quantum computation},
  journal   = {Proceedings of the Royal Society A: Mathematical, Physical and Engineering Sciences},
  volume    = {465},
  pages     = {1413--1439},
  year      = {2009},
doi = {https://doi.org/10.1098/rspa.2008.0443}
}

@article{Bremner2011,
  author    = {M. J. Bremner and R. Jozsa and D. J. Shepherd},
  title     = {Classical simulation of commuting quantum computations implies collapse of the polynomial hierarchy},
  journal   = {Proceedings of the Royal Society A: Mathematical, Physical and Engineering Sciences},
  volume    = {467},
  pages     = {459--472},
  year      = {2011},
doi = {https://doi.org/10.1098/rspa.2010.0301}
}

@article{Boixo2018,
  author    = {S. Boixo and S. V. Isakov and V. N. Smelyanskiy and R. Babbush and N. Ding and Z. Jiang and M. J. Bremner and J. M. Martinis and H. Neven},
  title     = {Characterizing quantum supremacy in near-term devices},
  journal   = {Nature Physics},
  volume    = {14},
  pages     = {595--600},
  year      = {2018},
doi = {https://doi.org/10.1038/s41567-018-0124-x},
}

@article{Mezher2019,
  author    = {R. Mezher and J. Ghalbouni and J. Dgheim and D. Markham},
  title     = {Efficient approximate unitary t-designs from partially invertible universal sets and their application to quantum speedup},
  eprint    = {arXiv:1905.01504},
  year      = {2019},
doi = {https://doi.org/10.48550/arXiv.1905.01504}
}

@article{Arute2019,
  author    = {Frank Arute and Kunal Arya and Ryan Babbush and et al.},
  title     = {Quantum supremacy using a programmable superconducting processor},
  journal   = {Nature},
  volume    = {574},
  pages     = {505--510},
  year      = {2019},
  doi       = {10.1038/s41586-019-1666-5}
}

@article{Eisert2020,
  author    = {Jens Eisert and Dominik Hangleiter and Nathan Walk and Ingo Roth and Damian Markham and Rhea Parekh and Ulysse Chabaud and Elham Kashefi},
  title     = {Quantum certification and benchmarking},
  journal   = {Nature Reviews Physics},
  pages     = {1--9},
  year      = {2020},
  doi = {https://doi.org/10.1038/s42254-020-0186-4}
}

@article{Renner2007,
  author    = {R. Renner},
  title     = {Symmetry of large physical systems implies independence of subsystems},
  journal   = {Nature Physics},
  volume    = {3},
  pages     = {645--649},
  year      = {2007},
doi = {https://doi.org/10.1038/nphys684}
}

@article{Renner2009,
  title = {de Finetti Representation Theorem for Infinite-Dimensional Quantum Systems and Applications to Quantum Cryptography},
  author = {Renner, R. and Cirac, J. I.},
  journal = {Phys. Rev. Lett.},
  volume = {102},
  issue = {11},
  pages = {110504},
  numpages = {4},
  year = {2009},
  month = {Mar},
  publisher = {American Physical Society},
  doi = {10.1103/PhysRevLett.102.110504},
}

@article{Gheorghiu2019,
  author    = {A. Gheorghiu and T. Kapourniotis and E. Kashefi},
  title     = {Verification of quantum computation: An overview of existing approaches},
  journal   = {Theory of Computing Systems},
  volume    = {4},
  pages     = {715--808},
  year      = {2019},
doi = {https://doi.org/10.1007/s00224-018-9872-3}
}

@inproceedings{Mahadev2018,
  author    = {U. Mahadev},
  title     = {Classical verification of quantum computations},
  booktitle = {2018 IEEE 59th Annual Symposium on Foundations of Computer Science (FOCS)},
  pages     = {259--267},
  year      = {2018},
  organization = {IEEE},
  doi={10.1109/FOCS.2018.00033}
}

@article{KahanamokuMeyer2023forgingquantumdata,
  doi = {10.22331/q-2023-09-11-1107},
  title = {Forging quantum data: classically defeating an {IQP}-based quantum test},
  author = {Kahanamoku-Meyer, Gregory D.},
  journal = {{Quantum}},
  issn = {2521-327X},
  publisher = {{Verein zur F{\"{o}}rderung des Open Access Publizierens in den Quantenwissenschaften}},
  volume = {7},
  pages = {1107},
  month = sep,
  year = {2023}
}

@article{Yung2020,
  author       = {M.-H. Yung and B. Cheng},
  title        = {Anti-Forging Quantum Data: Cryptographic Verification of Quantum Cloud Computing},
  year         = {2020},
  eprint       = {2005.01510},
  archivePrefix= {arXiv},
  primaryClass = {quant-ph},
  doi = {https://doi.org/10.48550/arXiv.2005.01510},
}

@inproceedings{Brakerski2018,
  author       = {Z. Brakerski and P. Christiano and U. Mahadev and U. Vazirani and T. Vidick},
  title        = {A cryptographic test of quantumness and certifiable randomness from a single quantum device},
  booktitle    = {2018 IEEE 59th Annual Symposium on Foundations of Computer Science (FOCS)},
  pages        = {320--331},
  year         = {2018},
  organization = {IEEE},
doi={10.1109/FOCS.2018.00038}
}

@article{Brakerski2020,
  author       = {Z. Brakerski and V. Koppula and U. Vazirani and T. Vidick},
  title        = {Simpler Proofs of Quantumness},
  year         = {2020},
  eprint       = {2005.04826},
  archivePrefix= {arXiv},
  primaryClass = {quant-ph},
  doi = {https://doi.org/10.48550/arXiv.2005.04826}
}

@article{Spagnolo2014,
  author    = {N. Spagnolo and C. Vitelli and M. Bentivegna and D. J. Brod and A. Crespi and F. Flamini and S. Giacomini and G. Milani and R. Ramponi and P. Mataloni and others},
  title     = {Experimental validation of photonic Boson Sampling},
  journal   = {Nature Photonics},
  volume    = {8},
  pages     = {615--620},
  year      = {2014},
doi = {https://doi.org/10.1038/nphoton.2014.135}
}

@article{Drummond2021,
  title = {Simulating complex networks in phase space: Gaussian boson sampling},
  author = {Drummond, Peter D. and Opanchuk, Bogdan and Dellios, A. and Reid, M. D.},
  journal = {Phys. Rev. A},
  volume = {105},
  issue = {1},
  pages = {012427},
  numpages = {11},
  year = {2022},
  month = {Jan},
  publisher = {American Physical Society},
  doi = {10.1103/PhysRevA.105.012427}
}

@article{Aaronson2013_2,
  author       = {S. Aaronson and A. Arkhipov},
  title        = {Boson Sampling is far from uniform},
  year         = {2013},
  eprint       = {1309.7460},
  archivePrefix = {arXiv},
  primaryClass = {quant-ph},
doi = {https://doi.org/10.48550/arXiv.1309.7460}
}

@article{Fuchs1999,
  author    = {Christopher A. Fuchs and Jeroen Van De Graaf},
  title     = {Cryptographic distinguishability measures for quantum-mechanical states},
  journal   = {IEEE Transactions on Information Theory},
  volume    = {45},
  pages     = {1216--1227},
  year      = {1999},
  doi={10.1109/18.761271}
}

@article{Braunstein2005,
  author = {Braunstein, Samuel L. and van Loock, Peter},
  title = {Quantum information with continuous variables},
  journal = {Rev. Mod. Phys.},
  volume = {77},
  pages = {513--577},
  month = jun,
  year = {2005},
    doi = {10.1103/RevModPhys.77.513}}

@article{Weedbrook2012,
  title = {Gaussian quantum information},
  author = {Weedbrook, Christian and Pirandola, Stefano and Garc\'{\i}a-Patr\'on, Ra\'ul and Cerf, Nicolas J. and Ralph, Timothy C. and Shapiro, Jeffrey H. and Lloyd, Seth},
  journal = {Rev. Mod. Phys.},
  volume = {84},
  issue = {2},
  pages = {621--669},
  numpages = {0},
  year = {2012},
  month = {May},
  publisher = {American Physical Society},
  doi = {10.1103/RevModPhys.84.621}
}

@article{Ferracin2019,
  author    = {S. Ferracin and T. Kapourniotis and A. Datta},
  title     = {Accrediting outputs of noisy intermediate-scale quantum computing devices},
  journal   = {New Journal of Physics},
  volume    = {21},
  pages     = {113038},
  year      = {2019},
  doi       = {10.1088/1367-2630/ab5099}
}

@article{Zhong2020,
  author    = {H. S. Zhong and H. Wang and Y. H. Deng and M. C. Chen and L. C. Peng and Y. H. Luo and J. Qin and D. Wu and X. Ding and Y. Hu and others},
  title     = {Quantum computational advantage using photons},
  journal   = {Science},
  volume    = {370},
  pages     = {1460--1463},
  year      = {2020},
  doi       = {10.1126/science.abe8770}
}

@article{AnnaMele2024,
      title={Learning quantum states of continuous variable systems},
  author={Mele, Francesco Anna and Mele, Antonio Anna and Bittel, Lennart and Eisert, Jens and Giovannetti, Vittorio and Lami, Ludovico and Leone, Lorenzo and Oliviero, Salvatore FE},
  eprint={arXiv:2405.01431},
  year={2024}
}

@article{Chabaud2021efficient,
  doi = {10.22331/q-2021-11-15-578},
  title = {Efficient verification of {B}oson {S}ampling},
  author = {Chabaud, Ulysse and Grosshans, Fr{\'{e}}d{\'{e}}ric and Kashefi, Elham and Markham, Damian},
  journal = {{Quantum}},
  issn = {2521-327X},
  publisher = {{Verein zur F{\"{o}}rderung des Open Access Publizierens in den Quantenwissenschaften}},
  volume = {5},
  pages = {578},
  month = nov,
  year = {2021}
}

@article{Arvind_1995,
   title={The real symplectic groups in quantum mechanics and optics},
   volume={45},
   ISSN={0973-7111},
   DOI={10.1007/bf02848172},
   number={6},
   journal={Pramana},
   publisher={Springer Science and Business Media LLC},
   author={Arvind and Dutta, B and Mukunda, N and Simon, R},
   year={1995},
   month=dec, pages={471–497} }

@article{paris1996quantum,
	Author = {Paris, Matteo GA},
	Journal = {Physical Review A},
	Number = {4},
	Pages = {2658},
	Publisher = {APS},
	Title = {Quantum state measurement by realistic heterodyne detection},
	Volume = {53},
	Year = {1996},
doi = {https://doi.org/10.1103/PhysRevA.53.2658}}

@article{pevrina1968regular,
  title={“Regular” form of the Sudarshan-Glauber representation of the density matrix},
  author={Pe{\v{r}}ina, J and Mi{\v{s}}ta, L},
  journal={Physics Letters A},
  volume={27},
  number={4},
  pages={217--218},
  year={1968},
  publisher={Elsevier},
doi = {https://doi.org/10.1016/0375-9601(68)91100-6}
}

@article{Gandhari2023,
  title = {Precision Bounds on Continuous-Variable State Tomography Using Classical Shadows},
  author = {Gandhari, Srilekha and Albert, Victor V. and Gerrits, Thomas and Taylor, Jacob M. and Gullans, Michael J.},
  journal = {PRX Quantum},
  volume = {5},
  issue = {1},
  pages = {010346},
  numpages = {13},
  year = {2024},
  month = {Mar},
  publisher = {American Physical Society},
  doi = {10.1103/PRXQuantum.5.010346},
}

@article{menicucci2006universal,
  title={Universal quantum computation with continuous-variable cluster states},
  author={Menicucci, Nicolas C and Van Loock, Peter and Gu, Mile and Weedbrook, Christian and Ralph, Timothy C and Nielsen, Michael A},
  journal={Physical review letters},
  volume={97},
  number={11},
  pages={110501},
  year={2006},
  publisher={APS}
}

@article{Aaronson2013,
	Author = {Aaronson, S. and Arkhipov, A.},
	Journal = {Theory of Computing},
	Pages = {143},
	Title = {The computational Complexity of Linear Optics},
	Volume = {9},
	Year = {2013},
	Eprint = {arXiv:1011.3245},
	Doi = {10.1145/1993636.1993682}
	}

@book{NielsenChuang,
	Address = {New York, NY, USA},
	Author = {Nielsen, Michael A. and Chuang, Isaac L.},
	Publisher = {Cambridge University Press},
	Title = {Quantum Computation and Quantum Information: 10th Anniversary Edition},
	Year = {2011},
	doi={10.1017/CBO9780511976667}
	}

@article{wunsche1998laguerre,
	Author = {W{\"u}nsche, Alfred},
	Journal = {Journal of Physics A: Mathematical and General},
	Number = {40},
	Pages = {8267},
	Publisher = {IOP Publishing},
	Title = {Laguerre 2D-functions and their application in quantum optics},
	Volume = {31},
	Year = {1998},
	Doi = {10.1088/0305-4470/31/40/017}
}

@book{gould1972combinatorial,
	Author = {Gould, Henry Wadsworth},
	Publisher = {Morgantown, W Va},
	Title = {Combinatorial Identities: A standardized set of tables listing 500 binomial coefficient summations},
	Year = {1972}
}

@article{eisert2020quantum,
    title={Quantum certification and benchmarking},
    author={Eisert, Jens and Hangleiter, Dominik and Walk, Nathan and Roth, Ingo and Markham, Damian and Parekh, Rhea and Chabaud, Ulysse and Kashefi, Elham},
    journal={Nature Reviews Physics},
    volume={2},
    number={7},
    pages={382--390},
    year={2020},
    publisher={Nature Publishing Group},
    doi={10.1038/s42254-020-0186-4}
}

@article{madsen2022quantum,
	author = {Madsen, Lars S. and Laudenbach, Fabian and Askarani, Mohsen Falamarzi. and Rortais, Fabien and Vincent, Trevor and Bulmer, Jacob F. F. and Miatto, Filippo M. and Neuhaus, Leonhard and Helt, Lukas G. and Collins, Matthew J. and Lita, Adriana E. and Gerrits, Thomas and Nam, Sae Woo and Vaidya, Varun D. and Menotti, Matteo and Dhand, Ish and Vernon, Zachary and Quesada, Nicol{\'a}s and Lavoie, Jonathan},
	doi = {10.1038/s41586-022-04725-x},
	id = {Madsen2022},
	isbn = {1476-4687},
	journal = {Nature},
	number = {7912},
	pages = {75--81},
	title = {Quantum computational advantage with a programmable photonic processor},
	volume = {606},
	year = {2022}}

@article{Yokoyama2013,
  author    = {Shota Yokoyama and Ryuji Ukai and Seiji C Armstrong and Chanond Sornphiphatphong and Toshiyuki Kaji and Shigenari Suzuki and Jun-ichi Yoshikawa and Hidehiro Yonezawa and Nicolas C Menicucci and Akira Furusawa},
  title     = {Ultra-large-scale continuous-variable cluster states multiplexed in the time domain},
  journal   = {Nature Photonics},
  volume    = {7},
  number    = {12},
  pages     = {982},
  year      = {2013},
  doi       = {10.1038/nphoton.2013.287}
}

@article{Gottesman2001,
  title = {Encoding a qubit in an oscillator},
  author = {Gottesman, Daniel and Kitaev, Alexei and Preskill, John},
  journal = {Phys. Rev. A},
  volume = {64},
  issue = {1},
  pages = {012310},
  numpages = {21},
  year = {2001},
  month = {Jun},
  publisher = {American Physical Society},
  doi = {10.1103/PhysRevA.64.012310}
}

@article{Yoshiakawa2016,
  author = {Yoshiakawa, J.-I. and Yokoyama, S. and Kaji, T. and Sornphiphatphong, C. and Shiozawa, Y. and Makino, K. and Furusawa, A.},
  title = {Invited article: Generation of one-million-mode continuous-variable cluster state by unlimited time-domain multiplexing},
  journal = {APL Photonics},
  volume = {1},
  pages = {060801},
  year = {2016},
    doi = {https://doi.org/10.1063/1.4962732}
}

@article{Hoeffding1963,
  author = {Hoeffding, W.},
  title = {Probability inequalities for sums of bounded random variables},
  journal = {Journal of the American Statistical Association},
  volume = {58},
  pages = {13--30},
  year = {1963},
doi = {https://doi.org/10.1080/01621459.1963.10500830},
}

@article{Yuen1980,
  author    = {H. Yuen and J. Shapiro},
  title     = {Optical communication with two-photon coherent states-Part III: Quantum measurements realizable with photoemissive detectors},
  journal   = {IEEE Transactions on Information Theory},
  volume    = {26},
  pages     = {78--92},
  year      = {1980},
doi={10.1109/TIT.1980.1056132}
}

@article{artiles2005invitation,
  title={An invitation to quantum tomography},
  author={Artiles, Luis M and Gill, Richard D and Gut{\c{}} {\u{a}}, MI},
  journal={Journal of the Royal Statistical Society Series B: Statistical Methodology},
  volume={67},
  number={1},
  pages={109--134},
  year={2005},
  publisher={Oxford University Press},
doi = {https://doi.org/10.48550/arXiv.quant-ph/0303020}
}

@article{huang2024certifying,
  title={Certifying almost all quantum states with few single-qubit measurements},
  author={Huang, Hsin-Yuan and Preskill, John and Soleimanifar, Mehdi},
  eprint={arXiv:2404.07281},
  year={2024},
doi = {https://doi.org/10.48550/arXiv.2404.07281}
}

@article{wu2021efficient,
  title={Efficient verification of continuous-variable quantum states and devices without assuming identical and independent operations},
  author={Wu, Ya-Dong and Bai, Ge and Chiribella, Giulio and Liu, Nana},
  journal={Physical Review Letters},
  volume={126},
  number={24},
  pages={240503},
  year={2021},
  publisher={APS},
  doi={10.1103/PhysRevLett.126.240503}
}

@article{Aubry_2009,
doi = {10.1088/0266-5611/25/1/015003},
year = {2008},
month = {nov},
publisher = {},
volume = {25},
number = {1},
pages = {015003},
author = {Jean-Marie Aubry and Cristina Butucea and Katia Meziani},
title = {State estimation in quantum homodyne tomography with noisy data},
journal = {Inverse Problems},
}

@article{fawzi2024optimal,
  title={Optimal Fidelity Estimation from Binary Measurements for Discrete and Continuous Variable Systems},
  author={Fawzi, Omar and Oufkir, Aadil and Salzmann, Robert},
  journal={arXiv preprint arXiv:2409.04189},
  year={2024},
  doi = {https://doi.org/10.48550/arXiv.2409.04189}
}

@article{MauroDAriano2003,
  title={Quantum tomography},
  author={D'Ariano, G Mauro and Paris, Matteo GA and Sacchi, Massimiliano F},
  journal={Advances in imaging and electron physics},
  volume={128},
  number={10.1016},
  pages={S1076--5670},
  year={2003},
  publisher={San Diego: Academic Press, c1995-}
}

@article{tian2023direct,
  title={Direct characteristic-function tomography of the quantum states of quantum fields},
  author={Tian, Zehua and Jing, Jiliang and Du, Jiangfeng},
  journal={Science China Physics, Mechanics \& Astronomy},
  volume={66},
  number={11},
  pages={110412},
  year={2023},
  publisher={Springer}
}

@article{Aolita2015,
  author    = {Leandro Aolita and Christian Gogolin and Martin Kliesch and Jens Eisert},
  title     = {Reliable quantum certification of photonic state preparations},
  journal   = {Nature Communications},
  volume    = {6},
  pages     = {8498},
  year      = {2015},
  doi       = {10.1038/ncomms9498}
}

@article{Adesso2014,
author = {Adesso, Gerardo and Ragy, Sammy and Lee, Antony R.},
title = {Continuous Variable Quantum Information: Gaussian States and Beyond},
journal = {Open Systems \& Information Dynamics},
volume = {21},
number = {01n02},
pages = {1440001},
year = {2014},
doi = {10.1142/S1230161214400010},
}

@article{Bartlett2002,
	Author = {Bartlett, Stephen D. and Sanders, Barry C. and Braunstein, Samuel L. and Nemoto, Kae},
	Issue = {9},
	Journal = {Physical Review Letters},
	Numpages = {4},
	Pages = {097904},
	Publisher = {American Physical Society},
	Title = {Efficient Classical Simulation of Continuous Variable Quantum Information Processes},
	Volume = {88},
	Year = {2002},
	Doi = {10.1103/PhysRevLett.88.097904}
}

@article{d2003quantum,
	Author = {D'Ariano, G Mauro and Paris, Matteo GA and Sacchi, Massimiliano F},
	Journal = {Advances in Imaging and Electron Physics},
	Pages = {206--309},
	Publisher = {San Diego: Academic Press, c1995-},
	Title = {Quantum tomography},
	Volume = {128},
	Year = {2003},
	ArchivePrefix = {arXiv},
	ePrint = {quant-ph/0302028}
}

@article{knill2005,
  title={Quantum computing with realistically noisy devices},
  author={Knill, Emanuel},
  journal={Nature},
  volume={434},
  number={7029},
  pages={39--44},
  year={2005},
  publisher={Nature Publishing Group UK London}
}

@article{Knill2001,
  author = {Knill, E. and Laflamme, R. and Milburn, G.},
  title = {A scheme for efficient quantum computation with linear optics},
  journal = {Nature},
  volume = {409},
  pages = {46--52},
  year = {2001},
  doi = {10.1038/35051009}
}

@article{sivak2023,
  author = {Sivak, V.V. and Eickbusch, A. and Royer, B. and others},
  title = {Real-time quantum error correction beyond break-even},
  journal = {Nature},
  volume = {616},
  pages = {50--55},
  year = {2023},
  doi = {10.1038/s41586-023-05782-6}
}

@article{Leverrier2013,
  title = {Security of Continuous-Variable Quantum Key Distribution Against General Attacks},
  author = {Leverrier, Anthony and Garc\'{\i}a-Patr\'on, Ra\'ul and Renner, Renato and Cerf, Nicolas J.},
  journal = {Phys. Rev. Lett.},
  volume = {110},
  issue = {3},
  pages = {030502},
  numpages = {5},
  year = {2013},
  month = {Jan}       ,
  publisher = {American Physical Society},
  doi = {10.1103/PhysRevLett.110.030502}
}

@inproceedings{chabaud2020building,
    title={Building Trust for Continuous Variable Quantum States},
    author={Chabaud, Ulysse and Douce, Tom and Grosshans, Fr{\'e}d{\'e}ric and Kashefi, Elham and Markham, Damian},
    booktitle={15th Conference on the Theory of Quantum Computation, Communication and Cryptography},
    year={2020},
    eprint={arXiv:1905.12700}
}

@article{sater2025efficient,
      title={Efficient certification of intractable quantum states with few Pauli measurements}, 
      author={Sami Abdul Sater and Maxime Garnier and Thierry Martinez and Harold Ollivier and Ulysse Chabaud},
      year={2025},
      eprint={2511.07300},
      archivePrefix={arXiv},
      primaryClass={quant-ph},
      doi = {https://doi.org/10.48550/arXiv.2511.07300}
}

@article{chabaud2020certification,
    title = {Certification of Non-{G}aussian States with Operational Measurements},
    author = {Chabaud, Ulysse and Roeland, Gana\"el and Walschaers, Mattia and Grosshans, Fr\'ed\'eric and Parigi, Valentina and Markham, Damian and Treps, Nicolas},
    journal = {PRX Quantum},
    volume = {2},
    issue = {2},
    pages = {020333},
    numpages = {19},
    year = {2021},
    month = {Jun},
    publisher = {American Physical Society},
    doi = {10.1103/PRXQuantum.2.020333}
}

@article{chabaud2020stellar,
	Author = {Chabaud, Ulysse and Markham, Damian and Grosshans, Fr{\'e}d{\'e}ric},
	Journal = {Physical Review Letters},
	Number = {6},
	Pages = {063605},
	Publisher = {APS},
	Title = {Stellar representation of non-{G}aussian quantum states},
	Volume = {124},
	Year = {2020},
	Doi = {10.1103/PhysRevLett.124.063605}
}

@article{Wang2019,
  title = {Boson Sampling with 20 Input Photons and a 60-Mode Interferometer in a $1{0}^{14}$-Dimensional Hilbert Space},
  author = {Wang, Hui and Qin, Jian and Ding, Xing and Chen, Ming-Cheng and Chen, Si and You, Xiang and He, Yu-Ming and Jiang, Xiao and You, L. and Wang, Z. and Schneider, C. and Renema, Jelmer J. and H\"ofling, Sven and Lu, Chao-Yang and Pan, Jian-Wei},
  journal = {Phys. Rev. Lett.},
  volume = {123},
  issue = {25},
  pages = {250503},
  numpages = {7},
  year = {2019},
  month = {Dec},
  publisher = {American Physical Society},
  doi = {10.1103/PhysRevLett.123.250503},
}

@phdthesis{Leichtle2024,
  TITLE = {{Security and Efficiency of Delegated Quantum Computing}},
  AUTHOR = {Leichtle, Dominik},
  URL = {https://theses.hal.science/tel-04718138},
  NUMBER = {2024SORUS183},
  SCHOOL = {{Sorbonne Universit{\'e}}},
  YEAR = {2024},
  MONTH = Feb,
  TYPE = {Theses},
  HAL_ID = {tel-04718138},
  HAL_VERSION = {v1},
}

@article{Giordani2018,
   title={Experimental statistical signature of many-body quantum interference},
   volume={12},
   ISSN={1749-4893},
   DOI={10.1038/s41566-018-0097-4},
   number={3},
   journal={Nature Photonics},
   publisher={Springer Science and Business Media LLC},
   author={Giordani, Taira and Flamini, Fulvio and Pompili, Matteo and Viggianiello, Niko and Spagnolo, Nicolò and Crespi, Andrea and Osellame, Roberto and Wiebe, Nathan and Walschaers, Mattia and Buchleitner, Andreas and Sciarrino, Fabio},
   year={2018},
   month=feb, pages={173–178} }

@article{sen2024,
  author    = {Sen, I.},
  title     = {Physical interpretation of non-normalizable harmonic oscillator states and relaxation to pilot-wave equilibrium},
  journal   = {Scientific Reports},
  year      = {2024},
  volume    = {14},
  number    = {1},
  pages     = {669},
  month     = {January},
  doi       = {10.1038/s41598-023-50814-w},
}

@article{yokoyama2013ultra,
	Author = {Yokoyama, Shota and Ukai, Ryuji and Armstrong, Seiji C and Sornphiphatphong, Chanond and Kaji, Toshiyuki and Suzuki, Shigenari and Yoshikawa, Jun-ichi and Yonezawa, Hidehiro and Menicucci, Nicolas C and Furusawa, Akira},
	Journal = {Nature Photonics},
	Number = {12},
	Pages = {982},
	Publisher = {Nature Publishing Group},
	Title = {Ultra-large-scale continuous-variable cluster states multiplexed in the time domain},
	Volume = {7},
	Year = {2013},
	Doi = {10.1038/nphoton.2013.287}
	}

@article{ferraro2005,
	Address = {Napoli},
	Author = {A. Ferraro and S. Olivares and M. G. A. Paris},
	Publisher = {Bibliopolis},
	Title = {Gaussian States in Quantum Information},
	eprint = {arxiv:quant-ph/0503237},
	Year = {2005}}


\onecolumn\newpage
\appendix


\begin{center}
    {\huge Appendix}
\end{center}


\section{Robust fidelity witnesses: Proof of Lemma \ref{lemma_mm_fid_witness_2}}
\label{proof_lemma_2}

We start by the proof of Eq.~(\ref{eqn_better_k_mode}) in Lemma \ref{lemma_mm_fid_witness_2}.
The proof breaks down the $m$-mode system into blocks of $k$ modes and combines the known inequality in each of the blocks to obtain the final inequality. 

Let's consider a $k$-mode system with the density matrix $\rho_{1\dots k} = \mathrm{Tr}_{\{1,\dots,m\} \backslash \{1, \dots, k\}} [\rho]$ and target state $\ket{\psi_{1}} \otimes \dots \otimes \ket{\psi_{k}}$. For this k-mode state,
\begin{equation}
     W^{(k)} (\rho,\psi) = 1 - \left( 1 - F (\rho_{1,\dots,k}, \ket{\psi_{1}} \otimes \dots \otimes \ket{\psi_{k}}) \right) = F (\rho_{1 ,\dots, k}, \ket{\psi_{1}} \otimes \dots \otimes \ket{\psi_{k}}).
\end{equation}
Using Eq.~(\ref{eq1.1})
\begin{eqnarray}
    W^{(1)} (\rho_{1 \dots k}, \ket{\psi_1} \otimes \dots \otimes \ket{\psi_k} ) &\leq& W^{(k)} (\rho_{1 \dots k}, \ket{\psi_1} \otimes \dots \otimes \ket{\psi_k} ), \\
    1 - \underset{i = 1}{\overset{k}{\sum}} \left( 1 - F(\rho_i,\ket{\psi_i}) \right) &\leq& 1 - \left( 1 - F (\rho_{1\dots k}, \ket{\psi_{1}} \otimes \dots \otimes \ket{\psi_{k}}) \right), \\
    \underset{i = 1}{\overset{k}{\sum}} \left( 1 - F(\rho_i,\ket{\psi_i}) \right) &\geq& \left( 1 - F (\rho_{1\dots k}, \ket{\psi_{1}} \otimes \dots \otimes \ket{\psi_{k}}) \right).
\end{eqnarray}
Similarly,
\begin{equation}
    \underset{i = k+1}{\overset{2k}{\sum}} \left( 1 - F(\rho_i,\ket{\psi_i}) \right) \geq \left( 1 - F (\rho_{k+1\dots 2k}, \ket{\psi_{k+1}} \otimes \dots \otimes \ket{\psi_{2k}}) \right),
\end{equation}
\begin{equation}
    \vdots
\end{equation}
\begin{equation}
    \underset{i = m-(k-1)}{\overset{m}{\sum}} \left( 1 - F(\rho_i,\ket{\psi_i}) \right) \geq \left( 1 - F (\rho_{m-(k-1)\dots m}, \ket{\psi_{m-(k-1)}} \otimes \dots \otimes \ket{\psi_{m}}) \right).
\end{equation}
Summing over,
\begin{eqnarray}
    \underset{i=1}{\overset{m}{\sum}} \left( 1 - F(\rho_i,\ket{\psi_i}) \right) &\geq& \underset{i = 1}{\overset{m/k}{\sum}} \left( 1 - F (\rho_{ki - (k - 1)\dots ki}, \ket{\psi_{ki - (k-1)}} \otimes \dots \otimes \ket{\psi_{ki}}) \right), \\
    1 - \underset{i=1}{\overset{m}{\sum}} \left( 1 - F(\rho_i,\ket{\psi_i}) \right) &\leq& 1 - \underset{i = 1}{\overset{m/k}{\sum}} \left( 1 - F (\rho_{ki - (k - 1)\dots ki}, \ket{\psi_{ki - (k-1)}} \otimes \dots \otimes \ket{\psi_{ki}}) \right), \\
    W^{(1)} (\rho,\psi) &\leq& W^{(k)} (\rho,\psi).
\end{eqnarray}

\medskip

We now turn to the proof of Eq.~(\ref{eqn_inequality_k-mode}) in Lemma \ref{lemma_mm_fid_witness_2}
This proof relies on relating the fidelity between two states to the fidelity between two corresponding substates of the given states.

The fidelity between two states $\rho$ and $\ket{\psi} = \ket{\psi_1}\otimes \dots \ket{\psi_m}$ is given by
\begin{eqnarray}
    F(\rho,\ket{\psi_1}\otimes\dots \otimes\ket{\psi_m}) &=& \mathrm{Tr}\left[ \rho \ket{\psi_1} \bra{\psi_1} \otimes \dots \otimes \ket{\psi_m}\bra{\psi_m}  \right] \\
    &\leq& \mathrm{Tr} \left[\rho \I \otimes \dots \otimes \ket{\psi_{ki - (k-1)}} \bra{\psi_{ki - (k-1)}} \otimes \dots \otimes \ket{\psi_{ki}} \bra{\psi_{ki}} \otimes \dots \otimes \I \right] \\
    &=& F(\rho_{ki-(k-1) \dots ki}, \ket{\psi_{ki - (k-1)}}\otimes \dots \otimes \ket{\psi_{ki}}).
\end{eqnarray}
Therefore,
\begin{eqnarray}
    \frac{m}{k} F(\rho,\ket{\psi}) &\leq& \underset{i = 1}{\overset{m/k}{\sum}}   F(\rho_{ki-(k-1) \dots ki}, \ket{\psi_{ki - (k-1)}}\otimes \dots \otimes \ket{\psi_{ki}}),\\
    1 - \frac{m}{k} + \frac{m}{k} F(\rho,\ket{\psi}) &\leq& 1 - \frac{m}{k} + \underset{i = 1}{\overset{m/k}{\sum}}   F(\rho_{ki-(k-1) \dots ki}, \ket{\psi_{ki - (k-1)}}\otimes \dots \otimes \ket{\psi_{ki}}), \\
     1 - \frac{m}{k} (1 - F(\rho,\ket{\psi})) &\leq& W^{(k)} (\rho,\psi).
\end{eqnarray}
Also
\begin{equation}
\hspace{-55mm}F(\rho,\ket{\psi}) = \mathrm{Tr}\left[ \rho \ket{\psi_1} \bra{\psi_1} \otimes \dots \otimes \ket{\psi_m} \bra{\psi_m}  \right]
\end{equation}

\begin{equation}
    \hspace{-10mm} = \mathrm{Tr} [\rho \I] - \mathrm{Tr} [\rho (\I - \ket{\psi_1}\bra{\psi_1}\otimes \dots \otimes \ket{\psi_k}\bra{\psi_k} ) \otimes I]
\end{equation}

\begin{equation}
    \hspace{35 mm} - \mathrm{Tr} [\rho \ket{\psi_1}\bra{\psi_1}\otimes \dots \otimes \ket{\psi_k}\bra{\psi_k} \otimes (1 - \ket{\psi_{k+1}}\bra{\psi_{k+1}}\otimes \dots \otimes \ket{\psi_{2k}}\bra{\psi_{2k}}) \otimes I ]
\end{equation}

\begin{equation}
    \vdots
\end{equation}

\begin{equation}
    \hspace{35 mm} - \mathrm{Tr} [\rho \ket{\psi_1}\bra{\psi_1}\otimes \dots \otimes \ket{\psi_{m - k}}\bra{\psi_{m - k}} \otimes (1 - \ket{\psi_{m-(k-1)}}\bra{\psi_{m-(k-1)}}\otimes \dots \otimes \ket{\psi_{m}}\bra{\psi_{m}}) ]
\end{equation}

\begin{equation}
    \hspace{11 mm} \geq 1 - \underset{i =1}{\overset{m/k}{\sum}} \mathrm{Tr} [\rho \I \otimes (\I - \ket{\psi_{ki - (k-1)}}\bra{\psi_{ki - (k-1)}} \otimes \dots \otimes \ket{\psi_{ki}}\bra{\psi_{ki}} ) \otimes \I]
\end{equation}

\begin{equation}
    \hspace{25 mm} = 1 -  \underset{i = 1}{\overset{m/k}{\sum}} \left( 1 - F (\rho_{ki - (k - 1)\dots ki}, \ket{\psi_{ki - (k-1)}} \otimes \dots \otimes \ket{\psi_{ki}}) \right) = W^{(k)} (\rho,\psi).
\end{equation}


\section{Normalizable and unnormalizable eigenfunctions of the harmonic oscillator} \label{normal_unnormal_HO}
For an eigenvalue $j$ of the harmonic oscillator, let 
$K = 2j+1$. We define the functions
\begin{eqnarray}
    h_0^{K} (x) &:=& 1 + \sum_{n=1}^{\infty} \frac{\Pi_{i=0}^{n-1}(4i+1 - K)}{(2n)!} x^{2n}, \\
    h_1^{K} (x) &:=& x + \sum_{n=1}^{\infty} \frac{\Pi_{i=0}^{n-1}(4i+3 - K)}{(2n+1)!} x^{2n+1}.
\end{eqnarray}
Then, the normalizable eigenfunctions of the harmonic oscillator are given by \cite{sen2024}:
\begin{equation}
\psi_j^{(\mathrm{nor})}(x) = 
\begin{cases} 
    N_j e^{-x^2/2}  h_0^{K} (x), & \text{if } j \text{ is even}, \\
    N_j e^{-x^2/2}  h_1^{K} (x), & \text{if } j \text{ is odd},
\end{cases}
\end{equation}
where $N_j$ is a normalization constant. Similarly, the unnormalizable eigenfunctions are given by
\begin{equation}
\psi_j^{(\mathrm{unnor})}(x) = 
\begin{cases} 
    e^{-x^2/2}  h_1^{K} (x), & \text{if } j \text{ is even}, \\
    e^{-x^2/2}  h_0^{K} (x), & \text{if } j \text{ is odd}.
\end{cases}
\end{equation}

\section{Efficiency of homodyne single-mode fidelity estimation: Proof of Theorem \ref{thrm_single_mode_homodyne}} \label{proof_thm_3.1}
The proof of Theorem \ref{thrm_single_mode_homodyne} combines Eq.~(\ref{equation_perfect_homodyne_estimation}) with Hoeffding's statistical sampling inequality (Lemma \ref{lem_hoeffding}) to estimate the statistical error in calculating the mean of $g^{(\mathrm{hom})}_C$ with a finite number of samples.

If $(x^{(1)},\theta^{(1)}),\dots,(x^{(N)},\theta^{(N)}) \in \R^2$ are the samples obtained by homodyne detection, then
\begin{equation}
    F_C = \frac{1}{N} \sum_{i=1}^N g^{(\mathrm{hom})}_C (x^{(i)},\theta^{(i)}).
\end{equation}
Also, if $\rho_{kl}$ are the elements of the density matrix $\rho$ in the fock basis, then the actual fidelity between $\rho$ and $\ket{C}$ 
\begin{equation}
    F (\rho,\ket{C}) =\sum_{k,l = 0}^{c-1} c_k^* c_l \rho_{kl} =\sum_{k,l = 0}^{c-1} c_k^* c_l \underset{P_{\rho} \mapsto (x,\theta)}{\E}[f_{lk}(x) e^{i(l-k) \theta}].
\end{equation}
From Eq.~(\ref{equation_perfect_homodyne_estimation}). Therefore,
\begin{equation}
    F_C = \sum_{k,l = 0}^{c-1} c_k^* c_l \frac{1}{N} \sum_{i=1}^N (f_{lk}(x^{(i)}) e^{i(l-k) \theta^{(i)}}),
\end{equation}
\begin{equation}
    \left| F (\rho,\ket{C}) -  F_C \right| = \left|  \sum_{k,l = 0}^{c-1} c_k^* c_l \left( \underset{P_{\rho} \mapsto (x,\theta)}{\E}[f_{lk}(x) e^{i(l-k) \theta}] - \frac{1}{N} \sum_{i=1}^N (f_{lk}(x^{(i)}) e^{i(l-k) \theta^{(i)}}) \right) \right|.
\end{equation}
Using Hoeffding's inequality,
\begin{equation}
   \textrm{Pr}\left[ \left|  F (\rho,\ket{C}) -  F_C \right| \geq \lambda \right] \leq 2 \textrm{exp} \left[ - \frac{2N\lambda^2}{G^2} \right],
\end{equation}
where $G = \textrm{max}(g^{\mathrm{hom}}_C(x,\theta)) - \textrm{min}(g^{\mathrm{hom}}_C(x,\theta))$.
\begin{equation}
    \left| g^{\mathrm{hom}}_C (x,\theta) \right| = \left| \sum_{k,l = 0}^{c-1} c_k^* c_l f_{lk}(x) e^{i(l-k) \theta}  \right| \leq \sum_{k,l = 0}^{c-1} |c_k^* c_l f_{lk}(x) e^{i(l-k) \theta}| \leq \sum_{k,l = 0}^{c-1} | f_{lk}(x) e^{i(l-k) \theta}|.
\end{equation}
Using Lemma 1 in \cite{Aubry_2009},
\begin{equation}
    |g^{(\mathrm{hom})}_C(x,\theta)| \leq K_\infty C^{\frac{10}{3}},
\end{equation}
where C is the support size of the core state and $K_\infty$ is a constant.
Therefore, $G \leq 2 K_\infty C^{\frac{10}{3}} $, and
\begin{equation}
   \textrm{Pr}\left[ \left|  F (\rho,\ket{C}) -  F_C \right| \geq \lambda \right] \leq 2 \textrm{exp} \left[ - \frac{N\lambda^2}{2(K_\infty C^{10/3})^2} \right].
\end{equation}
Therefore, for a given error $\epsilon$,
\begin{equation}
    \left| F (\rho,\ket{C}) -  F_C \right| \leq \epsilon,
\end{equation}
with the probability 1 - $\delta$, where $\delta = 2 \textrm{exp} \left[ - \frac{N\epsilon^2}{2(K_\infty C^{10/3})^2} \right]$. For fixed $\epsilon$ and $\delta$, therefore, $\left| F_C^{\mathrm{est}} (\rho) - F_C^{\mathrm{real}} \right| \leq \epsilon$, for $N \geq N_1$, where
\begin{equation}
    N_1 = \mathcal{O}\left( \left(\frac{C^\frac{10}{3}}{\epsilon}\right)^{2} \log \left(\frac{1}{\delta}\right) \right).
\end{equation}

\section{Back-propagation of synchronous homodyne measurement: Proof of Lemma \ref{lemma_prop_homodyne}}\label{proof_lemma_3.2}
The proof of Lemma \ref{lemma_prop_homodyne} uses the effect of a unitary on the quadrature operators to determine the properties the unitary should have for the transformed state to be an eigenstate of the non-rotated quadrature operators.

The action of $\hat{O}$ on the quadrature operators can be described as

\begin{equation}\label{eqn_action_quadrature}
    \hat{O} \begin{bmatrix}
\hat{X}^{(1)} \\
\hat{X}^{(2)} \\
\vdots \\
\hat{P}^{(1)}\\
\vdots \\
\hat{P}^{(m)} 
\end{bmatrix} \hat{O}^\dagger = \mathrm{S}_O \begin{bmatrix}
\hat{X}^{(1)} \\
\hat{X}^{(2)} \\
\vdots \\
\hat{P}^{(1)}\\
\vdots \\
\hat{P}^{(m)} 
\end{bmatrix}.
\end{equation}
Where $\mathrm{S}_O$ is a $2m \times 2m$ unitary matrix. Therefore, 

\begin{equation}
    \hat{X}^{(1)} \hat{O} \ket{\boldsymbol{x}} = \hat{O} \left(\mathrm{S}_O \begin{bmatrix}
\hat{X}^{(1)} \\
\hat{X}^{(2)} \\
\vdots \\
\hat{P}^{(1)}\\
\vdots \\
\hat{P}^{(m)} 
\end{bmatrix}\right)_1 \ket{\boldsymbol{x}}.
\end{equation}
If $\mathrm{S}_O$ is a block diagonal matrix with two $m \times m$ blocks, i.e.\ if $\mathrm{S}_O$ can be expressed as
\begin{equation}\label{eqn_S_O}
\mathrm{S}_O = \left[ {\begin{array}{cc}
   (\mathrm{S}_O)_1 & \boldsymbol{0} \\
   \boldsymbol{0} & (\mathrm{S}_O)_2 \\
  \end{array} } \right]   , 
\end{equation}
where $(S_O)_1$ and $(S_O)_2$ are $m \times m$ matrices. Then 
\begin{equation}
   \left( \mathrm{S}_O \begin{bmatrix}
\hat{X}^{(1)} \\
\hat{X}^{(2)} \\
\vdots \\
\hat{P}^{(1)}\\
\vdots \\
\hat{P}^{(m)} 
\end{bmatrix}\right)_{1} = \left((\mathrm{S}_O)_1 \begin{bmatrix}
\hat{X}^{(1)} \\
\hat{X}^{(2)} \\
\vdots \\
 \hat{X}^{(m)}
\end{bmatrix}\right)_1.
\end{equation}
And
\begin{equation}
     \hat{X}^{(1)} \hat{O} \ket{\boldsymbol{x}} = ((\mathrm{S}_O)_1 \boldsymbol{x})_1 \hat{O} \ket{\boldsymbol{x}},
\end{equation}
where $\boldsymbol{x}$ is the vector of eigenvalues of $\ket{\boldsymbol{x}}$.Therefore $\hat{O}\ket{\boldsymbol{x}}_\theta$ is an eigenvector of $\hat{\boldsymbol{X}}$ with eigenvalue $(\mathrm{S}_O)_1 \boldsymbol{x}_\theta$.
\begin{equation}
    \hat{O}\ket{\boldsymbol{x}} = \ket{(\mathrm{S}_O)_1 \boldsymbol{x}}.
\end{equation}
Furthermore, it can be proven \cite{Arvind_1995} that if $\mathrm{S}_O$ is block diagonal, it must be orthogonal, therefore $(\mathrm{S}_O)_1$ and $(\mathrm{S}_O)_2$ are orthogonal matrices. Now suppose that we rotate all the $m$-modes by an angle $\theta$. So
\begin{equation}
    \hat{X}_{\theta}^{(i)} = \hat{X}^{(i)} \cos \theta + \hat{P}^{(i)} \sin \theta,
\end{equation}
$\forall i \in  \{1,\ldots,m\}$. After the application of $\hat{O}$, the quadratures transform as
\begin{equation}
    \begin{bmatrix}
\hat{X}_{\theta}^{(1)} \\
\hat{X}_{\theta}^{(2)} \\
\vdots \\
 \hat{X}_{\theta}^{(m)}
\end{bmatrix} = (S_O)_1  \begin{bmatrix}
\hat{X}^{(1)} \\
\hat{X}^{(2)} \\
\vdots \\
 \hat{X}^{(m)}
\end{bmatrix} \cos \theta + (S_O)_2 \begin{bmatrix}
\hat{P}^{(1)} \\
\hat{P}^{(2)} \\
\vdots \\
 \hat{P}^{(m)}
\end{bmatrix} \sin \theta.
\end{equation}
If $(S_O)_1$ = $(S_O)_2$, then
\begin{equation}
    \begin{bmatrix}
\hat{X}_{\theta}^{(1)} \\
\hat{X}_{\theta}^{(2)} \\
\vdots \\
 \hat{X}_{\theta}^{(m)}
\end{bmatrix} = (S_O)_1 \begin{bmatrix}
\hat{X}^{(1)} \cos \theta + \hat{P}^{(1)} \sin \theta \\
\hat{X}^{(2)} \cos \theta + \hat{P}^{(2)} \sin \theta \\
\vdots \\
 \hat{X}^{(m)} \cos \theta + \hat{P}^{(m)} \sin \theta  
\end{bmatrix} = (S_O)_1 \begin{bmatrix}
\hat{X}_{\theta}^{(1)} \\
\hat{X}_{\theta}^{(2)} \\
\vdots \\
 \hat{X}_{\theta}^{(m)}
\end{bmatrix}.
\end{equation}
Therefore,
\begin{equation}
   \left( \mathrm{S}_O \begin{bmatrix}
\hat{X}_{\theta}^{(1)} \\
\hat{X}_{\theta}^{(2)} \\
\vdots \\
\hat{X}_{\theta + \pi/2}^{(1)}\\
\vdots \\
\hat{X}_{\theta + \pi/2}^{(m)} 
\end{bmatrix}\right)_{1} = \left((\mathrm{S}_O)_1 \begin{bmatrix}
\hat{X}_{\theta}^{(1)} \\
\hat{X}_{\theta}^{(2)} \\
\vdots \\
 \hat{X}_{\theta}^{(m)}
\end{bmatrix}\right)_1.
\end{equation}
And finally,
\begin{equation}\label{eqn_homodyne_prop}
    \hat{O}\ket{\boldsymbol{x}}_\theta = \ket{(\mathrm{S}_O)_1 \boldsymbol{x}}_\theta.
\end{equation}
Therefore, Eq.~(\ref{eqn_homodyne_prop}) only holds when $S_O$ is a block diagonal matrix with two $m \times m$ blocks, and the two blocks being equal $(S_O)_1 = (S_O)_2$. Furthermore, if all the modes are not rotated by the same angle, then Eq.~(\ref{eqn_homodyne_prop}) does not hold. \\
Next,
\begin{equation}
    \hat{D}(\boldsymbol{\beta}) \ket{\boldsymbol{x}}_\theta = \ket{\boldsymbol{x} + \boldsymbol{\beta}_\theta}_\theta,
\end{equation}
where $\boldsymbol{\beta}_\theta$ is the component of $\boldsymbol{\beta}$ along $\theta$. Finally,

\begin{equation}
    \hat{V} \Pi_{\boldsymbol{x}_\theta} \hat{V}^\dagger = \hat{V} \ket{\boldsymbol{x}}_\theta \bra{\boldsymbol{x}}_\theta \hat{V}^\dagger = \hat{D}(\boldsymbol{\beta}) \hat{O} \ket{\boldsymbol{x}}_\theta \bra{\boldsymbol{x}}_\theta \hat{O}^\dagger \hat{D}(\boldsymbol{\beta})^\dagger,
\end{equation}

\begin{equation}
    \hat{V} \Pi_{\boldsymbol{x}_\theta} \hat{V}^\dagger = \hat{D}(\boldsymbol{\beta}) \ket{(\mathrm{S}_O)_1 \boldsymbol{x}}_\theta \bra{(\mathrm{S}_O)_1 \boldsymbol{x}}_\theta \hat{D}(\boldsymbol{\beta})^\dagger = \ket{(\mathrm{S}_O)_1 \boldsymbol{x} + \boldsymbol{\beta}_\theta}_\theta \bra{(\mathrm{S}_O)_1 \boldsymbol{x} + \boldsymbol{\beta}_\theta}_\theta.
\end{equation}
Let $\boldsymbol{x'} = (\mathrm{S}_O)_1 \boldsymbol{x} + \boldsymbol{\beta}_\theta$. Then,
\begin{equation}
    \hat{V} \Pi_{\boldsymbol{x}_\theta} \hat{V}^\dagger = \ket{\boldsymbol{x'}}_\theta \bra{\boldsymbol{x'}}_\theta = \Pi_{\boldsymbol{x'}_\theta}.
\end{equation}

\section{Efficiency of homodyne verification: Proof of Theorem \ref{thm_multimode_homodyne}}\label{proof_thm_3.2}
We restate Theorem \ref{thm_multimode_homodyne}:
\begin{theo}[Efficiency of Protocol \ref{prot_multimode_fidelity_estimation_homodyne}]
Protocol \ref{prot_multimode_fidelity_estimation_homodyne} solves Verify($\ket{\psi},\rho,N,c,s$) for the class of target states
\begin{equation}
    \ket{\psi} = \hat{D}(\boldsymbol{\beta}) \hat{O} \underset{i=1}{\overset{m}{\bigotimes}} \ket{C_i}
\end{equation}
and probability greater than $1 - \delta$, with
\begin{eqnarray}
    N &=& \mathcal{O}\left( \frac{m^2}{\epsilon^2} C^{\frac{20}{3}} \mathrm{log} \left(\frac{m}{\delta} \right) \right), \nonumber \\
    c &=& \varepsilon, \nonumber \\
    s &=& m\varepsilon,
\end{eqnarray}
where $C = \underset{i \in \{1,\dots,m \}}{\mathrm{max}} c_i$, where $c_i$ is the support size for the core state in the mode i.
\end{theo}
\noindent To prove Theorem \ref{thm_multimode_homodyne}, we first prove the following result:
\begin{theo}[Efficiency of Protocol \ref{prot_multimode_fidelity_estimation_homodyne}]
     Let $\epsilon,\delta > 0$. With the notations of Protocol \ref{prot_multimode_fidelity_estimation_homodyne}, the estimate $W_{\psi}$ satisfies
\begin{equation}\label{appeq:witnes_fidelity}
    1 - m(1 - F(\rho,\ket{\psi})) - \epsilon \leq W_{\psi} \leq F(\rho,\ket{\psi}) + \epsilon
\end{equation}
with probability greater than $1 - \delta$, whenever $N \geq N_2$, with
\begin{equation}
    N_2 = \mathcal{O}\left( \frac{m^2}{\epsilon^2} C^{\frac{20}{3}} \mathrm{log} \left(\frac{m}{\delta} \right) \right),
\end{equation}
where $C = \underset{i \in \{1,\dots,m \}}{\mathrm{max}} c_i$, where $c_i$ is the support size for the core state in the mode i.
\end{theo}
\begin{proof}
    We first consider the case where $\hat{D}(\boldsymbol{\beta}) \hat{O} = \I$ and the target state is $\ket{\psi} = \underset{i=1}{\overset{m}{\bigotimes}} \ket{C_i}$, where $\ket{C_i} = \sum_{n = 0}^{c_i - 1} c_{i,n} \ket{n}$. If $\rho_i$ is the reduced state of $\rho$ in the $i^{th}$ mode, then from Theorem \ref{thrm_single_mode_homodyne}
\begin{equation}
    |F (\rho_i,\ket{C_i}) -  F_{C_i}| \leq \frac{\epsilon}{m},
\end{equation}
with probability greater than 1 - $\delta_i$, where $\delta_i = 2 \mathrm{exp} \left[ - \frac{N\epsilon^2}{2m^2(C^{10/3})^2} \right]$. Therefore, taking the union bound on probabilities,
\begin{eqnarray}
    \left| W_\psi - \sum_{i=1}^m (1 - F (\rho_i,\ket{C_i})) \right| &=& \left|\sum_{i=1}^m(F (\rho_i,\ket{C_i}) - F_{C_i})\right| \\
    &\leq&  \sum_{i=1}^m \left| F (\rho_i,\ket{C_i}) - F_{C_i} \right| \\
    &\leq& \epsilon.
\end{eqnarray}
with probability 1 - $P_W^{i.i.d.}$, where 
\begin{equation}\label{eqn_prob_homodyne}
    P_W^{i.i.d.} = 2 \sum_{i=1}^m \mathrm{exp}\left( - \frac{N \epsilon^2}{m^2 ( C_i^{\frac{10}{3}})^2} \right),
\end{equation}
where $C_i$ is the support size for the core state in the mode $i$. Combining this result with lemma 2,
\begin{equation}
    1 - m(1 - F(\rho,\ket{\psi})) - \epsilon \leq W_\psi \leq F(\rho,\ket{\psi}) + \epsilon,
    \label{eq2}
\end{equation}
with probability 1 - $P_W^{i.i.d}$, where $P_W^{i.i.d}$ is given by Eq.~(\ref{eqn_prob_homodyne}).
If $C := \underset{i \in \{1,\dots,m \}}{\mathrm{max}} c_i$, $\delta \leq (P_W^{i.i.d.})_\mathrm{max}$, where
\begin{equation}\label{eqn_prob_homodyne_max} 
(P_W^{i.i.d.})_\mathrm{max} = 2m \mathrm{exp}\left( - \frac{N\epsilon^2}{m^2 (K_\infty C^{\frac{10}{3}})^2} \right).
\end{equation}
 So, for a given precision $\epsilon$ and probability $1 - \delta$, Eq.~(\ref{eq2}) is satisfied when $N \geq N_2$, with $N_2$ given by Eq.~(\ref{eq1}).
We now generalize the proof to the case where the target state is of the form
\begin{equation}\label{eqn_target_homodyne}
    \ket{\psi} =  \hat{D}(\boldsymbol{\beta}) \hat{O} \underset{i =1}{\overset{m}{\bigotimes}} \ket{C_i},
\end{equation}
where for all $i \in \{1,\ldots,m\}, \ket{C_i} = \sum_{n=0}^{c_i - 1} c_{i,n} \ket{n}$ is a core state, $\hat{O}$ is an $m$-mode passive linear transformation with $2m \times 2m$ block diagonal orthogonal matrix $S_O$, and $\boldsymbol{\beta} \in \C^m$. Next, writing $\hat{V} := \hat{D}(\boldsymbol{\beta}) \hat{O}$ we make use of lemma \ref{lemma_prop_homodyne} to write
\begin{equation}
    \mathrm{Tr} (\hat{V}^\dagger \rho \hat{V} \Pi_{\boldsymbol{x}_\theta}) = \mathrm{Tr} (\rho \hat{V} \Pi_{\boldsymbol{x}_\theta} \hat{V}^\dagger ) = \mathrm{Tr}(\rho \Pi_{(S_O\boldsymbol{x}_\theta + \boldsymbol{\beta}_\theta)}).
\end{equation}
In essence, by homodyne sampling, and classical post-processing $\boldsymbol{x}_\theta = S_O^{-1} (\boldsymbol{x}'_\theta - \boldsymbol{\beta})$, we can estimate the lower bound on the fidelity between $\hat{V}^\dagger \rho \hat{V}$ and $\underset{i = 1}{\overset{m}{\bigotimes}} \ket{C_i}$:
\begin{equation}
    1 - m(1 - F(\hat{V}^\dagger \rho \hat{V}, \underset{i = 1}{\overset{m}{\bigotimes}} \ket{C_i}\bra{C_i})) - \epsilon \leq W_\psi \leq F(\hat{V}^\dagger \rho \hat{V}, \underset{i = 1}{\overset{m}{\bigotimes}} \ket{C_i}\bra{C_i}) + \epsilon,
    \label{eq2other}
\end{equation}
with probability greater than $1 - (P_W^{i.i.d.})_\mathrm{max}$, where $(P_W^{i.i.d.})_\mathrm{max}$ is given by Eq.~(\ref{eqn_prob_homodyne_max}). Given that 
\begin{equation}
    F(\hat{V}^\dagger \rho \hat{V}, \underset{i = 1}{\overset{m}{\bigotimes}} \ket{C_i}\bra{C_i}) = F (\rho,\ket{\psi}).
\end{equation}
We conclude 
\begin{equation}
     1 - m(1 - F( \rho , \ket{\psi})) - \epsilon \leq W_\psi \leq F( \rho , \ket{\psi}) + \epsilon,
\end{equation}
with probability greater than $1 - (P_W^{i.i.d.})_\mathrm{max}$, where $(P_W^{i.i.d.})_\mathrm{max}$ is given by Eq.~(\ref{eqn_prob_homodyne_max}).
\end{proof}
\noindent Now, to understand the precision with which we need to approximate, we note from Eq.\ref{appeq:witnes_fidelity} that
\begin{eqnarray}
    F(\rho,\ket{\psi}) &\geq& W_{\psi} - \epsilon, \nonumber \\
    F(\rho,\ket{\psi}) &\leq& 1 + \frac{1}{m}(W_{\psi} - 1 + \epsilon).
\end{eqnarray}
Therefore, we accept the state $\rho$ if
\begin{eqnarray}
    W_{\psi} - \epsilon &>& 1 - c, \nonumber \\
    W_{\psi} &>& 1 - (c-\epsilon).
\end{eqnarray}
Note that this would imply $F(\rho,\ket{\psi}) > 1 - c$. On the other hand, we reject the state $\rho$ if
\begin{eqnarray}
    1 + \frac{1}{m}(W_{\psi} - 1 + \epsilon) &<& 1 - s, \nonumber \\
    W_{\psi} &<& 1 - (ms+\epsilon).
\end{eqnarray}
This, in turn, implies $F(\rho,\ket{\psi}) < 1 - s$. Choosing $\epsilon = \varepsilon/4$, $c = \varepsilon$ and $s = m \varepsilon$, we get the statement of Theorem \ref{thm_multimode_homodyne}.

\section{Controlling the bias of heterodyne multimode fidelity estimators: Proof of Lemma \ref{lemma_density_estimator}} \label{proof_lemma_2.3}

Given an operator $\rho = \sum_{0 \leq k,l \leq c-1}$ $\rho_{kl} \ket{k}\bra{l}$ (we use the notation ${\rho}$ since we are generally considering density matrices, but what follows also holds for operators which are not necessarily density matrices), it can be proven that \cite{Chabaud2021efficient}
\begin{equation}\label{eqn_single_mode_estimator}
  E_{\alpha \leftarrow Q_{\rho}}[g_{m,n}^{(p)}(\alpha, \tau)] = \rho_{mn} + (-1)^{p+1} \sum_{q=p}^{\infty} \rho_{m+q,n+q}\tau^q \binom{q-1}{p-1} \sqrt{\binom{m+q}{m} \binom{n+q}{n}}.
\end{equation}
Therefore, we have the expression for the bias error in a single-mode estimator. Using this result, we want to find the bias error when we multiply $k$ of these estimators for the multimode heterodyne estimator. Defining
\begin{equation}
    \Pi_{mn}^{(p)} := \int g_{m,n}^{(p)} (\alpha) Q_{\rho} (\alpha) d\alpha \ket{\alpha}\bra{\alpha},
\end{equation}
Eq.~(\ref{eqn_single_mode_estimator}) can be reinterpreted as
\begin{equation}
    \Pi_{mn}^{(p)} = \ket{m}\bra{n} + (-1)^{p+1}\sum_{q=p}^{\infty} \ket{m}\bra{n} \tau^q \binom{q-1}{p-1} \sqrt{\binom{m+q}{m} \binom{n+q}{n}}.
\end{equation}
We define
\begin{equation}
    \Pi^{\mathrm{err}}_{m,n,p} := (-1)^{p+1}\sum_{q=p}^{\infty}\ket{m}\bra{n} \tau^q \binom{q-1}{p-1} \sqrt{\binom{m+q}{m} \binom{n+q}{n}}.
\end{equation}
Let us first consider $k=2$.
\begin{equation}
\underset{\boldsymbol{\alpha} \leftarrow Q_{\rho}(\boldsymbol{\alpha})}{\E} [g_{m_1,n_1}^{(p_1)}g_{m_2,n_2}^{(p_2)}] = \mathrm{Tr} \left[\Pi_{m_1 n_1}^{(p_1)}\otimes\Pi_{m_2 n_2}^{(p_2)} \rho \right] = \mathrm{Tr}\left[(\ket{m_1}\bra{n_1} + \Pi^{\mathrm{err}}_{m_1,n_1,p_1}) \otimes (\ket{m_2}\bra{n_2} + \Pi^{\mathrm{err}}_{m_2,n_2,p_2})\rho \right]
\end{equation}
\begin{equation}
    \hspace{10 mm} = \rho_{\substack{\\m_1 n_1\\m_2 n_2}} +  \mathrm{Tr}\left[\ket{m_1}\bra{n_1} \otimes \Pi^{\mathrm{err}}_{m_2,n_2,p_2} \rho  \right] + \mathrm{Tr}\left[\Pi^{\mathrm{err}}_{m_1,n_1,p_1} \otimes \ket{m_2}\bra{n_2} \rho  + \right] + \mathrm{Tr} \left[\Pi^{\mathrm{err}}_{m_1,n_1,p_1} \otimes \Pi^{\mathrm{err}}_{m_2,n_2,p_2} \rho \right].
\end{equation}
Therefore,
\begin{equation}\label{eqn_B43}
    \left| \rho_{\substack{\\m_1 n_1\\m_2 n_2}} -  \underset{\boldsymbol{\alpha} \leftarrow Q_{\rho}(\boldsymbol{\alpha})}{\E} [g_{m_1,n_1}^{(p_1)}g_{m_2,n_2}^{(p_2)}]\right| = |\mathrm{Tr}\left[\ket{m_1}\bra{n_1} \otimes \Pi^{\mathrm{err}}_{m_2,n_2,p_2} \rho  \right]| + |\mathrm{Tr}\left[\Pi^{\mathrm{err}}_{m_1,n_1,p_1} \otimes \ket{m_2}\bra{n_2} \rho  + \right]| 
\end{equation}
\begin{equation}
    \hspace{60mm} + |\mathrm{Tr} \left[\Pi^{\mathrm{err}}_{m_1,n_1,p_1} \otimes \Pi^{\mathrm{err}}_{m_2,n_2,p_2} \rho \right]|.
\end{equation}
Now,
\begin{equation}
    |\mathrm{Tr}\left[\ket{m_1}\bra{n_1} \otimes \Pi^{\mathrm{err}}_{m_2,n_2,p_2} \rho  \right]| \leq ||\ket{m_1}\bra{n_1} \otimes \Pi^{\mathrm{err}}_{m_2,n_2,p_2}||_{\mathrm{HS}} ||\rho||_{\mathrm{HS}},
\end{equation}
where $||.||_{\mathrm{HS}}$, refers to the Hilbert-Schmidt norm. 
\begin{equation}
    ||\rho||_{\mathrm{HS}} = \sqrt{\mathrm{Tr}[\rho^2]} \leq 1,
\end{equation}
\begin{equation}
    ||\ket{m_1}\bra{n_1} \otimes \Pi^{\mathrm{err}}_{m_2,n_2,p_2}||_{\mathrm{HS}} \leq \sqrt{\mathrm{Tr}_1 [\ket{m_1}\bra{m_1}] \mathrm{Tr}_2 [(\Pi^{\mathrm{err}}_{m_2,n_2,p_2})^{\dagger} \Pi^{\mathrm{err}}_{m_2,n_2,p_2}] }.
\end{equation}
Now,
\begin{equation}
\mathrm{Tr}_2[(\Pi^{\mathrm{err}}_{m_2,n_2,p_2})^{\dagger} \Pi^{\mathrm{err}}_{m_2,n_2,p_2}] = \sqrt{\sum_{q_2=p_2}^{\infty} \tau^{2q_2} \binom{m_2 +q_2}{m_2} \binom{n_2 +q_2}{n_2} }.
\end{equation}
Further,
\begin{equation}
    \sum_{q_2=p_2}^{\infty} \tau^{2q_2} \binom{m_2 +q_2}{m_2} \binom{n_2 +q_2}{n_2} = \tau^{2p_2} \sum_{q_2=0}^{\infty} \tau^{2q_2} \binom{m_2 +q_2+p_2}{m_2} \binom{n_2 +q_2+p_2}{n_2}.
\end{equation}
Using,
\begin{equation}
    \binom{m_p+q}{m} \leq (1+m)^{p+q},
\end{equation}
\begin{equation}
    \sum_{q_2=p_2}^{\infty} \tau^{2q_2} \binom{m_2 +q_2}{m_2} \binom{n_2 +q_2}{n_2} \leq \tau^{2p_2} (1+m_2)^{p_2} (1+n_2)^{p_2} \sum_{q_2=0}^{\infty} (\tau^2(1+m_2)(1+n_2))^{q_2} .
\end{equation}
If 
\begin{equation}
    \tau \leq \frac{1}{\sqrt{(1+m_2)(1+n_2)}}.
\end{equation}
Then
\begin{equation}
    \sum_{q_2=p_2}^{\infty} \tau^{2q_2} \binom{m_2 +q_2}{m_2} \binom{n_2 +q_2}{n_2} \leq \frac{\tau^{2p_2}(1+m_2)^{p_2} (1+n_2)^{p_2}}{1 - \tau^2(1+m_2)(1+n_2)}.
\end{equation}
Using this result in Eq.~(\ref{eqn_B43}),
\begin{equation}
    \left| \rho_{\substack{\\m_1 n_1\\m_2 n_2}} -  \underset{\boldsymbol{\alpha} \leftarrow Q_{\rho}(\boldsymbol{\alpha})}{\E} [g_{m_1,n_1}^{(p_1)}g_{m_2,n_2}^{(p_2)}]\right| \leq \frac{\tau^{p_2}(1+m_2)^{p_2/2} (1+n_2)^{p_2/2}}{\sqrt{1 - \tau^2(1+m_2)(1+n_2)}} + \frac{\tau^{p_1}(1+m_1)^{p_1/2} (1+n_1)^{p_1/2}}{\sqrt{1 - \tau^2(1+m_1)(1+n_1)}}
\end{equation}
\begin{equation}
    \hspace{50mm} + \frac{\tau^{p_1}(1+m_1)^{p_1/2} (1+n_1)^{p_1/2}}{\sqrt{1 - \tau^2(1+m_1)(1+n_1)}} \times \frac{\tau^{p_2}(1+m_2)^{p_2/2} (1+n_2)^{p_2/2}}{\sqrt{1 - \tau^2(1+m_2)(1+n_2)}} .
\end{equation}
Calling 
\begin{equation}\label{eqn_b55}
 E^{p_1}_{m_1,n_1} := \frac{\tau^{p_1}(1+m_1)^{p_1/2} (1+n_1)^{p_1/2}}{\sqrt{1 - \tau^2(1+m_1)(1+n_1)}},
\end{equation}
\begin{equation}
    \left| \rho_{\substack{\\m_1 n_1\\m_2 n_2}} -  \underset{\boldsymbol{\alpha} \leftarrow Q_{\rho}(\boldsymbol{\alpha})}{\E} [g_{m_1,n_1}^{(p_1)}g_{m_2,n_2}^{(p_2)}]\right| \leq E^{p_1}_{m_1,n_1} + E^{p_2}_{m_2,n_2} + E^{p_1}_{m_1,n_1} E^{p_2}_{m_2,n_2} =: \epsilon^{(p_1,p_2)}(2) .
\end{equation}
By taking $\tau \rightarrow 0$, we can make $E^{p_1}_{m_1,n_1}$ and hence $\epsilon^{(p_1,p_2)}(2)$ arbitrarily close to zero.

By induction, it can be similarly proven that 
\begin{eqnarray}\label{eq:needthatone}
    \left| \rho_{\boldsymbol{m,n}} - \underset{\boldsymbol{\alpha} \leftarrow Q_{\rho}}{\E}\left[ g_{m_1 n_1}^{(p_1)} (\alpha_1,\tau) \ldots g_{m_k n_k}^{(p_k)} (\alpha_k,\tau)  \right]  \right| &\leq& E^{p_1}_{m_1,n_1} \epsilon^{(p_2,\ldots,p_{k-1})}(k-1) + \epsilon^{(p_2,\ldots,p_{k-1})}(k-1) + E^{p_1}_{m_1,n_1} \nonumber \\
    &:=& \epsilon_{\mathrm{bias}}(\bm p, k).
\end{eqnarray}
Similar to the two-mode case, $\epsilon_{\mathrm{bias}}(\bm p, k)$ can be made to go to 0 by taking $\tau \rightarrow 0$. 
\section{Efficiency of heterodyne multimode fidelity estimation: Proof of Theorem \ref{theorem_k_mode_fidelity_est}}\label{proof_theorem_2.1}
The proof of Theorem \ref{theorem_k_mode_fidelity_est} combines Lemma \ref{lemma_density_estimator} with Hoeffding inequality (Lemma \ref{lem_hoeffding}) to determine the maximum possible error in fidelity estimation.

The true fidelity between the mixed $m$-mode state $\rho$ and the target state  $\ket{C} = \sum_{\boldsymbol{0} \leq \boldsymbol{m,n} \leq \boldsymbol{c - 1}} c_{\boldsymbol{n}} \ket{\boldsymbol{n}}$ is given by
\begin{equation}
    F_C(\rho) = \sum_{\boldsymbol{0} \leq \boldsymbol{m,n} \leq \boldsymbol{c - 1}} c_{\boldsymbol{m}}^* c_{\boldsymbol{n}} \rho_{\boldsymbol{m n}}.
\end{equation}
Let us denote
\begin{equation}
    \underset{\alpha_1 \leftarrow Q_{\rho^{(1)(m_2,n_2)}}}{\E}\left[ g_{m_1 n_1}^{(p_1)} (\alpha_1,\tau) \right] \dots \underset{\alpha_k \leftarrow Q_{\rho_k}}{\E}\left[ g_{m_k n_k}^{(p_k)} (\alpha_k,\tau) \right] := \underset{\boldsymbol{\alpha} \leftarrow Q_{\rho}}{\E} \left[ g_{\boldsymbol{m n}}^{(\boldsymbol{p})} (\boldsymbol{\alpha},\tau)\right].
\end{equation}
Therefore,
\begin{equation}
    F_C(\rho) -  \underset{\boldsymbol{\alpha} \leftarrow Q_{\rho}}{\E} \left[\sum_{\boldsymbol{0} \leq \boldsymbol{m,n} \leq \boldsymbol{c - 1}}  c_{\boldsymbol{m}}^* c_{\boldsymbol{n}}  g_{\boldsymbol{m n}}^{(\boldsymbol{p})} (\boldsymbol{\alpha},\tau)  \right]  = \sum_{\boldsymbol{0} \leq \boldsymbol{m,n} \leq \boldsymbol{c - 1}}  c_{\boldsymbol{m}}^* c_{\boldsymbol{n}} \left(\rho_{\boldsymbol{m n}} - \underset{\boldsymbol{\alpha} \leftarrow Q_{\rho}}{\E} \left[ g_{\boldsymbol{m n}}^{(\boldsymbol{p})} (\boldsymbol{\alpha},\tau)\right] \right).
\end{equation}
From Lemma \ref{lemma_density_estimator}, $\left|\rho_{\boldsymbol{m n}} - \underset{\boldsymbol{\alpha} \leftarrow Q_{\rho}}{\E} \left[ g_{\boldsymbol{m n}}^{(\boldsymbol{p})} (\boldsymbol{\alpha},\tau)\right]  \right| \leq \epsilon_{\mathrm{bias}}(\bm p,k)$.
Therefore,
\begin{equation}
\left| F_C(\rho) -  \underset{\boldsymbol{\alpha} \leftarrow Q_{\rho}}{\E} \left[\sum_{\boldsymbol{0} \leq \boldsymbol{m,n} \leq \boldsymbol{c - 1}}  c_{\boldsymbol{m}}^* c_{\boldsymbol{n}}  g_{\boldsymbol{m n}}^{(\boldsymbol{p})} (\boldsymbol{\alpha},\tau)  \right] \right| \leq \sum_{\boldsymbol{0} \leq \boldsymbol{m,n} \leq \boldsymbol{c - 1}}  |c_{\boldsymbol{m}}^* c_{\boldsymbol{n}}| \epsilon_{\mathrm{bias}}(\bm p,k) = \epsilon_{\mathrm{(bias)}}(\boldsymbol{C}, \boldsymbol{p}, \tau).
\end{equation}
Then, defining
\begin{equation}
    F_C^{\mathrm{k-est}}(\rho,\tau) = \sum_{\boldsymbol{0} \leq \boldsymbol{m,n} \leq \boldsymbol{c - 1}} \frac{1}{N}\underset{l=1}{\overset{N}{\sum}} c_{\boldsymbol{m}}^* c_{\boldsymbol{n}} g_{\boldsymbol{mn}}^{\boldsymbol{p}} (\boldsymbol{\alpha}^{(l)},\tau).
\end{equation}
Using Hoeffding's inequality,
\begin{equation}
    \mathrm{Pr} \left[ \left|  \sum_{\boldsymbol{0} \leq \boldsymbol{m,n} \leq \boldsymbol{c - 1}} c_{\boldsymbol{m}}^* c_{\boldsymbol{n}} \left(\underset{\boldsymbol{\alpha} \leftarrow Q_{\rho}}{\E} \left[ g_{\boldsymbol{m n}}^{(\boldsymbol{p})} (\boldsymbol{\alpha},\tau)\right] -  \frac{1}{N} \sum_{l=1}^N \left( g_{\boldsymbol{m n}}^{\boldsymbol{p}} (\boldsymbol{\alpha}^{(l)},\tau) \right) \right)   \right| \geq \lambda \right] = 2 \mathrm{exp}\left ( - \frac{2N\lambda^2}{R^2}\right),
\end{equation}
where $R = \mathrm{max}(\underset{\boldsymbol{0} \leq \boldsymbol{m,n} \leq \boldsymbol{c - 1}}{\sum}  c_{\boldsymbol{m}}^* c_{\boldsymbol{n}} g_{\boldsymbol{m n}}^{\boldsymbol{p}}) - \mathrm{min}(\underset{\boldsymbol{0} \leq \boldsymbol{m,n} \leq \boldsymbol{c - 1}}{\sum}  c_{\boldsymbol{m}}^* c_{\boldsymbol{n}} g_{\boldsymbol{m n}}^{\boldsymbol{p}})$.
\[ \hspace{-80mm}\left| \sum_{\boldsymbol{0} \leq \boldsymbol{m,n} \leq \boldsymbol{c - 1}}  c_{\boldsymbol{m}}^* c_{\boldsymbol{n}} g_{\boldsymbol{m n}}^{\boldsymbol{p}} \right| \leq \sum_{\boldsymbol{0} \leq \boldsymbol{m,n} \leq \boldsymbol{c - 1}}  |c_{\boldsymbol{m}}^* c_{\boldsymbol{n}}| |g_{\boldsymbol{m n}}^{\boldsymbol{p}}|  \]
\begin{equation}
    \hspace{39 mm} \leq  \sum_{\boldsymbol{0} \leq \boldsymbol{m,n} \leq \boldsymbol{c - 1}}  |c_{\boldsymbol{m}}^* c_{\boldsymbol{n}}| \frac{1}{\tau^{1 + \frac{\boldsymbol{m+n}}{2}}} \binom{\boldsymbol{\mathrm{max(m,n)}+p}}{\boldsymbol{p - 1}} \sqrt{2^{|\boldsymbol{m - n}| }\binom{\mathrm{max}\boldsymbol{(m,n)}}{\mathrm{min}\boldsymbol{(m,n)}}} = R(\boldsymbol{C},\boldsymbol{p},\tau).
\end{equation}
And, $R \leq 2R(\boldsymbol{C},\boldsymbol{p},\tau)$. Therefore,
\begin{equation}
    \mathrm{Pr} \left[ \left|  \sum_{\boldsymbol{0} \leq \boldsymbol{m,n} \leq \boldsymbol{c - 1}}  c_{\boldsymbol{m}}^* c_{\boldsymbol{n}} \left(\underset{\boldsymbol{\alpha} \leftarrow Q_{\rho}}{\E} \left[ g_{\boldsymbol{m n}}^{(\boldsymbol{p})} (\boldsymbol{\alpha},\tau)\right] -  \frac{1}{N} \sum_{l=1}^N \left( g_{\boldsymbol{m n}}^{\boldsymbol{p}} (\boldsymbol{\alpha}^{(l)},\tau) \right) \right)   \right| \geq \lambda \right]\leq  2 \mathrm{exp}\left ( - \frac{N\lambda^2}{2(R(\boldsymbol{C},\boldsymbol{p},\tau))^2}\right).
\end{equation}
And
   \[\hspace{-50mm}|F_C(\rho) - F_C^{\mathrm{k-est}}(\tau)| = \left|\sum_{\boldsymbol{0} \leq \boldsymbol{m,n} \leq \boldsymbol{c - 1}}  c_{\boldsymbol{m}}^* c_{\boldsymbol{n}} \left(\rho_{\boldsymbol{m n}} -  \frac{1}{N} \sum_{l=1}^N \left( g_{\boldsymbol{m n}}^{\boldsymbol{p}} (\boldsymbol{\alpha}^{(l)},\tau) \right) \right)  \right|  \]
   \begin{equation}
       \hspace{-20mm} \leq \left|\sum_{\boldsymbol{0} \leq \boldsymbol{m,n} \leq \boldsymbol{c - 1}} c_{\boldsymbol{m}}^* c_{\boldsymbol{n}} \left(\rho_{\boldsymbol{m n}} - \underset{\boldsymbol{\alpha} \leftarrow Q_{\rho}}{\E} \left[ g_{\boldsymbol{m n}}^{(\boldsymbol{p})} (\boldsymbol{\alpha},\tau)\right] \right)  \right| 
   \end{equation}
   \begin{equation}
      \hspace{20mm} + \left|\sum_{\boldsymbol{0} \leq \boldsymbol{m,n} \leq \boldsymbol{c - 1}}  c_{\boldsymbol{m}}^* c_{\boldsymbol{n}} \left(\underset{\boldsymbol{\alpha} \leftarrow Q_{\rho}}{\E} \left[ g_{\boldsymbol{m n}}^{(\boldsymbol{p})} (\boldsymbol{\alpha},\tau)\right] - \frac{1}{N} \sum_{l=1}^N \left( g_{\boldsymbol{m n}}^{\boldsymbol{p}} (\boldsymbol{\alpha}^{(l)},\tau) \right) \right)  \right|.
   \end{equation}
   And therefore,
   \begin{equation}
    |F_C(\rho) - F_C^{\mathrm{k-est}}(\tau)| \leq \lambda + \epsilon_{(\mathrm{bias})}(\boldsymbol{C},\boldsymbol{p},\tau) := \epsilon.
   \end{equation}
with probability $1 - \delta$, where $\delta = 2\mathrm{exp}\left ( - \frac{N\lambda^2}{2(R(\boldsymbol{C},\boldsymbol{p},\tau))^2}\right)$.

\noindent Now, given 
\begin{equation}
    \epsilon = \lambda + \epsilon_{(\mathrm{bias})}(\boldsymbol{C},\boldsymbol{p},\tau),
\end{equation}
and
\begin{equation}
    \delta = 2\mathrm{exp}\left ( - \frac{N\lambda^2}{2(R(\boldsymbol{C},\boldsymbol{p},\tau))^2}\right).
\end{equation}
We want to find out that given $\epsilon$ and $\delta$, how does the number of samples $N$ scales with the number of modes that are grouped together, $k$. We can see from the expression for $\epsilon$ and $\delta$ that in general the scaling will be a complicated function depending on a lot of parameters. Hereafter, we focus only on the scaling with $k$, and we thus drop the dependency of $\epsilon_{\mathrm{bias}}(\bm p,k), g_{\boldsymbol{mn}}^{\boldsymbol{p}}$ and $\epsilon_{(\mathrm{bias})}(\boldsymbol{C},\boldsymbol{p},\tau)$ on $\boldsymbol{m,n,p}$ and $c$. 

We have,
\begin{equation}
    \epsilon_{(\mathrm{bias})}(\boldsymbol{C},\tau) = \sum_{0\leq \boldsymbol m,\boldsymbol n \leq \boldsymbol{c-1}} |c_{\boldsymbol{m}}^* c_{\boldsymbol{n}}| \epsilon(k) \le {c}^2 \epsilon (k).
\end{equation}
Further, we denote $E^{p_1}_{m_1,n_1}$ from Eq.~(\ref{eqn_b55}) as $\epsilon(1)$ and similarly $E^{p_k}_{m_k,n_k}$ as $\epsilon(k)$. From Eq.~(\ref{eq:needthatone}):
\begin{equation}
    \epsilon(k) = \epsilon(1)\epsilon(k-1) + \epsilon(1) + \epsilon(k-1).
\end{equation}
so $(\epsilon(k)+1)= (\epsilon(k-1)+1)(\epsilon(1) + 1)$  and thus
\begin{equation}
    \epsilon(k) = (1+\epsilon(1))^k-1.
\end{equation}
Now,
\begin{equation}
    \epsilon(1) = \frac{\tau^{p}(1+m)^{p/2} (1+n)^{p/2}}{\sqrt{1 - \tau^2(1+m)(1+n)}}.
\end{equation}
For small $\tau$ we obtain:
\begin{equation}
    \epsilon(k) =\mathcal O(k\tau^p),
\end{equation}
and taking $\lambda=\mathcal O(k\tau^p)$ we have 
\begin{equation}
    \delta = 2 \exp\left(- \frac{N \lambda^2}{2  (R(\boldsymbol{C},\boldsymbol{p},\tau))^2}\right) =   2 \exp\left(- \frac{N \mathcal{O}(k^2\tau^{2p})}{2 (R(\boldsymbol{C},\boldsymbol{p},\tau))^2}\right).
\end{equation}
and 
\begin{equation}
    \epsilon =\mathcal{O}(k\tau^p).
\end{equation}
Note that, by adjusting $\tau$, we can make $\epsilon$ arbitrarily small. Equivalently, we can pick 
\begin{equation}
    \tau =\mathcal{O}\left(\frac{\epsilon^{1/p}} {k^{1/p}}\right).
\end{equation}

Now,
\begin{equation}
    R(\boldsymbol{C},\boldsymbol{p},\tau) = \sum_{\boldsymbol{0} \leq \boldsymbol{m,n} \leq \boldsymbol{c - 1}}  c_{\boldsymbol{m}}^* c_{\boldsymbol{n}} \frac{1}{\tau^{1 + \frac{\boldsymbol{m+n}}{2}}} \binom{\boldsymbol{\mathrm{max(m,n)}+p}}{\boldsymbol{p - 1}} \sqrt{2^{|\boldsymbol{m - n}| }\binom{\mathrm{max}\boldsymbol{(m,n)}}{\mathrm{min}\boldsymbol{(m,n)}}}.
\end{equation}
For $\tau =\mathcal{O}\left(\frac{\epsilon^{1/p}} {k^{1/p}}\right)$ and small $\epsilon$, this expression scales as 
\begin{equation}\label{appeq:range_constant}
    R(\boldsymbol{C},\boldsymbol{p},\tau) = \mathcal{O} \left(\frac{K^kk^{kc/p}}{\epsilon^{kc/p}}\right),
\end{equation}
where $K$ depends on the target core state and the free parameter $p$, but is independent of $\tau$ and $k$. Finally, the number of samples scales as 
\begin{equation}
    N = \mathcal{O} \left(\frac{R(\boldsymbol{C},\boldsymbol{p},\tau)^2}{\lambda^2} \log \left(\frac{1}{\delta}\right)  \right) = \mathcal{O} \left(\frac{K^kk^{2kc/p}}{\epsilon^{2 + \frac{2kc}{p}}} \log \left(\frac{1}{\delta}\right) \right).
\end{equation}
Therefore, the sufficient number of samples $N$ scales at most super-exponentially with the number of modes $k$. For $k=1$, we recover the polynomial scaling derived in \cite{Chabaud2021efficient}.

\section{Removing the i.i.d.\ assumption}
\label{proof_no_iid}
 
\begin{figure}
    \centering
    \includegraphics[width=\linewidth]{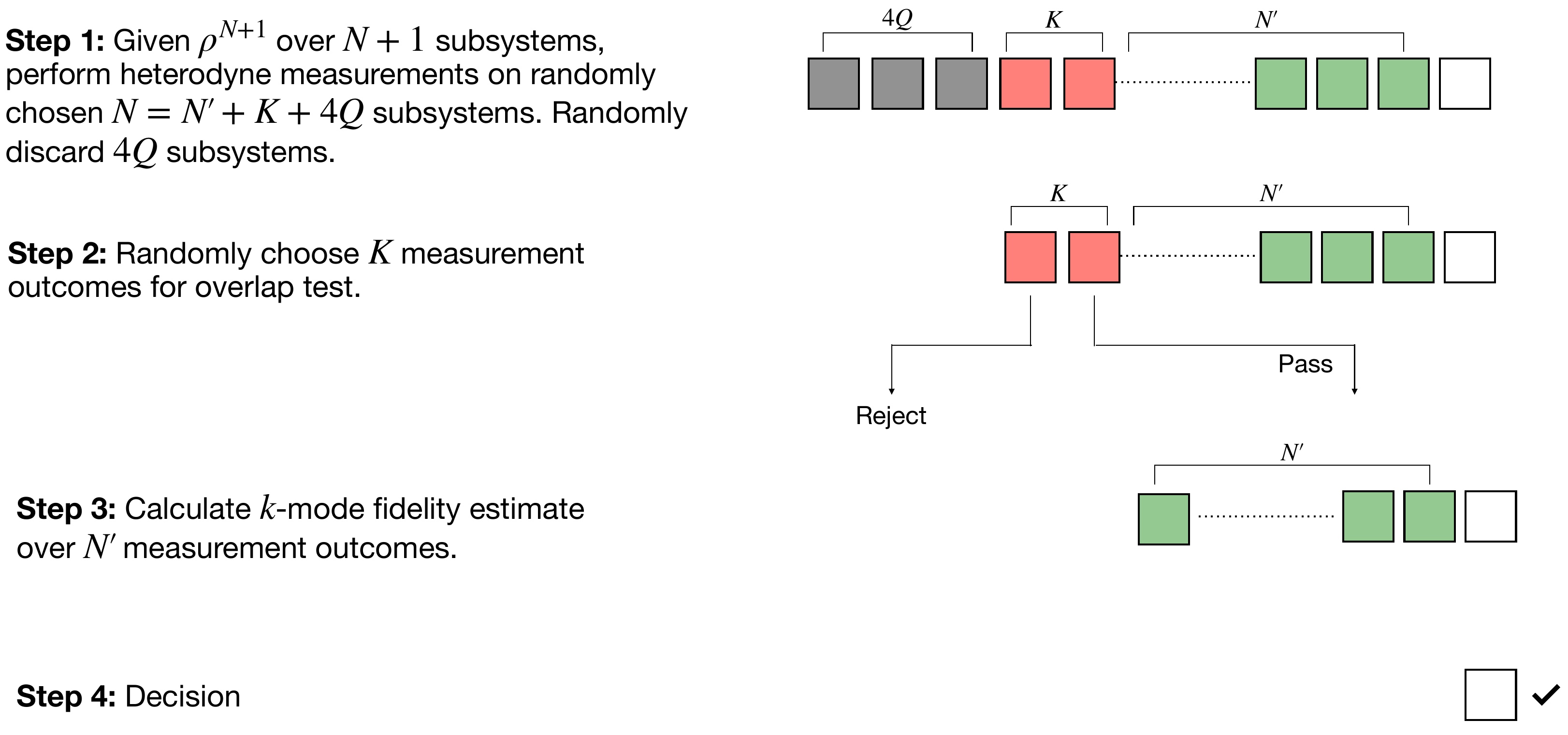}
    \caption{Schematic representation of Protocol \ref{protocol_k_mode_fidelity_no_i.i.d}. Given $\rho^{N+1}$, performing heterodyne measurements over randomly chosen $N$ subsystems and after appropriate processing over the obtained measurement outcomes, we can find a lower bound on the fidelity between the target core state and the remaining subsystem $\rho$. Note that we refer to Step 3 of Protocol \ref{protocol_k_mode_fidelity_no_i.i.d} as the \textit{overlap test} in the Figure. This figure is inspired by \cite[Fig. 1]{wu2021efficient}.}
    \label{fig_no_iid}
\end{figure}

In this section, we sketch how to remove the i.i.d.\ assumption in Protocol \ref{protocol_k_mode_fidelity} by modifying it as follows:
\begin{protocol} \label{protocol_k_mode_fidelity_no_i.i.d}
    (General t-mode fidelity estimation with heterodyne detection) Let $\ket{C} = \underset{\boldsymbol{0} \leq \boldsymbol{n} \leq \boldsymbol{c-1}}{\sum} c_{\boldsymbol{n}} \ket{\boldsymbol{n}}$ be a core state, where $c_1,\dots,c_t \in \N^*$. Let also $N \in \mathbb{N}^*$, and let $p_1,\dots, p_t \in \mathbb{N}^*$, $E, S \in \mathbb{N}$ and $0 < \tau < 1$ be free parameters. Let $\rho^{N+1}$ be an unknown t-mode quantum state over $N+1$ subsystems. We write $N = N' + K + 4Q$, for $N', K, Q \in \mathbb{N}$.
    
\begin{enumerate}
    \item Measure $N = N' + K + 4Q$ subsystems of $\rho^{N+1}$ chosen at random with heterodyne detection, obtaining the measurement outcomes $\boldsymbol{\alpha_1}, \dots, \boldsymbol{\alpha_{N'}}, \boldsymbol{\beta_1}, \dots, \boldsymbol{\beta_K}, \boldsymbol{\gamma_1}, \dots, \boldsymbol{\gamma_{4Q}} \in \mathbb{C}^t$. Let $\rho$ be the remaining state.
    \item Discard the $4Q$ outcomes $\boldsymbol{\gamma_1}, \dots, \boldsymbol{\gamma_{4Q}}$.
    \item Record the number $R$ of measurement outcomes $\boldsymbol{\beta_i}$ such that $\boldsymbol{\beta_i}^\dagger \cdot \boldsymbol{\beta_i} + 1 > E$, for $i = 1, \dots, K$. The protocol aborts if $R \geq S$ (For the purpose of Figure \ref{fig_no_iid}, we call this the \textit{overlap test}).
    \item Otherwise, compute the mean $F_C^{k-est}(\rho)$ of the function $\boldsymbol{z} \mapsto g_C^{\mathrm{k-est}}(\boldsymbol{z}, \tau)$ (defined in Eq.~\ref{mm_fidelity_estimate_eqn}) over the measurement outcomes $\boldsymbol{\alpha_1}, \dots, \boldsymbol{\alpha_{N'}} \in \mathbb{C}^t$.
    \item This gives the fidelity estimate $F_C^{\mathrm{k-est}}$.
\end{enumerate}
\end{protocol}
\noindent Protocol \ref{protocol_k_mode_fidelity_no_i.i.d} is also depicted in Figure \ref{fig_no_iid}. Note that this protocol differs from Protocol \ref{protocol_k_mode_fidelity} due to the inclusion of two additional classical post-processing steps, specifically steps 2 and 3. These steps are integral to a de Finetti reduction for infinite-dimensional systems as detailed in \cite{Renner2009}. Notably, step 3 involves an energy test, where the energy parameters $E$ and $S$ must be selected to ensure completeness. This means that if the perfect core state is sent, it will pass the energy test with high probability.

The value $F_C^{\mathrm{k-est}}(\tau)$ obtained is an estimate of the fidelity between the remaining state $\rho$ and the target core state $\ket{C}$. The efficiency of the protocol is summarised by the following result:
\begin{theo}\label{thm_noiid_multi}

Let $\epsilon > 0$. With the notations of Protocol \ref{protocol_k_mode_fidelity_no_i.i.d},

\begin{equation}
    \left|F_C(\rho) - F_C^{k-est}(\tau)\right| \leq \epsilon + A_C^{\boldsymbol{p}}(\tau) +  P_{\mathrm{deFinetti}}^C,
\end{equation}

or the protocol aborts in step 3, with probability greater than $1 - (P_{\mathrm{support}}^C + P_{\mathrm{deFinetti}}^C + P_{\mathrm{choice}}^C + P_{\text{Hoeffding}}^C)$, where

\begin{equation}
    P_{\mathrm{support}}^C = 8 K^{3/2} \mathrm{exp}\left[ - \frac{K}{9} \left(\frac{Q}{N' + 1 + 4Q} - \frac{2S}{K}  \right)^2 \right],
    \label{P_support}
\end{equation}

\begin{equation}
    P_{\text{deFinetti}}^C = Q^{\frac{(E+1)^{2t}}{2}} \exp\left[-\frac{2Q(Q+1)}{N'+1+4Q}\right],
\end{equation}

\begin{equation}
    P_{\mathrm{choice}}^C = \frac{4Q}{N' + 1},
\end{equation}

\begin{equation}
    P_{\text{Hoeffding}}^C = 2 \binom{N'+1}{4Q} \exp \left(-\frac{N' + 1 - 4Q}{2} \left( \frac{\epsilon}{R(\boldsymbol{C},\boldsymbol{p},\tau)} -  8Q\right)^2\right).
\end{equation}
\end{theo}

\begin{proof}
    The proof follows these three steps:

\begin{itemize}
    \item \textbf{Support estimation:} with probability arbitrarily close to 1, most of the subsystems of the permutation-invariant state $\rho^{N' + 4Q}$ lie in a lower dimensional subspace, or the score of the state $\rho^{N' + 4Q + K}$ at the support estimation step is high
    \item \textbf{De Finetti reduction:} any permutation-invariant state with most of its subsystems in a lower dimensional subspace admits a purification in the symmetric subspace that still has most of its subsystems in a lower dimensional subspace. This purification is well approximated by a mixture of almost-i.i.d.\ states;
    \item \textbf{Hoeffding inequality for almost-i.i.d.\ states:} mixture of almost-i.i.d.\ states can be certified in a similar fashion as i.i.d.\ states.
\end{itemize}

\subsection{Support estimation for permutation invariant states}
The proof essentially follows the proof in section E.1 in the appendix of \cite{chabaud2020building}, where Eq.~(124) in the paper is modified using the following lemma \cite{Leverrier2013}
\begin{lem}
    Let $T_n$ and $U_n$ be defined as
\[
T_t := \frac{1}{\pi^t} \int_{\sum_{i=1}^t |\alpha_i|^2 \geq E} \ket{\alpha_1}\bra{\alpha_1} \cdots \ket{\alpha_t}\bra{\alpha_t} \, d\alpha_1 \cdots d\alpha_t,
\]
and
\[
U_t := \sum_{m=E+1}^{\infty} \Pi_m^t \text{ with } \Pi_m^t = \sum_{m_1 + \cdots + m_t = m} \ket{m_1 \cdots m_t} \bra{m_1 \cdots m_t}.
\]
Then, the following inequality holds:
\[
U_t \leq 2 T_t.
\]
\end{lem}
Following the proof in section E.1 of \cite{chabaud2020building}, with $U \mapsto U_t, T \mapsto T_t, n \mapsto N'+4Q+1, s \mapsto S, q \mapsto Q$ and  $k \mapsto K$, $\rho^{N'+4Q+1}$ will be successfully projected into 
\begin{equation}
    S^{N'+ 1 + 4Q}_{\bar{\mathcal{H}}^{\otimes N'+ 1 + 4Q - Q}} := \mathrm{span} \bigcup_{\pi} \pi (\bar{\mathcal{H}}^{\otimes N'+ 1 + 3Q} \otimes \mathcal{H}^{\otimes Q}) \pi^{-1},
\end{equation}
or the protocol aborts (R $\geq$ S) with probability $P_{\mathrm{support}}^C$, given by Eq.~(\ref{P_support}). Note that 
\begin{eqnarray}
    \bar{\mathcal{H}} &=& \bar{\mathcal{H}_1} \otimes \dots \otimes \bar{\mathcal{H}_t}, \\
    \mathcal{H} &=& \mathcal{H}_1 \otimes \dots \otimes \mathcal{H}_t,
    \label{H,Hbar}
\end{eqnarray}
where $\bar{\mathcal{H}}_i$ is a single mode Hilbert space of at most E photons, of dimension $E+1$ $ \forall i \in \{ 1,\dots,t \}$ 

\subsection{De Finetti reduction}

We modify Lemma 9 and Theorem 5 from \cite{chabaud2020building} as follows

\begin{lem} \label{purification}
    Any permutation-invariant state $\rho^{N' + 1 + 4Q} \in  S^{N' + 1 + 4Q}_{\bar{\mathcal{H}}^{\otimes N' + 1 + 4Q - Q}} $ has a purification  $\tilde{\rho}^{N'+1+4Q}$ in 
    $ \operatorname{Sym}^{N' + 1 + 4Q}(\mathcal{H}\otimes\mathcal{H})\cap S^n_{(\bar{\mathcal{H}} \otimes \bar{\mathcal{H}})^{\otimes N'+1+4Q - 2Q}} $.

\end{lem}

\noindent where $\operatorname{Sym}^n (\mathcal{K}) = \{ \phi \in \mathcal{K}^{\otimes n}, \pi \phi = \phi\, (\forall \pi) \}$ and $\mathcal{H}$ and $\bar{\mathcal{H}}$ are defined in Eq.~(\ref{H,Hbar}). Here $\pi$ refers to a permutation of subsystems. \\

For all \( n, r \geq 0 \) and all $\ket{v} \in \bar{\mathcal{H}} \otimes \bar{\mathcal{H}}$, the set of \textit{almost-i.i.d.\ states along} $\ket{v}$, \( S^n_{v^{\otimes n-r}} \), is defined as the span of all vectors that are, up to reorderings, of the form $\ket{v}^{\otimes n-r} \otimes \ket{\phi}$, for an arbitrary $\ket{\phi} \in (\mathcal{H}\otimes\mathcal{H})^{\otimes r}$. In the following, we simply refer to these states as \textit{almost-i.i.d.\ states} (which becomes relevant when \( r \ll n \)).

\begin{theo} \label{definetti}

Let $\tilde{\rho}^{N'+1+4Q} \in \operatorname{Sym}^{N' + 1 + 4Q}(\mathcal{H} \otimes \mathcal{H}) \cap S^n_{(\bar{\mathcal{H}} \otimes \bar{\mathcal{H}})^{\otimes N'+1+4Q - 2Q}}$ and let $ \tilde{\rho}^{N'+1} = \mathrm{Tr}_{4Q}[\tilde{\rho}^{N'+1+4Q}] $. Then, there exist a finite set $\mathcal{V}$ of unit vectors $\ket{v} \in \bar{\mathcal{H}} \otimes \bar{\mathcal{H}}$, a probability distribution \( \{p_v\}_{v \in \mathcal{V}} \) over \( \mathcal{V} \), and almost-i.i.d.\ states $\tilde{\rho}_v^{N'+1} \in S^{N'+1}_{v^{\otimes N'+1+4Q - 8Q}}$ such that
\begin{equation}
F\left( \tilde{\rho}^{N'+1}, \sum_{v \in \mathcal{V}} p_v \tilde{\rho}_v^{N'+1} \right) > 1 - Q^{(E+1)^{2t}} \exp\left[-\frac{4Q(Q+1)}{N'+1+4Q}\right] .  
\end{equation}

\end{theo}

Given a state $\rho^{N' + 1 + 4Q} \in  S^{N' + 1 + 4Q}_{\bar{\mathcal{H}}^{\otimes (N' + 1 + 4Q - Q)}}$, applying Theorem \ref{definetti} to the purification $\tilde{\rho}^{N' + 1 + 4Q}$ provided by Lemma \ref{purification} shows that the reduced state $\tilde{\rho}^{N' + 1}$ is close in fidelity to a mixture of states that are i.i.d.\ on $N' + 1 - 4Q$ subsystems. These results follow from the proof in the paper \cite{Renner2009}.

\subsection{Hoeffding inequality for almost-i.i.d states}
Given an almost i.i.d.\ state $\ket{\Phi} \in \mathcal{S}^t_{v^{\otimes t - r}}$, if we do heterodyne measurements over $t-m$ of the subsystems of $\ket{\Phi}$ and find the average over a function $f(\alpha)$, for our problem it is important to know how close will this statistical average be to the expectation value of this function over the Husimi $Q$ function defined by $\ket{v}$. The following lemma gives this result \cite{chabaud2020building}

\begin{lem}
Let $M > 0 \in \mathbb{R}$ and let $f : \mathbb{C} \to \mathbb{R}$ be a function bounded as $|f(\boldsymbol{\alpha})| \leq M$ for all $\boldsymbol{\alpha} \in \mathbb{C}^t$. Let $\mu > 0$ and $1 \leq m \leq r < t$ such that
\[
(t - m)\mu > 2Mr. \tag{137}
\]
Let also $\ket{v} \in \mathcal{H}$ and $\ket{\Phi} \in \mathcal{S}^t_{v^{\otimes t - r}}$. Let $\mathcal{M} = \{ \mathcal{M}_\alpha \}_{\alpha \in \mathbb{C}^t}$ be a POVM on $\mathcal{H}$ and let $D_{\ket{v}}$ be the probability density function of the outcomes of the measurement $\mathcal{M}$ applied to $\ket{v}\bra{v}$. Then
\[
\Pr_\alpha \left[
    \left| \frac{1}{t - m} \underset{i = 1}{\overset{t-m}{\sum}} f(\boldsymbol{\alpha}_i) - \left( \mathbb{E}_{\boldsymbol{\beta} \leftarrow D_{\ket{v}}} \left[ f(\boldsymbol{\beta}) \right] \right) \right| > \mu
\right] \leq 2 \binom{t}{r} \exp \left[ -\frac{t - r}{2} \left( \frac{\mu}{M} - \frac{2r}{t - m} \right)^2 \right], \tag{138}
\]
where the probability is taken over the outcomes $\boldsymbol{\alpha} = (\boldsymbol{\alpha_1, \ldots, \alpha_{t-m}})$ of the product measurement $\mathcal{M}^{\otimes t-m}$ applied to $\ket{\Phi}\bra{\Phi}$.

\end{lem}

Following the steps of Appendix 3.4 of \cite{chabaud2020building}, it is tedious but straightforward to prove Theorem \ref{thm_noiid_multi} from here.

\end{proof}

\section{Efficiency of heterodyne verification: Proof of Theorem \ref{theorem_mm_witness_inequality}}\label{proof_theorem_2.2}
To prove Theorem \ref{theorem_mm_witness_inequality}, we use the following result:
\begin{theo}[Efficiency of Protocol \ref{protocol_multimode_k_fidelity witness}]
    Let $\lambda_i > 0 \hspace{1mm} \forall i \in \{1,\ldots,m\}$. With the notations of Protocol \ref{protocol_multimode_k_fidelity witness}, there exists $N_2 = {\mathcal{O}} \left(\frac{(K_1 m^{K_2})^{k}}{\epsilon^{K_2 k}} \log \left(\frac{m}{k\delta} \right)\right)$, up to $\poly\!m$ factors, where $K_1,K_2 > 0$ are constants depending on the support size of the core state, such that for the number of samples $N \geq N_2$, the obtained estimate $W_\psi^{(k)}$ follows the bound

    \begin{equation}\label{appeq:fidelity_mm_witness_inequality}
         1 - \frac{m}{k} (1 - F(\rho,\ket{\psi})) - \epsilon  \leq W_\psi^{(k)}   \leq F(\rho,\ket{\psi}) + \epsilon
    \end{equation}
    with probability 1 - $\delta$. Here $\epsilon$ and $\delta$ are given by

    \begin{equation}
        \epsilon  = \underset{i=1}{\overset{m/k}{\sum}} (\lambda_i + \epsilon_{(\mathrm{bias})}(\boldsymbol{C_i},\boldsymbol{p_i},\tau_i)),
    \end{equation}

    \begin{equation}
        \delta = 2 \underset{i=1}{\overset{m/k}{\sum}} \mathrm{exp}\left ( - \frac{N\lambda_i^2}{2(R(\boldsymbol{C_i},\boldsymbol{p_i},\tau_i))^2}\right),
    \end{equation}
where $\epsilon_{\mathrm{(bias)}}(\boldsymbol{C_i},\boldsymbol{p_i},\tau_i)$ and $R(\boldsymbol{C_i},\boldsymbol{p_i},\tau_i)$ represents the bias error and range of the estimator, as explained in Theorem \ref{theorem_k_mode_fidelity_est}.
\end{theo}
\begin{proof}
    The proof of Theorem \ref{theorem_mm_witness_inequality} uses $k$-mode fidelity estimation as a subroutine with $k$-mode fidelity witness and statistical sampling inequality. 

Let's first consider the case $\hat{S}(\boldsymbol{\xi})\hat{D}(\boldsymbol{\beta}) \hat{U} = \I$, and therefore $\boldsymbol{\alpha} = \boldsymbol{\gamma}$, and our target state is $\ket{\psi} = \underset{i=1}{\overset{m}{\bigotimes}} \ket{C_i}$. For all $i \in \{1,\ldots,m/k\}$, we write $\ket{\boldsymbol{C_i}} = \sum_{\boldsymbol{n} = \boldsymbol{0}}^{\boldsymbol{c_i - 1}} \boldsymbol{c_{i,n}} \ket{\boldsymbol{n}}$, where $\boldsymbol{c_i} \in (\N^*)^k$, and $\rho_i = \mathrm{Tr}_{\{1,\dots,m \} \backslash {ki - (k-1),\dots,ik}} (\rho)$ the reduced state of $\rho$ over the selected $k$ modes. As it is proved in Theorem \ref{theorem_k_mode_fidelity_est},
\begin{equation}
    |F_{C_i}^{\mathrm{real}} (\rho_i) - F_{C_i}^{\mathrm{k-est}}(\tau_i)| \leq \lambda_i + \epsilon_{(\mathrm{bias})}(\boldsymbol{C_i},\boldsymbol{p_i},\tau_i) = \mathcal O(k \tau_i^{p_i}),
\end{equation}
by choosing $\lambda_i = \mathcal O(k\tau_i^{p_i})$, $\boldsymbol{p_i} = \{p_i,\ldots, p_{i} \}$ (for simplicity, we are choosing the free parameters associated to a particular block to be the same), with probability $1 - \delta_i$, where
\begin{equation}
    \delta_i = 2\mathrm{exp}\left ( - \frac{N\lambda_i^2}{2(R(\boldsymbol{C_i},\boldsymbol{p_i},\tau_i))^2}\right) = 2\exp\left(- \frac{N}{2} \mathcal{O}\left(\frac{k^2 \tau_i^{2kc_i +p_i}}{K_i^{2k}}\right)\right),
\end{equation} 
where the constant $K_i>0$ is given in Eq.~\ref{appeq:range_constant}. Therefore, taking the union bound on probabilities 

    \[\hspace{-30mm} \left|  1 -  \sum_{i=1}^{m/k} \left( 1 - F (\rho_{ki - (k - 1)\dots ki}, \ket{\psi_{ki - (k-1)}} \otimes \dots \otimes \ket{\psi_{ki}}) \right) - W_\psi^{(k)}  \right| \]
    
    \[= \left| \sum_{i=1}^{m/k} \left(  F (\rho_{ki - (k - 1)\dots ki}, \ket{\psi_{ki - (k-1)}} \otimes \dots \otimes \ket{\psi_{ki}}) - F_{C_i}^{(k-est)} (\tau_i)  \right) \right| \]

    \[\hspace{-5mm}\leq \sum_{i=1}^{m/k} \left|  F (\rho_{ki - (k - 1)\dots ki}, \ket{\psi_{ki - (k-1)}} \otimes \dots \otimes \ket{\psi_{ki}}) - F_{C_i}^{(k-est)}(\tau_i) \right| \]

    \begin{equation}
        \hspace{-75mm}\leq  \sum_{i=1}^{m/k} (\lambda_i + \epsilon_{(\mathrm{bias})}(\boldsymbol{C_i},\boldsymbol{p_i},\tau_i)) = \sum_{i=1}^{m/k} \mathcal O(k\tau_i^{p_i}).
    \end{equation}
with probability higher than 1 - $P_W^{k-iid}$, where
\begin{equation}
    P_W^{k-iid} := 2 \underset{i=1}{\overset{m/k}{\sum}} \mathrm{exp}\left ( - \frac{N\lambda_i^2}{2(R(\boldsymbol{C_i},\boldsymbol{p_i},\tau_i))^2}\right) =  2 \sum_{i = 1}^{m/k}\mathcal{O}\left(\exp\left(-\frac{k^2 \tau_i^{2kc_i +p_i}}{K_i^{2k}}\right)\right).
    \label{eqn_P_W^{k-iid}}
\end{equation}
Furthermore, using Eq.~(\ref{eqn_inequality_k-mode}), 
\begin{equation}
     1 - \frac{m}{k} (1 - F(\rho,\ket{\psi})) \leq 1 -  \sum_{i=1}^{m/k} \left( 1 - F (\rho_{ki - (k - 1)\dots ki}, \ket{\psi_{ki - (k-1)}} \otimes \dots \otimes \ket{\psi_{ki}}) \right)   \leq F(\rho,\ket{\psi}).
\end{equation}
For the total error to be $\epsilon$, we take $k\tau_i^{p_i}= \mathcal O(\frac{k \epsilon}{m})$ and equivalently, $\tau = \mathcal O(\epsilon^{1/p_i}/m^{1/p_i})$. Then the probability of failure
\begin{eqnarray}\label{appeq:failure_prob_multi_heterodyne}
      P_W^{k-iid} &=&  2 \sum_{i = 1}^{m/k}\mathcal{O}\left(\exp\left(-\frac{k^2 \tau_i^{2kc_i +p_i}}{K_i^{2k}}\right)\right) = 2 \sum_{i = 1}^{m/k} \mathcal O\left(\exp\left(-N\frac{k^2 \epsilon^{2kc_i/p_i + 2}}{K_i^{2k}m^{2k c_i/p_i + 2}}\right)\right) \nonumber \\
      &=& \mathcal{O}\left(\frac{m}{k}\exp\left(-N \frac{k^2 \epsilon^{2kc/p + 2}}{K^{2k}m^{2kc/p + 2}}\right)\right),
\end{eqnarray}
where $K = \max(K_1,\dots,K_{m/k}), c = \max(c_1,\dots,c_{m/k})$ and $p = \min(p_1,\dots,p_{m/k})$. Note that in the statement of the Theorem, we write $K_1 = K^2$ and $K_2 = 2c/p$. Therefore,
\begin{equation}
     1 - \frac{m}{k} (1 - F(\rho,\ket{\psi})) - \epsilon  \leq W_\psi^{(k)}   \leq F(\rho,\ket{\psi}) + \epsilon.
\end{equation}
with probability higher than 1 - $P_W^{k-iid}$, where $P_W^{k-iid}$ is given by Eq.~\ref{appeq:failure_prob_multi_heterodyne}. Using the free parameters $p_1,\ldots,p_{m/k}$ and $\tau_1,\ldots,\tau_{m/k}$, the errors and probabilities can be optimized.

We now generalize the previous proof to the case where the target state is of the form

\begin{equation}
    \ket{\psi} = \hat{S} (\boldsymbol{\xi}) \hat{D} (\boldsymbol{\beta}) \hat{U} \underset{i=1}{\overset{m}{\bigotimes}} \ket{C_i},
\end{equation}
where for all $i \in \{1,\ldots,m \}$ each of the states $\ket{C_i} = \sum_{n = 0}^{c_i - 1} c_{i,n} \ket{n}$ is a core state, where $\hat{U}$ is an $m$-mode passive linear transformation with $m \times m$ unitary matrix U and where $\boldsymbol{\xi,\beta} \in \C^m$. Using Lemma \ref{lemma_prop_heterodyne},
\begin{equation}
    \mathrm{Tr}(\hat{V}^\dagger \rho \hat{V} \Pi_{\boldsymbol{\alpha}}^{\boldsymbol{0}}) = \mathrm{Tr}(\rho \hat{V} \Pi_{\boldsymbol{\alpha}}^{\boldsymbol{0}}\hat{V}^\dagger) = \mathrm{Tr}(\rho\Pi_{\mathrm{U}\boldsymbol{\alpha}+\boldsymbol{\beta}}^{\boldsymbol{\xi}}).
\end{equation}
Therefore, measurement outcomes obtained from the unbalanced heterodyne sampling of $\rho$,  $\boldsymbol{\gamma^{(1)}},\ldots,\boldsymbol{\gamma^{(N)}}$ follow the same probability distribution as measurement outcomes obtained from the balanced heterodyne sampling of $\hat{V}^\dagger \rho \hat{V}$, $\boldsymbol{\alpha^{(1)}},\ldots,\boldsymbol{\alpha^{(N)}}$. Therefore, with classical post-processing $\boldsymbol{\alpha} = \mathrm{U}^\dagger (\boldsymbol{\gamma} - \boldsymbol{\beta})$, we obtain a fidelity witness $W_{\psi}^{(k)}$, such that
\begin{eqnarray}
     1 - \frac{m}{k} (1 - F(\hat{V}^\dagger \rho \hat{V},\underset{i = 1}{\overset{m}{\bigotimes}} \ket{C_i}\bra{C_i})) - \epsilon \leq W_\psi^{(k)}  \leq F(\hat{V}^\dagger \rho \hat{V},\underset{i = 1}{\overset{m}{\bigotimes}} \ket{C_i}\bra{C_i}) + \epsilon,
\end{eqnarray}
with probability higher than 1 - $P_W^{k-iid}$, where $P_W^{k-iid}$ is given by Eq.~\ref{appeq:failure_prob_multi_heterodyne}. Given that 
\begin{equation}
    F(\hat{V}^\dagger \rho \hat{V},\underset{i = 1}{\overset{m}{\bigotimes}} \ket{C_i}\bra{C_i}) = F(\rho,\ket{\psi}\bra{\psi}),
\end{equation}
We have
\begin{equation}
    1 - \frac{m}{k} (1 - F(\rho,\ket{\psi}\bra{\psi})) - \epsilon  \leq W_\psi^{(k)}   \leq F(\rho,\ket{\psi}\bra{\psi}) + \epsilon.
\end{equation}
\end{proof}
\noindent Now, to understand the precision with which we need to approximate, we note from Eq.~(\ref{appeq:fidelity_mm_witness_inequality}) that
\begin{eqnarray}
    F(\rho,\ket{\psi}) &\geq& W_{\psi}^{(k)} - \epsilon, \nonumber \\
    F(\rho,\ket{\psi}) &\leq& 1 + \frac{k}{m}(W_{\psi}^{(k)} - 1 + \epsilon).
\end{eqnarray}
Therefore, we accept the state $\rho$ if
\begin{eqnarray}
    W_{\psi} - \epsilon &>& 1 - c, \nonumber \\
    W_{\psi} &>& 1 - (c-\epsilon).
\end{eqnarray}
Note that this would imply $F(\rho,\ket{\psi}) > 1 - c$. On the other hand, we reject the state $\rho$ if
\begin{eqnarray}
    1 + \frac{k}{m}(W_{\psi} - 1 + \epsilon) &<& 1 - s, \nonumber \\
    W_{\psi} &<& 1 - (\frac{m}{k}s+\epsilon).
\end{eqnarray}
This, in turn, implies $F(\rho,\ket{\psi}) < 1 - s$. Choosing $\epsilon = \varepsilon/4$, $c = \varepsilon$ and $s = m \varepsilon/k$, we get the statement of Theorem \ref{theorem_mm_witness_inequality}.
\section{Verification of doped Gaussian states}\label{sec:verify_t-doped}
To verify doped Gaussian states with the Gaussian unitaries restricted to passive linear transformations and displacement operators, we use a fidelity witness that is a combination of $\kappa t$-mode fidelity witness and single mode witness $W_{\psi}^{(\kappa t, 1)}$,
\begin{equation}
    W^{(\kappa t,1)} := 1 - (1 - F(\rho_{1,\dots,\kappa t},\phi)) - \sum_{i = \kappa t + 1}^m (1 - F(\rho_i,\psi_i)).
\end{equation}
where $\ket{\phi}$ is a $\kappa t$-mode quantum state.

Following the steps of the proof of Lemma \ref{lemma_mm_fid_witness_2}, we can prove
\begin{eqnarray}\label{eqn_t-doped_witness_inequality}
    1 - (m - \kappa t + 1) (1 - F(\rho, \ket{\phi}\otimes \ket{\psi_{\kappa t + 1}} \otimes \dots \otimes \ket{\psi_m} )) \nonumber && \\ && \hspace{-80mm}\leq W^{(\kappa t,1)} \leq  F(\rho, \ket{\phi}\otimes \ket{\psi_{\kappa t + 1}} \otimes \dots \otimes \ket{\psi_m} ).
\end{eqnarray}
Therefore, estimating $W^{(\kappa t,1)}$ can give a lower bound on the fidelity between the output state $\rho$ and the target state $\ket{\psi} = D(\boldsymbol{\beta}) \hat{U} \ket{\psi} \otimes \ket{0}^{\otimes m - \kappa t}$. Using this fidelity witness, a protocol to verify $\ket{\psi}$ can be described as \\

\begin{protocol}\label{protocol_t_doped_Gaussian}
    (Fidelity witness estimation for t-doped Gaussian states). We write \(\ket{\psi} = D(\boldsymbol{\beta}) \hat{U} \ket{\phi} \otimes \ket{0}^{\otimes m - \kappa t} \) as the \(m\)-mode target pure state. Let \(N \in \mathbb{N}^*\) and let \(p_1, \ldots, p_m \in \mathbb{N}^*\) and \(0 < \tau_{\kappa t}, \tau_{\kappa t + 1} \dots, \tau_{m} < 1\) be free parameters. Let \(\rho^{\otimes N+1} = N + 1\) copies of an unknown \(m\)-mode (mixed) quantum state \(\rho\).

\begin{enumerate}
    \item Measure all \(m\) subsystems of \(N\) copies of \(\rho\) with unbalanced heterodyne detection with unbalancing parameters $\boldsymbol{\xi} =$ \( \{\xi_1, \ldots, \xi_m\}\), obtaining the vectors of measurement outcomes \( \boldsymbol{ \gamma^{(1)}, \ldots, \gamma^{(N)}}\in \mathbb{C}^m\).
    
    \item For all \(k \in \{1, \ldots, N\}\), compute the vectors $\boldsymbol{\alpha}^{(k)} = U^\dagger (\boldsymbol{\gamma}^{(k)} - \boldsymbol{\beta})$. We write $\boldsymbol{\alpha}^{(k)} = (\alpha_1^{(k)},\ldots,\alpha_m^{(k)} ) \in \C^m$
    
    \item Group the first $\kappa t$-modes together and compute the mean $F_\phi^{(\kappa t-est)}$ of the function $\boldsymbol{z} \mapsto g_{\phi}^{\kappa t-est} (\boldsymbol{z},\tau_{\kappa t})$ (defined in Eq.~(\ref{mm_fidelity_estimate_eqn})) over the samples $\boldsymbol{\alpha^{(1)}_{\kappa t}},\dots,\boldsymbol{\alpha^{(N)}_{\kappa t}} \in \C^{\kappa t}$, where $\boldsymbol{\alpha^{(i)}_{\kappa t}} = \{ \alpha_1^{(i)},\ldots, \alpha_{\kappa t}^{(i)}  \}$.
    
    \item For all \(i \in \{\kappa t + 1, \dots, m\}\), compute the mean $F_{0_i}$ of the function $z \mapsto g_{0,0}^{(p_i)}(z,\tau_i)$ over the measurement outcomes \(\alpha^{(1)}_i, \ldots, \alpha^{(N)}_i \in \mathbb{C}\).
    
    \item Compute the fidelity witness estimate
    
    \begin{equation}
        W_{\psi}^{(\kappa t,1)} = 1 - (1 - F_\phi^{(t-est)}) - \sum_{i = \kappa t + 1}^m (1 - F_{0_i}).
    \end{equation}
 \item \textbf{Accept} the remaining state $\rho$ if $W_{\psi}^{(\kappa t,1)} > 1 - 3\epsilon/4$ and $\textbf{Reject}$ if $W_{\psi}^{(\kappa t,1)} < 1 - \epsilon((m-\kappa t + 1)^2 + 1/4)$.
\end{enumerate}
\end{protocol}
\noindent Then we have the following Theorem:
\begin{theo}[Efficient verification of doped states]\label{apptheo:verification_doped}
Protocol \ref{protocol_t_doped_Gaussian} solves \textsf{Verify}($\ket{\psi},c,s,\delta$)  (Problem \ref{pb:verification_problem}) for the class of $t$-doped Gaussian states
\begin{equation}
    \ket{\psi} = D(\boldsymbol{\beta}) \hat{U} \ket{\phi} \otimes \ket{0}^{\otimes m - \kappa t},
\end{equation}
with
\begin{eqnarray}
    N_{\mathrm{threshold}} &=&  \tilde{\mathcal{O}}\left(\frac{K_3^{\kappa t}m^{2K_4(\kappa t)}}{\epsilon^{K_4 \kappa t}}\log\left(\frac{m - \kappa t +1}{\delta}\right)\right), \nonumber \\
    c &=& \varepsilon, \nonumber \\
    s &=& (m - \kappa t + 1)\varepsilon.
\end{eqnarray}
where $K_3, K_4>0$ are constants depending on the core state support size of $\ket{\phi}$ and where $\tilde{\mathcal{O}}$ represents ordering up to terms polynomial in $m, \kappa t$ and $\varepsilon$.
\end{theo}
\noindent To prove Theorem \ref{apptheo:verification_doped}, we first have the following result:
\begin{theo} \label{theorem_t_doped_Gaussian_verification}
    Let $\epsilon,\delta > 0$. With the notations of Protocol \ref{protocol_t_doped_Gaussian}, using a number of samples $N \geq N_{\mathrm{threshold}}$, with 
    \begin{equation}
        N_{\mathrm{threshold}} = \tilde{\mathcal{O}}\left(\frac{K_3^{\kappa t}m^{2K_4(\kappa t)}}{\epsilon^{K_4 \kappa t}}\log\left(\frac{m - \kappa t +1}{\delta}\right)\right),
    \end{equation}
where $K_3,K_4 > 0 $ are constants depending on the core state support size of $\ket{\phi}$ and where $\tilde{\mathcal{O}}$ represents ordering up to terms polynomial in $m, \kappa t$ and $\varepsilon$, we can find a fidelity witness estimate that satisfies
\begin{equation}\label{appeq:dope_relation_fidelity_witness}
         1 - (m - \kappa t + 1) (1 - F(\rho, \ket{\phi}\otimes \ket{\psi_{\kappa t + 1}} \otimes \dots \otimes \ket{\psi_m})) - \epsilon  \leq W_\psi^{(\kappa t,1)}   \leq  F(\rho, \ket{\phi}\otimes \ket{\psi_{\kappa t + 1}} \otimes \dots \otimes \ket{\psi_m} ) + \epsilon,
    \end{equation}
with probability $1 - \delta$.
\end{theo}
\begin{proof}
    The proof of Theorem \ref{theorem_t_doped_Gaussian_verification} combines Theorem \ref{theorem_k_mode_fidelity_est} with Eq.~(\ref{eqn_t-doped_witness_inequality}) and Hoeffding inequality (Lemma \ref{lem_hoeffding}).

We will prove the Theorem for $\hat{D}(\boldsymbol{\beta})\hat{U} = \I$, the extension to any arbitrary $\hat{D}(\boldsymbol{\beta})\hat{U} (\ket{\phi} \otimes \ket{0}^{\otimes (m - \kappa t)})$ can be done following the same steps (using classical post-processing) as in the proof of Theorem \ref{theorem_mm_witness_inequality}.

Our target state is $\ket{\psi} = \ket{\phi} \otimes \ket{0}^{ \otimes (m - \kappa t)}$. For the first $\kappa t$ modes, if $\rho_{\kappa t}$ denotes the reduced state over the first $\kappa t$ modes, then from Theorem \ref{theorem_k_mode_fidelity_est}, by choosing $\lambda_{\kappa t} = \mathcal O((\kappa t)\tau_{\kappa t}^{p_{\kappa t}})$,

\begin{equation}
|F_{\phi}(\rho_{\kappa t}) - F_{\phi}^{k-est}(\tau_{\kappa t})| \leq \lambda_{\kappa t} + \epsilon_{\mathrm{bias}} (\phi,\boldsymbol{p}_{\kappa t},\tau_{\kappa t}) = \mathcal{O}((\kappa t)\tau_{\kappa t}^{p_{\kappa t}}),
\end{equation}
with probability $1 - \delta_{kt}$, where $\delta_{kt} = 2\exp\left(-\frac{N\lambda_{kt}^2}{2(R(\phi, \boldsymbol{p}_{\kappa t},\tau_{\kappa t}))^2}\right)$, and $\boldsymbol{p}_{\kappa t} = \{ p_{\kappa t}, \dots, p_{\kappa t} \}$ (for simplicity, we take the free parameters to be the same). Furthermore, for $i = \{ \kappa t+1, \dots, m \}$, if $\rho_i$ denotes the reduced state over the $i^{th}$ mode, then from appendix C of \cite{Chabaud2021efficient}, by choosing $\lambda_i = \mathcal O(\tau^{p_i})$
\begin{equation}
    |F(\ket{0_i},\rho_i) - F_{0_i}(\tau_i)| \leq \lambda_i + \tau^{p_i} = \mathcal{O}(\tau^{p_i}) ,
\end{equation}
with probability $1 - \delta_i$, where $\delta_i = 2\exp\left(-\frac{N\lambda_i^2 \tau_i^2}{2(B_{0_i}^{(p_i)}(\tau_i))^2}\right)$, where
\begin{equation}
    B_{0_i}^{(p_i)}(\tau_i) = \binom{p_i+1}{p_i-1} =: K_0.
\end{equation}
Therefore, taking the union bound on probabilities,
    \[\hspace{-60mm}\left|1 - (1 - F(\ket{\phi},\rho_{1,\dots,\kappa t})) - \sum_{i=\kappa t+1}^{m} 1 - F(\ket{0}_i,\rho_i)) - W_{\psi}^{\kappa t,1}\right|\]
    \[= \left|(F(\ket{\phi},\rho_{1,\dots,\kappa t}) - F_{\phi}^{(t-est)}(\tau_{\kappa t})) + \sum_{i=k+1}^m (F(\ket{0}_i,\rho_i) - F_{0_i}(\tau_i))\right|\]
    \[
    \leq |F(\ket{\phi},\rho_{1,\dots,\kappa t}) - F_{\phi}^{(t-est)}(\tau_{\kappa t})| + \sum_{i=k+1}^m |F(\ket{0}_i,\rho_i) - F_{0_i}(\tau_i)|
    \]
    \begin{equation}
        \hspace{-35mm}\leq \lambda_{\kappa t} + \epsilon_{\mathrm{bias}} (\phi,\boldsymbol{p}_{\kappa t},\tau_{\kappa t}) + \sum_{i=k+1}^m (\lambda_i + \tau^{p_i} A_{0_i}^{(p_i)}) = \mathcal O\left((\kappa t)\tau_{\kappa t}^{p_{\kappa t}} + \sum_{i = \kappa t+1}^m \tau_{i}^{p_i}\right).
    \end{equation}
Therefore, to make the total error equal to $\epsilon$, we take
\begin{eqnarray}
    (\kappa t)\tau_{\kappa t}^{p_{\kappa t}} &=& \frac{\epsilon}{m - \kappa t  + 1}, \nonumber \\
    \tau_{\kappa t} &=& \frac{\epsilon^{1/p_{\kappa t}}}{(m - \kappa t + 1)^{1/p_{\kappa t}}(\kappa t)^{1/p_{\kappa t}}},
\end{eqnarray}
and $\forall i \in (\kappa \tau +1,\dots,m)$, we take
\begin{eqnarray}
    \tau_{i}^{p_i} &=& \frac{\epsilon}{m-\kappa t+1}, \nonumber \\
    \tau_i &=& \frac{\epsilon^{1/p_i}}{(m-\kappa t + 1)^{1/p_i}}.
\end{eqnarray}
Combined with the fact that from Eq.~\ref{appeq:range_constant},
\begin{equation}
    R(\phi, \boldsymbol{p}_{\kappa t},\tau_{\kappa t}) = \mathcal O\left(\frac{K_{\kappa t}^{\kappa t} (\kappa t)^{\kappa t c/p_{\kappa t}}(m - \kappa t +1)^{\kappa t c/p_{\kappa t}}}{\epsilon^{\kappa t c/p_{\kappa t}}}\right),
\end{equation}
where $c$ is the core support size of $\ket{\phi}$. We get that probability of failure,
\begin{eqnarray}\label{eqn_P_Wk i.i.d.}
    P_W^{k-i.i.d.} &=& 2\exp\left(-\frac{N\lambda_{\kappa t}^2}{2(R(\phi, \boldsymbol{p}_{\kappa t},\tau_{\kappa t}))^2}\right) + 2 \sum_{i=\kappa t+1}^m \exp\left(-\frac{N\tau_i^2\lambda_i^2}{2(B_{0_i}^{p_i}(\tau_i)^2))}\right) \nonumber \\
    &=& \mathcal O\left(\exp\left(-N\frac{\epsilon^{2 \kappa t c/p_{\kappa t} + 2}}{2K_{\kappa t}^{2\kappa t}(m-\kappa t +1)^{2 \kappa t c/p_{\kappa t} + 2} (\kappa t)^{2\kappa t c/p}} \right) + \sum_{i = \kappa t+1}^m \exp\left(-N\frac{\epsilon^{2/p_i + 2}}{2 K_0^2 (m - \kappa t+1)^{2/p_i + 2}}\right)\right) \nonumber \\
    &=& \mathcal O\left((m-\kappa t +1)\exp\left(-N \frac{\epsilon^{2\kappa t c/p + 2}}{2 K^{2\kappa t}(m - \kappa t+1)^{2\kappa t c/p + 2}(\kappa t)^{2\kappa t c/p}}\right)\right),
\end{eqnarray}
where $K = \max(K_{\kappa t},K_0), p = \min(p_{\kappa t},p_{\kappa t + 1},\dots,p_m)$. Furthermore, using Eq.~(\ref{eqn_t-doped_witness_inequality}),
\begin{equation}
    1 - (m - \kappa t + 1) F(\ket{\phi}\otimes \ket{\psi_{\kappa t + 1}} \otimes \dots \otimes \ket{\psi_m}, \rho ) \leq W^{(\kappa t,1)} \leq  F(\ket{\phi}\otimes \ket{\psi_{\kappa t + 1}} \otimes \dots \otimes \ket{\psi_m}, \rho ).
\end{equation}
Therefore,
\begin{eqnarray}
    1 - (m - \kappa t + 1) F(\ket{\phi}\otimes \ket{\psi_{\kappa t + 1}} \otimes \dots \otimes \ket{\psi_m}, \rho ) - \epsilon &\leq& W_{\psi}^{(\kappa t,1)} \leq W_{\psi}^{(\kappa t,1)} \leq F(\ket{\phi}\otimes \ket{\psi_{\kappa t + 1}} \otimes \dots \otimes \ket{\psi_m}, \rho ) + \epsilon, \nonumber \\
\end{eqnarray}
with probability higher than $1 - P_W^{k-i.i.d.}$, where $P_W^{k-i.i.d.}$ is given by Eq.~\ref{eqn_P_Wk i.i.d.} (for simplicity, we denote $P_W^{k-i.i.d.} =: \delta$ moving forward), as long as the number of samples
\begin{equation}
    N = \mathcal{O}\left(\frac{K^{2\kappa t}(m - \kappa t +1)^{2\kappa t c/p + 2}(\kappa t)^{2\kappa t c/p}}{\epsilon^{2\kappa t c/p+2}}\log\left(\frac{m - \kappa t +1}{\delta}\right)\right) = \tilde{\mathcal{O}}\left(\frac{K_3^{\kappa t}m^{2K_4(\kappa t)}}{\epsilon^{K_4 \kappa t}}\log\left(\frac{m - \kappa t +1}{\delta}\right)\right),
\end{equation}
where $K_3 := K^2 > 0, K_4:= 2c/p > 0$ are constants depending on the core state support size and the free parameters and where $\tilde{\mathcal{O}}$ represents ordering up to terms polynomial in $m, \kappa t$ and $\varepsilon$. Using the free parameters $p_{\kappa t}, p_{\kappa t + 1},\ldots,p_m$ and $\tau_{\kappa t}, \tau_{\kappa t + 1}\ldots,\tau_{m}$, the number of samples required to obtain the desired precision with the desired probability can be optimized.
\end{proof}

Now, to understand the precision with which we need to approximate, we note from Eq.~\ref{appeq:dope_relation_fidelity_witness} that (note that, for brevity, we write $\ket{\psi} = \ket{\phi}\otimes \ket{\psi_{\kappa t + 1}} \otimes \dots \otimes \ket{\psi_m}$)
\begin{eqnarray}
    F(\rho,\ket{\psi}) &\geq& W_{\psi}^{(\kappa t,1)} - \epsilon, \nonumber \\
    F(\rho,\ket{\psi}) &\leq& 1 + \frac{1}{m - \kappa t +1}(W_{\psi}^{(\kappa t,1)} - 1 + \epsilon).
\end{eqnarray}
Therefore, we accept the state $\rho$ if
\begin{eqnarray}
    W_{\psi} - \epsilon &>& 1 - c, \nonumber \\
    W_{\psi} &>& 1 - (c-\epsilon).
\end{eqnarray}
Note that this would imply $F(\rho,\ket{\psi}) > 1 - c$. On the other hand, we reject the state $\rho$ if
\begin{eqnarray}
    1 + \frac{1}{m-\kappa t +1}(W_{\psi} - 1 + \epsilon) &<& 1 - s, \nonumber \\
    W_{\psi} &<& 1 - ((m - \kappa t+1)s+\epsilon).
\end{eqnarray}
This, in turn, implies $F(\rho,\ket{\psi}) < 1 - s$. Choosing $\epsilon = \varepsilon/4$, $c = \varepsilon$ and $s = (m-\kappa t +1) \varepsilon$, we get the statement of Theorem \ref{apptheo:verification_doped}.

Therefore, Protocol $\ref{protocol_t_doped_Gaussian}$ allows us to verify a doped Gaussian state with heterodyne sampling using multimode fidelity witness. This certification is not possible using the witness based on single-mode fidelities ($k=1$), since the state $\ket{\phi}$ maybe possibly be entangled.
\section{Noisy heterodyne estimator} \label{noisy_hetero}
We denote by $\underset{\alpha\leftarrow D}{\mathbb E}[f(\alpha)]$ the expected value of a function $f$ for samples drawn from a distribution $D$. We write $\eta\in(0,1]$ the quantum efficiency of the heterodyne detection.  The corresponding probability distribution is related to the smoothed Husimi $Q$ function:
\begin{equation}
W(\alpha,s)=-\frac{2}{1+s}\int_{\beta\in\mathbb C}Q(\beta)e^{\frac2{1+s}|\alpha-\beta|^2}\frac{d^2\beta}\pi,
\end{equation}
for all $\alpha\in\mathbb  C$, where $s=1-2\eta^{-1}\le-1$~\cite{paris1996quantum}.  The noisy heterodyne probability distribution is then given by $\frac1{\eta}W(\frac\alpha{\sqrt{\eta}},s)$. In particular, dividing a sample of lossy heterodyne detection by $\sqrt{\eta}$ gives a complex number sampled from $W(\alpha,s)$.
Let us introduce for $k,l\ge0$ the polynomials
\be
\ba
\mathcal{L}_{k,l}(z)&:=e^{zz^*}\frac{(-1)^{k+l}}{\sqrt{k!}\sqrt{l!}}\frac{\partial^{k+l}}{\partial z^k\partial z^{*l}}e^{-zz^*}\\
&=\sum_{p=0}^{\min{(k,l)}}{\frac{\sqrt{k!}\sqrt{l!}(-1)^p}{p!(k-p)!(l-p)!}z^{l-p}z^{*k-p}},
\ea
\label{2DL}
\ee
for $z\in\mathbb C$, which are, up to a normalisation, the Laguerre $2$D polynomials. These polynomials allow us to write a developped expression for the generalised quasiprobability distribution of a state $\rho=\sum_{m,n\ge0}\rho_{mn}\ket m\!\bra n$~\cite{wunsche1998laguerre} (note that the generalised quasiprobability distributions in this reference are parametrized by a parameter $r$ which is related to our parametrization by $r=-s$): for all $s<1$ and all $\alpha\in\mathbb C$,
\be
W_\rho(\alpha,s)=\sum_{m,n\ge0}\rho_{mn}W_{\ket m\bra n}(\alpha,s),
\ee
where for all $m,n\in\mathbb N$,
\be
\ba
W_{\ket m\bra n}(\alpha,s)&=\frac2\pi\left(\frac{1+s}{1-s}\right)^{\frac{m+n}2}e^{-\frac2{1-s}|\alpha|^2}\mathcal{L}_{m,n}\left(\frac{2\alpha}{\sqrt{1-s^2}}\right)\\
&=\frac1\pi\left(\frac2{1-s}\right)^{m+n+1}e^{-\frac2{1-s}|\alpha|^2}\sum_{q=0}^{\min{(m,n)}}{\frac{\sqrt{m!}\sqrt{n!}(-1)^q}{q!(m-q)!(n-q)!}\left(\frac{1-s^2}4\right)^q\alpha^{n-q}\alpha^{*m-q}}.
\ea
\ee
Additionally,  for all $k,l\in\mathbb N$, we define with these polynomials the functions
\be
\ba
f_{k,l}(z)&:=\frac1{\eta^{k+l+1}\tau^{\frac{k+l}2+1}}e^{\left(\eta-\frac{1}{\tau}\right)zz^*}\mathcal{L}_{l,k}\left(\frac z{\sqrt{\tau}}\right)\\
&=\frac1{(\tau \eta)^{k+l+1}}e^{\left(\eta-\frac{1}{\tau}\right)zz^*}\sum_{p=0}^{\min{(k,l)}}{\frac{\sqrt{k!}\sqrt{l!}(-1)^p\tau^p}{p!(k-p)!(l-p)!}z^{k-p}z^{*l-p}},
\label{fkl}
\ea
\ee
for all $z\in\mathbb C$ and all $0<\tau<\frac1{\eta}$, which may be thought of as regularized Glauber--Sudarshan $P$ functions of Fock basis operators~\cite{pevrina1968regular,chabaud2020building}. The function $f_{k,l}$, being a polynomial multiplied by a converging Gaussian function, is bounded over $\mathbb C$.

We obtain the following result:

\begin{lem} \label{lem:f}
Let $s=1-2\eta^{-1}\le-1$, let $k,l\in\mathbb N$ and let $0<\tau<\frac1{\eta}$. Let $\rho=\sum_{m,n\ge0}{\rho_{mn}\ket m\!\bra n}$ be a density operator.
Then,
\be
\underset{\alpha\leftarrow W_\rho(\alpha,s)}{\mathbb E}[f_{k,l}]=\rho_{kl}+\sum_{j=1}^{+\infty}{\rho_{k+j,l+j}\left(1-\eta+\tau \eta^2\right)^j\sqrt{\binom{j+k}k\binom{j+l}l}}.
\label{Lemmaf}
\ee
where the function $f_{k,l}$ is defined in Eq.~(\ref{fkl}).
\end{lem}

\begin{proof}

We first prove the Lemma without caring about convergence of the sums, and justify these afterwards. We have
\be
\ba
\underset{\alpha\leftarrow W_\rho(\alpha,s)}{\mathbb E}[f_{k,l}]&=\int f_{k,l}(\alpha)W_\rho(\alpha,s)d^2\alpha\\
&=\sum_{m,n\ge0}\rho_{mn}\int f_{k,l}(\alpha)W_{\ket m\bra n}(\alpha,s)d^2\alpha\\
&=\sum_{m,n\ge0}\rho_{mn}\!\!\!\!\sum_{q=0}^{\min(m,n)}\!\!\left(\frac2{1-s}\right)^{m+n+1}\!\!\!\!\!\frac{\sqrt{m!}\sqrt{n!}(-1)^q}{q!(m-q)!(n-q)!}\left(\frac{1-s^2}4\right)^q\frac1{(\tau \eta)^{k+l+1}}\!\!\!\!\sum_{p=0}^{\min{(k,l)}}\frac{\sqrt{k!}\sqrt{l!}(-1)^p\tau^p}{p!(k-p)!(l-p)!}\\
&\quad\quad\quad\times\int\alpha^{n-q}\alpha^{*m-q}\alpha^{k-p}\alpha^{*l-p}e^{-\frac2{1-s}|\alpha|^2}e^{\left(\eta-\frac{1}{\tau}\right)|\alpha|^2}\frac{d^2\alpha}\pi.
\ea
\label{expr1}
\ee
With $s=1-2\eta^{-1}$, the integral in the last line can be computed as
\be
\int\alpha^{n-q+k-p}\alpha^{*m-q+l-p}e^{-\frac{|\alpha|^2}\tau}\frac{d^2\alpha}\pi=\left(\frac{m+n+k+l}2-p-q\right)!\tau^{\frac{m+n+k+l}2-p-q+1}\delta_{m-k,n-l}.
\ee
where $\delta$ is the Kronecker symbol, and where we used polar coordinates $d^2\alpha=rdrd\theta$ with $\int_0^{2\pi}e^{ia\theta}d\theta=2\pi\delta_{a,0}$ and $\int_{\mathbb R}r^{2N+1}e^{-\frac{r^2}\tau}dr=\frac12N!\tau^{N+1}$ for $N=\frac{m+n+k+l}2-p-q$. The expression in Eq.~(\ref{expr1}) thus reduces to:
\be
\ba
\underset{\alpha\leftarrow W_\rho(\alpha,s)}{\mathbb E}[f_{k,l}]&=\sum_{\substack{m,n\ge0\\m-k=n-l}}\rho_{mn}\!\!\!\!\sum_{q=0}^{\min(m,n)}\!\!\left(\frac2{1-s}\right)^{m+n+1}\!\!\!\!\!\frac{\sqrt{m!}\sqrt{n!}(-1)^q}{q!(m-q)!(n-q)!}\left(\frac{1-s^2}4\right)^q\frac1{(\tau \eta)^{k+l+1}}\\
&\quad\quad\quad\times\sum_{p=0}^{\min{(k,l)}}\frac{\sqrt{k!}\sqrt{l!}(-1)^p\tau^p}{p!(k-p)!(l-p)!}\left(\frac{m+n+k+l}2-p-q\right)!\tau^{\frac{m+n+k+l}2-p-q+1}\\
&=\sum_{\substack{m,n\ge0\\m-k=n-l}}\rho_{mn}\!\!\!\!\sum_{q=0}^{\min(m,n)}\!\!\left(\frac2{1-s}\right)^{m+n+1}\!\!\!\!\!\frac{\sqrt{m!}\sqrt{n!}(-1)^q}{q!(m-q)!(n-q)!}\left(\frac{1-s^2}4\right)^q\frac{\tau^{\frac{m+n-k-l}2-q}}{\eta^{k+l+1}}\\
&\quad\quad\quad\times\frac{\left(\frac{m+n+k+l}2-q\right)!}{\sqrt{k!}\sqrt{l!}}\sum_{p=0}^{\min{(k,l)}}(-1)^p\frac{\binom kp\binom lp}{\binom{\frac{m+n+k+l}2-q}p}\\
&=\sum_{\substack{m,n\ge0\\m-k=n-l}}\rho_{mn}\!\!\!\!\sum_{q=0}^{\min(m,n)}\!\!\left(\tau \eta^2\right)^{\frac{m+n-k-l}2-q}\!\!\!\!\!\frac{\sqrt{m!}\sqrt{n!}}{q!(m-q)!(n-q)!}(1-\eta)^q\\
&\quad\quad\quad\times\frac{\left(\frac{m+n+k+l}2-q\right)!}{\sqrt{k!}\sqrt{l!}}\sum_{p=0}^{\min{(k,l)}}(-1)^p\frac{\binom kp\binom lp}{\binom{\frac{m+n+k+l}2-q}p},
\ea
\label{expr2}
\ee
where we used $\frac2{1-s}=\eta$ and $\frac{1-s^2}4=-\frac{1-\eta}{\eta^2}$ in the fifth line. Now the last sum is given by (see result 7.1 of~\cite{gould1972combinatorial}):
\be
\sum_{p=0}^{\min{(k,l)}}(-1)^p\frac{\binom kp\binom lp}{\binom{\frac{m+n+k+l}2-q}p}=\begin{cases}\frac{\left(\frac{m+n+k-l}2-q\right)!\left(\frac{m+n-k+l}2-q\right)!}{\left(\frac{m+n+k+l}2-q\right)!\left(\frac{m+n-k-l}2-q\right)!}&\text{ for }\frac{m+n+k+l}2-q\ge k+l,\\&\\0&\text{ for }\frac{m+n+k+l}2-q<k+l.\end{cases}
\ee
In particular, with the condition $m-k=n-l$, having $q\ge0$ implies $m\ge k$ and $n\ge l$. We thus obtain
\be
\ba
\underset{\alpha\leftarrow W_\rho(\alpha,s)}{\mathbb E}[f_{k,l}]&=\!\!\!\!\!\sum_{\substack{m\ge k,n\ge l\\m-k=n-l}}\rho_{mn}\!\!\!\!\sum_{q=0}^{m-k=n-l}\!\!\!\!\left(\tau \eta^2\right)^{\frac{m+n-k-l}2-q}\!\!\!\frac{\sqrt{m!}\sqrt{n!}}{q!(m-q)!(n-q)!}(1-\eta)^q\frac{\left(\frac{m+n+k-l}2-q\right)!\left(\frac{m+n-k+l}2-q\right)!}{\sqrt{k!}\sqrt{l!}\left(\frac{m+n-k-l}2-q\right)!}\\
&=\sum_{j\ge0}\rho_{j+k,j+l}\sum_{q=0}^{j}\!\!\left(\tau \eta^2\right)^{j-q}\!\!\!\!\!\frac{\sqrt{(j+k)!}\sqrt{(j+l)!}}{q!(j+k-q)!(j+l-q)!}(1-\eta)^q\frac{\left(j+k-q\right)!\left(j+l-q\right)!}{\sqrt{k!}\sqrt{l!}\left(j-q\right)!}\\
&=\sum_{j\ge0}\rho_{j+k,j+l}\left(\tau \eta^2\right)^j\sqrt{\binom{j+k}k\binom{j+l}l}\sum_{q=0}^j\binom jq\left(\frac{1-\eta}{\tau \eta^2}\right)^q\\
&=\sum_{j\ge0}\rho_{j+k,j+l}\left(1-\eta+\tau \eta^2\right)^j\sqrt{\binom{j+k}k\binom{j+l}l}\\
&=\rho_{k,l}+\sum_{j\ge1}\rho_{j+k,j+l}\left(1-\eta+\tau \eta^2\right)^j\sqrt{\binom{j+k}k\binom{j+l}l},
\ea
\label{expr3}
\ee
where we have set $j=m-k=n-l$ in the second line.

We now consider the convergence of the above expressions. The interversion of sums and integral in Eq.~(\ref{expr1}) is valid if the series in Eq.~(\ref{expr3}) converges. Let $t:=1-\eta+\tau \eta^2$. Since $0<\eta\le1$ and $0<\tau<\frac1{\eta}$, we have $0< t<1$. The term $\sqrt{\binom{j+k}k\binom{j+l}l}$ is bounded by a polynomial in $j$, so the function $j\mapsto t^j\sqrt{\binom{j+k}k\binom{j+l}l}$ is bounded over $\mathbb N$. Writing $M$ its maximum, we obtain for all $N\in\mathbb N$:
\begin{equation}
\ba
\left|\sum_{j=0}^N\rho_{j+k,j+l}\left(1-\eta+\tau \eta^2\right)^j\sqrt{\binom{j+k}k\binom{j+l}l}\right|&\le M\sum_{j=0}^N|\rho_{j+k,j+l}|\\
&\le M\sum_{j=0}^N\sqrt{\rho_{j+k,j+k}}\sqrt{\rho_{j+l,j+l}}\\
&\le M\sqrt{\sum_{j=0}^N\rho_{j+k,j+k}\sum_{j=0}^N\rho_{j+l,j+l}}\\
&\le M,
\ea
\end{equation}
where we used the fact that $\rho$ is positive semidefinite in the second line, Cauchy-Schwartz inequality in the third line and $\Tr(\rho)=1$ in the last line. Taking $N\rightarrow+\infty$ completes the proof.

\end{proof}

\noindent We may thus use the bounded functions $f_{k,l}$ to estimate density matrix elements using noisy heterodyne detection, with an error quantified by Lemma~\ref{lem:f}. Noticing that the error term for the coefficient $(k,l)$ only depends on higher coefficients,  we take linear combinations of these estimators in order to ``push'' the error to higher coefficients.  Namely, for $0<\tau<\frac1{\eta}$, writing $t=1-\eta+\tau \eta^2$, we define for all $p\in\mathbb N^*$ and all $z\in\mathbb C$:
\be
g_{k,l}^{p(het)}(z,\eta,\tau):=\sum_{j=0}^{p-1}{(-1)^jf_{k+j,l+j}(z)t^j\sqrt{\binom{j+k}k\binom{j+l}l}}.
\label{gkl}
\ee
We obtain the following result:

\begin{lem} \label{lem:g}
Let $s=1-2\eta^{-1}\le-1$, let $k,l\in\mathbb N$, let  $p\in\mathbb N^*$, let $0<\tau<\frac1{\eta}$ and let $t=1-\eta+\tau \eta^2$. Let $\rho=\sum_{m,n\ge0}{\rho_{mn}\ket m\!\bra n}$ be a density operator.
Then,
\be
\underset{\alpha\leftarrow W_\rho(\alpha,s)}{\mathbb E}[g_{k,l}^{(p)}]=\rho_{kl}+(-1)^{p+1}\sum_{q=p}^{+\infty}{\rho_{k+q,l+q}t^q\binom{q-1}{p-1}\sqrt{\binom{k+q}k\binom{l+q}l}}.
\label{Lemmag}
\ee
where the function $g_{k,l}$ is defined in Eq.~(\ref{gkl}).
\end{lem}

\noindent The proof is identical to the case $\eta=1$, by replacing $\tau$ with $t=1-\eta+\tau \eta^2$ and $Q_\rho(\alpha)$ by $W_\rho(\alpha,s)$, with $s=1-2\eta^{-1}$. We reproduce it below for completeness:

\begin{proof}

We prove the lemma by induction on $p\in\mathbb N^*$.
By Lemma~\ref{lem:f} we have, for all single-mode (mixed) state $\rho=\sum_{i,j\ge0}{\rho_{ij}\ket i\!\bra j}$, for all $k,l\in\mathbb N$
\be
\underset{\alpha\leftarrow W_\rho(\alpha,s)}{\mathbb E}[f_{k,l}]=\rho_{kl}+\sum_{q=1}^{+\infty}{\rho_{k+q,l+q}t^q\sqrt{\binom{k+q}k\binom{l+q}l}}.
\label{Lemma3}
\ee
With Eq.~(\ref{gkl}), this gives
\be
\underset{\alpha\leftarrow W_\rho(\alpha,s)}{\mathbb E}[g_{k,l}^{(1)}]=\rho_{kl}+\sum_{q=1}^{+\infty}{\rho_{k+q,l+q}t^q\sqrt{\binom{k+q}k\binom{l+q}l}},
\ee
hence the lemma is proven for $p=1$ since $\binom{q-1}0=1$. 

Assuming the lemma holds for some $p\in\mathbb N^*$,
\be
\ba
\underset{\alpha\leftarrow W_\rho(\alpha,s)}{\mathbb E}[g_{k,l}^{(p)}]&=\rho_{kl}+(-1)^{p+1}\sum_{q=p}^{+\infty}{\rho_{k+q,l+q}t^q\binom{q-1}{p-1}\sqrt{\binom{k+q}k\binom{l+q}l}}\\
&=\rho_{kl}+(-1)^{p+1}\rho_{k+p,l+p}t^p\sqrt{\binom{k+p}k\binom{l+p}l}\\
&\quad\quad\quad+(-1)^{p+1}\sum_{q=p+1}^{+\infty}{\rho_{k+q,l+q}t^q\binom{q-1}{p-1}\sqrt{\binom{k+q}k\binom{l+q}l}}.
\ea
\ee
Now by Eq.~(\ref{Lemma3}), with $k=k+p$ and $l=l+p$, we have
\be
\underset{\alpha\leftarrow W_\rho(\alpha,s)}{\mathbb E}[f_{k+p,l+p}]=\rho_{k+p,l+p}+\sum_{q=1}^{+\infty}{\rho_{k+q+p,l+q+p}t^q\sqrt{\binom{k+q+p}{k+p}\binom{l+q+p}{l+q}}},
\ee
so that 
\be
\ba
\underset{\alpha\leftarrow W_\rho(\alpha,s)}{\mathbb E}\left[g_{k,l}^{(p+1)}\right]&=\underset{\alpha\leftarrow W_\rho(\alpha,s)}{\mathbb E}\left[g_{k,l}^{(p)}+(-1)^pt^p\sqrt{\binom{k+p}k\binom{l+p}l}f_{k+p,l+p}\right]\\
&=\rho_{kl}+(-1)^{p+1}\rho_{k+p,l+p}t^p\sqrt{\binom{k+p}k\binom{l+p}l}\\
&\quad\quad\quad+(-1)^{p+1}\sum_{q=p+1}^{+\infty}{\rho_{k+q,l+q}t^q\binom{q-1}{p-1}\sqrt{\binom{k+q}k\binom{l+q}l}}\\
&\quad\quad\quad+(-1)^pt^p\sqrt{\binom{k+p}k\binom{l+p}l}\rho_{k+p,l+p}\\
&\quad\quad\quad+(-1)^pt^p\sqrt{\binom{k+p}k\binom{l+p}l}\sum_{q=1}^{+\infty}{\rho_{k+q+p,l+q+p}t^q\sqrt{\binom{k+q+p}{k+p}\binom{l+q+p}{l+p}}}\\
&=\rho_{kl}+(-1)^{p+1}\sum_{q=p+1}^{+\infty}{\rho_{k+q,l+q}t^q\binom{q-1}{p-1}\sqrt{\binom{k+q}k\binom{l+q}l}}\\
&\quad\quad\quad+(-1)^pt^p\sqrt{\binom{k+p}k\binom{l+p}l}\sum_{q=1}^{+\infty}{\rho_{k+q+p,l+q+p}t^q\sqrt{\binom{k+q+p}{k+p}\binom{l+q+p}{l+p}}}\\
&=\rho_{kl}+(-1)^{p+1}\sum_{q=p+1}^{+\infty}{\rho_{k+q,l+q}t^q\binom{q-1}{p-1}\sqrt{\binom{k+q}k\binom{l+q}l}}\\
&\quad\quad\quad+(-1)^p\sqrt{\binom{k+p}k\binom{l+p}l}\sum_{q=p+1}^{+\infty}{\rho_{k+q,l+q}t^q\sqrt{\binom{k+q}{k+p}\binom{l+q}{l+p}}}\\
&=\rho_{kl}+(-1)^{p+2}\sum_{q=p+1}^{+\infty}\rho_{k+q,l+q}t^q\\
&\quad\quad\quad\times\left[\sqrt{\binom{k+p}k\binom{l+p}l\binom{k+q}{k+p}\binom{l+q}{l+p}}-\binom{q-1}{p-1}\sqrt{\binom{k+q}k\binom{l+q}l}\right].
\ea
\label{inductiong}
\ee
We have
\be
\ba
\binom{k+p}k\binom{l+p}l\binom{k+q}{k+p}\binom{l+q}{l+p}&=\frac{(k+p)!(l+p)!(k+q)!(l+q)!}{k!p!l!p!(k+p)!(q-p)!(l+p)!(q-p)!}\\
&=\frac{(k+q)!(l+q)!}{k!(k+p)!p!^2(q-p)!^2}\\
&=\binom qp^2\binom{k+q}k\binom{l+q}l,
\ea
\ee
so the last expression in Eq.~(\ref{inductiong}) simplifies as
\be
\ba
\underset{\alpha\leftarrow W_\rho(\alpha,s)}{\mathbb E}\left[g_{k,l}^{(p+1)}\right]&=\rho_{kl}+(-1)^{p+2}\sum_{q=p+1}^{+\infty}\rho_{k+q,l+q}t^q\sqrt{\binom{k+q}k\binom{l+q}l}\left[\binom qp-\binom{q-1}{p-1}\right]\\
&=\rho_{kl}+(-1)^{p+2}\sum_{q=p+1}^{+\infty}\rho_{k+q,l+q}t^q\binom{q-1}p\sqrt{\binom{k+q}k\binom{l+q}l},
\ea
\ee
which completes the proof. Note that the convergence of the series in the right hand side directly follows from the convergence of the series in the error term of Lemma~\ref{lem:f}, since it is obtained as the sum of $p$ such series.

\end{proof}

\noindent The practical differences between the cases $\eta=1$ and $\eta<1$ stem from two points:

\begin{itemize}
\item the parameter $0<\tau<1$ is replaced by $t=1-\eta+\tau \eta^2$, with $1-\eta<t<1$, so the optimisation region for this parameter is smaller;
\item the definition of the functions $f$ and $g$ are modified, which affects their range.
\end{itemize}

\noindent For the functions $f$, this last point can be made more precise: writing explicitly the dependence in $\tau$, we obtain with Eq.~(\ref{fkl}):
\be
f^{\text{noisy}}_{k,l}\left(\frac z{\sqrt{\eta}},\frac\tau{\eta}\right)=\frac1{\eta^{\frac{k+l}2}}f^{\text{ideal}}_{k,l}(z,\tau),
\ee
which shows that the overhead due to the imperfect detection is given by a factor $\frac1{\eta^{\frac{k+l}2}}$ in the range of the estimator, and thus with Hoeffding inequality by a factor $\frac1{\eta^{k+l}}$ in the number of samples needed for the same confidence interval. In practice this only gives a rough estimate since the optimised parameter is $t$ and not $\tau$, and since the functions used are the $g$ functions and not the $f$ functions. In any case, we expect the number of samples needed to increase as the quantum efficiency decreases, since we are seeking the same information from a more noisy distribution.


An immediate application of Lemma~\ref{lem:g} gives
\be
\ba
\left|\underset{\alpha\leftarrow W_\rho(\alpha,s)}{\mathbb E}[g_{k,l}^{(p)}]-\rho_{kl}\right|&\le\sum_{q=p}^{+\infty}{|\rho_{k+q,l+q}|t^q\binom{q-1}{p-1}\sqrt{\binom{k+q}k\binom{l+q}l}}\\
&\le\max_{q\ge p}\left(t^q\binom{q-1}{p-1}\sqrt{\binom{k+q}k\binom{l+q}l}\right)\sum_{q=p}^{+\infty}{|\rho_{k+q,l+q}|}\\
&\le\max_{q\ge p}\left(t^q\binom{q-1}{p-1}\sqrt{\binom{k+q}k\binom{l+q}l}\right)\sum_{q=p}^{+\infty}{\sqrt{\rho_{k+q,k+q}}\sqrt{\rho_{l+q,l+q}}}\\
&\le\max_{q\ge p}\left(t^q\binom{q-1}{p-1}\sqrt{\binom{k+q}k\binom{l+q}l}\right)\sqrt{\sum_{q=p}^{+\infty}{\rho_{k+q,k+q}}\sum_{q=p}^{+\infty}{\rho_{l+q,l+q}}}\\
&\le\max_{q\ge p}\left(t^q\binom{q-1}{p-1}\sqrt{\binom{k+q}k\binom{l+q}l}\right),
\ea
\label{inductiong1}
\ee
where we used the fact that $\rho$ is positive semidefinite in the third line, Cauchy--Schwarz inequality in the fourth line, and $\Tr(\rho)=1$ in the last line. This last bound is tight: if $p_{k,l}\ge p$ is the integer achieving the maximum in the last equation, then the inequality is an equality for $\rho=\ket{p_{k,l}}\!\bra{p_{k,l}}$. We have
\be
\ba
t^{q+1}\binom q{p-1}\sqrt{\binom{k+q+1}k\binom{l+q+1}l}&\lesseqgtr t^q\binom{q-1}{p-1}\sqrt{\binom{k+q}k\binom{l+q}l}\quad\quad\quad\quad\quad\quad\\
\Leftrightarrow\quad t&\lesseqgtr\frac{q-p+1}q\sqrt{\left(\frac{q+1}{k+q+1}\right)\left(\frac{q+1}{l+q+1}\right)}\\
\Leftrightarrow\quad t&\lesseqgtr\left(1-\frac{p-1}q\right)\sqrt{1-\frac k{k+q+1}}\sqrt{1-\frac l{l+q+1}},
\ea
\ee
where the right hand side in the last line is an increasing function of $q$.  Hence, the function $q\mapsto t^q\binom{q-1}{p-1}\sqrt{\binom{k+q}k\binom{l+q}l}$ is monotonically increasing to its maximum and then monotonically decreasing, and its maximum is attained for the first integer $p_{k,l}$ greater or equal to $p$ such that such that $t\le\left(1-\frac{p-1}{p_{k,l}}\right)\sqrt{1-\frac k{k+p_{k,l}+1}}\sqrt{1-\frac l{l+p_{k,l}+1}}$. The bound above then gives
\be
\left|\underset{\alpha\leftarrow W_\rho(\alpha,s)}{\mathbb E}[g_{k,l}^{(p)}]-\rho_{kl}\right|\le t^{p_{k,l}}\binom{p_{k,l}-1}{p-1}\sqrt{\binom{k+p_{k,l}}k\binom{l+p_{k,l}}l}.
\label{betterboundEgkl}
\ee
\section{Verification of bosonic quantum states using characteristic function}\label{sec:characteristic_function}

In this section, we explain how our verification framework can be adapted for measurements of the characteristic function.  The main difference with the protocols presented in the main text is that for characteristic function measurements, we backpropagate expectation values of measurement outcomes rather than POVM elements. The corresponding verification protocol is detailed in section \ref{sec:protocol}.

\subsection{From multimode to single-mode fidelities}

For $i=1,\dots,m$, we denote the reduced state of the $i^{th}$ mode of an $m$-mode state $\sigma$ as $\sigma_i:=\Tr_{\{1,\dots,m\}\setminus\{i\}}[\sigma]$. 

Using the single-mode fidelity witness (Eq. (\ref{eq3.1}) for $k=1$) $W^{(1)}(\sigma,\psi) = 1 - \sum_{i=1}^m (1 - F(\sigma_i,\ket{\psi_i}))$, recall that it is sufficient to estimate the single-mode fidelities $F(\sigma_i,\ket{\psi_i})$ to obtain a witness for the $m$-mode fidelity between $\sigma$ and $\bigotimes_{i=1}^m\ket{\psi_i}$. Moreover, this witness is robust to errors in the tested state \cite{Chabaud2021efficient}.

\subsection{Undoing the entanglement}

Let $\ket{\psi}=\hat G\bigotimes_{i=1}^m\ket{\psi_i}$ be a target state, with $\hat G$ an $m$-mode Gaussian unitary with symplectic matrix and displacement vector $(S,\bm d)$. For any $m$-mode density matrix $\rho$,
\begin{equation}\label{eq:unentanglement}
    \begin{aligned}
    F(\rho,\ket{\psi})&=\Tr[\rho\ket{\psi}\!\bra{\psi}]\\
    &=\Tr\left[\rho\,\hat G\bigotimes_{i=1}^m\ket{\psi_i}\!\bra{\psi_i}\hat G^\dag\right]\\
    &=\Tr\left[\hat G^\dag\rho\hat G\,\bigotimes_{i=1}^m\ket{\psi_i}\!\bra{\psi_i}\right].\\
    &=F\left(\hat G^\dag\rho\hat G,\bigotimes_{i=1}^m\ket{\psi_i}\!\bra{\psi_i}\right).
    \end{aligned}
\end{equation}

Therefore, we can obtain a witness for the $m$-mode fidelity between $\rho$ and $\hat G\bigotimes_{i=1}^m\ket{\psi_i}$ by estimating the single-mode fidelities $F((\hat G^\dag\rho\hat G)_i,\ket{\psi_i})$ between the single-mode reduced states of $\hat G^\dag\rho\hat G$ and the corresponding single-mode states $\ket{\psi_i}$, namely:
\begin{equation}\label{eq:fid_witness_final}
    F(\rho,\psi)\ge1-\sum_{i=1}^m(1-F((\hat G^\dag\rho\hat G)_i,\ket{\psi_i})).
\end{equation}

\subsection{Estimates of single-mode fidelities from estimates of the characteristic function}

Measurements of the characteristic function allow us to estimate the value of its real and imaginary part separately (as e.g.\ in \cite{tian2023direct}), at any chosen phase-space point $\bm\alpha$. As the modulus of the characteristic function is bounded by $1$, the reconstruction is very efficient. In particular, Hoeffding's inequality ensures that an estimate can be obtained with additive precision $\epsilon$ and failure probability $\delta$, using a number of samples $N_Q:=\frac2{\epsilon^2}\log(\frac2\delta)$. 
In what follows, we assume that we are able to estimate the characteristic function of the state $\rho$ at any chosen point $\bm\alpha$.

Our goal is then to obtain an estimate of the single-mode fidelities between the single-mode reduced states of $\hat G^\dag\rho\hat G$ and the corresponding single-mode states $\ket{\psi_i}$. Writing the symplectic matrix and the displacement vector of the Gaussian unitary $\hat G$ as $S=(s_{kl})_{1\le k,l\le m}$ and $\bm d=(d_1,\dots,d_m)$, we obtain, for all $i=1,\dots,m$,
\begin{equation}\label{eq:from_char_to_fid}
    \begin{aligned}
        F((\hat G^\dag\rho\hat G)_i,\ket{\psi_i})&=\int_{\mathbb C}\frac{d^2\alpha}\pi\chi_{(\hat G^\dag\rho\hat G)_i}(\alpha)\chi_{\ket{\psi_i}}(-\alpha)\\
        &=\int_{\mathbb C}\frac{d^2\alpha}\pi\chi_{\hat G^\dag\rho\hat G}(0,\dots,0,\alpha,0,\dots,0)\chi_{\ket{\psi_i}}(-\alpha)\\
        &=\int_{\mathbb C}\frac{d^2\alpha}\pi e^{\alpha d_i^*-\alpha^*d_i}\chi_{\rho}(s_{i1}^*\alpha-s_{i+m\,1}\alpha^*,\dots,s_{im}^*\alpha-s_{i+m\,m}\alpha^*)\chi_{\ket{\psi_i}}(-\alpha)\\
        &\!\!\!\!\!\!\!\!\!\!\!\!\!\!\!\!\!\!\!\!\!\!\!\!\!\!\!\!\!\!\!\!\!\!\!\!\!\!\!\!\!\!\!\!\!\!\!=\int_{\mathbb C}\!\!\!\underset{\text{probability distribution}}{\underbrace{\frac{|\chi_{\ket{\psi_i}}(\alpha)|}{\|\chi_{\ket{\psi_i}}\|_1}\frac{d^2\alpha}\pi}}\!\underset{\text{estimator}}{\underbrace{\chi_{\rho}(s_{i+m\,1}\alpha^*-s_{i1}^*\alpha,\dots,s_{i+m\,m}\alpha^*-s_{im}^*\alpha)\|\chi_{\ket{\psi_i}}\|_1\frac{\chi_{\ket{\psi_i}}(\alpha)}{|\chi_{\ket{\psi_i}}(\alpha)|}e^{\alpha^*d_i-\alpha d_i^*}}},
    \end{aligned}
\end{equation}
where $\|\chi\|_1:=\int_{\mathbb C}\frac{d^2\alpha}\pi|\chi(\alpha)|$, and where we used the overlap formula 
\begin{equation}\label{eq:overlap}
    \Tr[\rho\sigma]=\int_{\mathbb C^m}\frac{d^{2m}\bm\alpha}{\pi^m}\chi_{\rho}(\bm\alpha)\chi_{\sigma}(-\bm\alpha).
\end{equation}
in the first line, the formula for the characteristic function of a reduced state
\begin{equation}\label{eq:char_reduced}
    \begin{aligned}
        \chi_{\Tr_{2\dots m}[\rho]}(\alpha)&=\Tr[\Tr_{2\dots m}[\rho]\hat D(\alpha)]\\
        &=\Tr[\rho\hat D(\alpha)\otimes\mathbb I\otimes\cdots\otimes\mathbb I]\\
        &=\Tr[\rho\hat D(\alpha)\otimes\hat D(0)\otimes\cdots\otimes\hat D(0)]\\
        &=\chi_{\rho}(\alpha,0,\dots,0).
    \end{aligned}
\end{equation}
in the second line, and the evolution of the characteristic function under a Gaussian unitary operation 
\begin{equation}\label{eq:Gaussian_on_char}
    \chi_{\hat G^\dag\rho\hat G}(\bm\alpha)=e^{\bm\alpha\cdot\bm d^*-\bm\alpha^*\cdot\bm d}\chi_{\rho}\!\left(\bm\alpha^S\right).
\end{equation}
in the third line, where $\bm\alpha^S:=(\alpha^S_1,\dots,\alpha^S_m)$ with
\begin{equation}\label{eq:alphaS}
    \alpha^S_l:=\sum_{k=1}^m\alpha_ks_{kl}^*-\alpha_k^*s_{k+m\,l}.
\end{equation}

The expression in Eq.~(\ref{eq:from_char_to_fid}) provides a statistical method for estimating the single-mode fidelities, by classically sampling from the probability distribution and averaging the estimator over the measurement outcomes. In practice, the estimator depends on values of the characteristic function of $\rho$ at the sampled points, which can be themselves estimated by measurements of the characteristic function of $\rho$. If those estimates are precise up to $\epsilon$, this translates to estimates of the fidelity that are precise up to $\|\chi_{\ket{\psi_i}}\|_1\epsilon$, since the term $\frac{\chi_{\ket{\psi_i}}(\alpha)}{|\chi_{\ket{\psi_i}}(\alpha)|}e^{\alpha^*d_i-\alpha d_i^*}$ is simply a phase factor.

\subsection{The verification protocol}
\label{sec:protocol}

We now put everything together. Let us recall a few notations: $m$ denotes the number of modes, $\rho$ the measured state, and $\ket{\psi}=\hat G\bigotimes_{i=1}^m\ket{\psi_i}$ the target state, where $\hat G$ is a Gaussian unitary described by its symplectic matrix and displacement vector $(S=(s_{kl})_{1\le k,l\le2m},\bm d=(d_1,\dots,d_m))$, and where $\ket{\psi_i}$ are single-mode pure states such that $\|\chi_{\psi_i}\|_1$ is bounded\footnote{For instance, this is the case for any finite superposition of Fock states; for a Fock state $\ket n$, this quantity scales as $\sqrt n$ \cite{fawzi2024optimal}.}. We also introduce $N_i^C,N_i^Q\in\mathbb N$ for $i=1,\dots,m$, which denote numbers of samples used in the protocol, and define the probability distribution
\begin{equation}
    p_{\psi_i}(\alpha):=\frac{|\chi_{\psi_i}(\alpha)|}{\|\chi_{\psi_i}\|_1},
\end{equation}
where $\|\chi\|_1:=\int_{\mathbb C}\frac{d^2\alpha}\pi|\chi(\alpha)|$, and the estimator
\begin{equation}
    f_{\psi_i}(\alpha,\kappa):=\kappa\,\|\chi_{\psi_i}\|_1\frac{\chi_{\psi_i}(\alpha)}{|\chi_{\psi_i}(\alpha)|}e^{\alpha^*d_i-\alpha d_i^*}.
\end{equation}
With these notations introduced, here is a step-by-step description of the verification protocol:
\begin{enumerate}
    \item For $i=1,\dots,m$:
    \begin{enumerate}
        \item for $j=1,\dots,N_i^C$:
        \begin{enumerate}
            \item Sample a complex number $\alpha_{ij}$ from the probability distribution $p_{\psi_i}$
            \item Compute an estimate $\kappa_{ij}$ of the characteristic function of $\rho$ at the point $(s_{i+m\,1}\alpha_{ij}^*-s_{i1}^*\alpha_{ij},\dots,s_{i+m\,m}\alpha_{ij}^*-s_{im}^*\alpha_{ij})$ using $N_i^Q$ measurement samples
        \end{enumerate}
        \item Compute $F_i:=\frac1{N_i^Q}\sum_{j=1}^{N_i^Q}f_{\psi_i}(\alpha_{ij},\kappa_{ij})$
    \end{enumerate}
    \item Compute $W:=1-\sum_{i=1}^m(1-F_i)$
\end{enumerate}

Note that the protocol combines a classical sampling step (sampling the right phase-space point) with a quantum sampling step (measuring the characteristic function at that point). The analysis from the previous sections (see Eqs.~(\ref{eq:fid_witness_final}) and~(\ref{eq:from_char_to_fid})) combined with Hoeffding's bound and the union bound ensures that the output $W$ of the verification protocol satisfies $W-\epsilon\le F(\rho,\ket{\psi})$ with failure probability at most $\delta$ as long as the total number of samples (classical and quantum) satisfies
\begin{equation}
    N=\Omega\left(\frac{m^2}{\epsilon^2}\log\left(\frac m\delta\right)\right),
\end{equation}
where the precise constants depend on $\|\chi_{\psi_i}\|_1$ for $i=1,\dots,m$.

\end{document}